%% file: ti_ff_main.tex
\documentclass[12pt,a4paper,notitlepage,fleqn]{article}

\usepackage{setspace}
\usepackage[margin=1in]{geometry}
\usepackage[symbol]{footmisc} 

\usepackage[T1]{fontenc}
\usepackage[utf8]{inputenc}
\usepackage{authblk}

\makeatletter

\newcommand{\fboxsubsec}[1]{
	\begin{flushleft}
		#1
	\end{flushleft}
	}
\renewcommand{\subsection}{\@startsection{subsection}{2}{0pt}
	{1ex}
	{0.5ex}
	{\reset@font\it\fboxsubsec}
	}
\makeatother

\title{Time Instability of the Fama-French Multifactor Models: An International Evidence}

\author{Koichiro Moriya$^{a}$\thanks{\scriptsize Corresponding Author. E-mail: moriya.koichiro@keio.jp, Tel. +81-3-5427-1517.} \ and \ Akihiko Noda$^{b,c}$

{\scriptsize ${}^{a}$ \it Graduate School of Economics, Keio University, 2-15-45 Mita, Minato-ku, Tokyo 108-8345, Japan} 

{\scriptsize ${}^{b}$ \it School of Commerce, Meiji University, 1-1 Kanda-Surugadai, Chiyoda-ku, Tokyo 101-8301, Japan}

{\scriptsize ${}^{c}$ \it Keio Economic Observatory, Keio University, 2-15-45 Mita, Minato-ku, Tokyo 108-8345, Japan}}

\date{This Version: \today}

\renewcommand\thefootnote{\arabic{footnote}}

\usepackage{graphicx}

\usepackage[]{natbib}%
\usepackage{amsmath}%
\usepackage{amssymb}%
\usepackage{ascmac}%
\usepackage{multirow}%
\usepackage{pdflscape}%
\usepackage{subfigmat}

\usepackage{pifont}%
\usepackage{arydshln}%
\usepackage[format=hang]{caption}
\usepackage[all]{xy}
\usepackage{url}

\PassOptionsToPackage{hyphens}{url}
\usepackage{color}
\usepackage[bookmarkstype=toc,
	    colorlinks=true, 
	    pdfborder={0 0 1}, 
	    bookmarks=true, 
	    bookmarksnumbered=true, 
	    citecolor=blue]{hyperref}
\bibpunct{(}{)}{;}{a}{}{,}

\def\hsymbu#1{\smash{\lower1.7ex\hbox{\huge$#1$}}}

\def\ve #1{{\mbox{\boldmath $#1$}}}

\newcommand{\citetapos}[1]{\citeauthor{#1}'s \citeyearpar{#1}}
\newcommand{\citeapos}[2]{\citeauthor{#1}'s (\citeyear{#2})}
\newcommand{\ex}{{\mathbb{E}}}

\def\ve #1{{\mbox{\boldmath $#1$}}}

\begin{document}

\begin{titlepage}

\renewcommand{\thepage}{}
\renewcommand{\thefootnote}{\fnsymbol{footnote}}

\maketitle

\vspace{-10mm}

\noindent
\hrulefill

\noindent
{\bfseries Abstract:} This paper investigates the time-varying structure of \citeapos{fama1993crf}{fama1993crf,fama2015ffa} multi-factor models using \citetapos{fama1973rre} two-step estimation based on the rolling window method. In particular, we employ the generalized GRS statistics proposed by \citet{kamstra2024tra} to examine whether the validity of the risk factors (or factor redundancy) in the FF3 and FF5 models remains stable over time, and investigate whether the manner of portfolio sorting affects the time stability of the validity of the risk factors. In addition, we examine whether the similar results are obtained even when we use different datasets by country and region. First, we find that the effectiveness of factors in the FF3 and FF5 models is not stable over time in all countries. Second, the effectiveness of factors is also affected by the manner of portfolio sorting. Third, the validity of the FF3, FF5, and their nested models do not remain stable over time except for Japan. This suggests that the efficient market hypothesis is supported in the Japanese stock market. Finally, the factor redundancy varies over time and is affected by the manner of portfolio sorting mainly in the U.S. and Europe. \\



\noindent
\hrulefill

\end{titlepage}

\bibliographystyle{asa}


\input{ti_ff_intro}

\input{ti_ff_model}

\input{ti_ff_data}

\input{ti_ff_empirical}

\input{ti_ff_conclusion}

\input{ti_ff_ack}

\input{ti_ff_main.bbl}
\input{ti_ff_table}

\end{document}

%% file: ti_ff_intro.tex
\section{Introduction}\label{sec:tv_cb_intro}
Financial economists have recognized \citetapos{fama1993crf} three-factor (FF3) model as a {\it{de facto}} standard framework for explaining cross-sectional equity returns. Many previous studies show that the FF3 model is effective in explaining stock returns, not only in the U.S. but also in other developed countries and regions (i.e., \citet{fama1993crf}; \citet{jagannathan1998rlr}; \citet{arshanapalli1998map}). However, some studies point out the existence of new empirical regularities, such as the profitability premium (\citet{novy2013osv}) and the investment premium (\citet{titman2004cis}; \citet{anderson2006eec}). Subsequently, \citet{fama2015ffa} propose a five-factor (FF5) model that adds two risk factors representing the new empirical regularities to the FF3 model. \citet{fama2017itf} examine whether the FF5 model is appropriate for explaining stock returns in 23 developed markets grouped into four regions (North America, Europe, Japan, and Asia-Pacific). They argue that the FF5 model works well to explain the stock returns in the regions. However, some literature has obtained opposite results to \citet{fama2017itf}, in that the FF5 model is not effective for explaining the stock returns in developed countries. \citet{foye2018tav} shows that the FF3 and FF5 models are unable to offer a convincing description of U.K. stock returns, while \citet{kubota2018dff} conclude that the FF5 is not the best benchmark pricing model of Japanese stock returns. Therefore, the effectiveness of Fama--French multi-factor models is controversial. 

To clarify the cause of this issue, some studies focus on the following two topics: the the manner of the portfolio sorting and factor redundancy. \citet{lewellen2010asa} point out that many studies have used portfolios sorted by size and book-to-market (Size-B/M) when testing the FF3 model, but that the FF3 model may not be valid when using portfolios sorted in the different manners. In response to \citet{lewellen2010asa}, \citet{harvey2021lf} propose a new statistical inference to test the validity of the risk factors in the FF3 and FF5 models, even when individual stocks are used instead of portfolios sorted by Size-B/M. As a result, they find that the validity of the risk factors in the models depends strongly on the differences in portfolio sorting. There are two important studies on the factor redundancy. \citet{fama2015ffa,fama2016daf} show that the value factor may be redundant in the FF5 model based on \citetapos{gibbons1989teg} (hereafter referred to as the GRS) statistics. Indeed, \citet{linnainmma2018hcs} find that the value premium does not appear after 1991 and the investment premium does not appear throughout the sample period. Furthermore, \citet{fama2021vp} revisit the existence of the value premium and conclude that it may not emerge over time. This implies that that the validity of the FF3 and FF5 models may also change over time.


In practice, some studies focus on whether the FF3 and FF5 models work as a benchmark over time. \citet{takehara2018tve} explores the time-stability of the FF3 model using \citetapos{fama1973rre} two-step regression based on the rolling-window method. He finds that the parameters of the FF3 model change over time in Japan; in particular, the value premium gradually disappears over time. \citet{horvath2021eff} estimate the time-varying parameters on the FF5 model using a rolling GMM estimation to evaluate the performance of the FF5 model on the U.S. stock market. They find significant impacts on the effectiveness of the model for the periods of the dot-com bubble in the early 2000s, the global financial crisis in 2008, and the latest COVID-19 outbreaks, although the estimates are not significant over time in the GMM model. Following on discussions in previous studies, we thus address two research questions. Firstly, we examine whether the validity of the risk factors (or factor redundancy) in the FF3 and FF5 models remains stable over time. If the validity is not stable over time, then the validity of the FF3 and FF5 models and factor redundancy may also be unstable over time. Secondly, we investigate whether the manner of portfolio sorting affects the time stability of the validity of the risk factors. If the validity and its stability over time differ depending on the manner of portfolio sorting, then the validity of the FF3 and FF5 models themselves and the stability of factor redundancy over time may also depend on the manner of portfolio sorting. In addition, we examine whether the similar results are obtained even when we use different datasets by country and region. If the FF3 and FF5 models are robust, the conclusions should be consistent across countries and regions.

Specifically, we first estimate the time-varying risk premium in the the FF3 and FF5 models using \citetapos{fama1973rre} two-step estimation based on the rolling window method. We consider the validity of the factors in the model to be compromised for periods when each risk premium is not statistically significantly different from zero. We then rank the models based on the generalized GRS statistics proposed by \citet{kamstra2024tra}. This allows us to examine the time stability of the validity of the FF3 and FF5 models and the factor redundancy, as well as whether they depend on the manner of portfolio sorting. The rest of this paper is organized as follows. Section 2 presents our methodology. Section 3 describes the data used in this paper. Section 4 shows our empirical results. Section 5 concludes this study.

%% file: ti_ff_model.tex
\section{Methodology}\label{sec:tv_cb_met}
This section presents the econometric method employed to examine the time-varying structure of Fama-French multi-factor models for the U.S., Japan, and Europe. We employ \citetapos{fama1973rre} two-step regression with rolling windows rather than testing the models using the entire dataset to identify how and when the significance of the factors changes in the above countries or regions. We also conduct statistical inference to examine the the time instability of the models using \citetapos{kamstra2024tra} generalized GRS test and model ranking with rolling windows.

\subsection{Preliminaries}\label{subsec:PL}
Following \citet{fama1993crf,fama2015ffa}, we introduce the Fama--French multi-factor models. Equation (\ref{eq:FF3}) represents the FF3 model, which is the most widely known multi-factor model:
\begin{equation}
 R_{i,t}-R_{f,t}=\alpha_i+\beta^{Mkt}_{i}(R_{m,t}-R_{f,t})+\beta^{SMB}_{i}SMB_t+\beta^{HML}_{i}HML_t+\varepsilon_{i,t},\label{eq:FF3}
\end{equation}
where $R_{i,t}$ is the returns on the $i$-th portfolio of Fama and French's 25 benchmark portfolios at time $t$, $R_{f,t}$ is the risk-free rate at time $t$, $R_{m,t}$ is the returns on the market portfolio at time $t$, and $\varepsilon_{i,t}$ is the $i$-th error term at time $t$. \citet{fama1993crf} expand on the CAPM by adding size and value risk factors to capture market anomalies. The size-risk factor ($SMB_t$) explains that stocks with small (or large) market caps earn higher returns, while the value-risk factor ($HML_t$) explains the superior performance of stocks with low (or high) prices-to-book. Following this, \citet{fama2015ffa} introduced the FF5 model, as shown in equation (\ref{eq:FF5}):
\begin{equation}
\begin{split}
 R_{i,t}-R_{f,t}=\alpha_i+\beta^{Mkt}_{i}(R_{m,t}-R_{f,t})+\beta^{SMB}_{i}SMB_t+\beta^{HML}_{i}HML_t\\
+\beta^{RMW}_{i}RMW_t+\beta^{CMA}_{i}CMA_t+\varepsilon_{i,t}.\label{eq:FF5}
\end{split}
\end{equation}
They added two risk factors, the profitability-risk ($RWM_t$) and investment-risk ($CMA_t$) factors, to the FF3 model.

\subsection{\citetapos{fama1973rre} Two-Step Regression with Rolling Windows}\label{subsec:TV-RW}
We adopt \citetapos{fama1973rre} two-step regression with rolling windows to examine the time-instability of the Fama--French multi-factor models. In the two-step regression, we employ the feasible generalized least squares (FGLS) estimator instead of the ordinary least squares (OLS) to reduce the estimation errors as shown in \citet{shanken1992ebp}. Assuming that the parameters of Equations (\ref{eq:FF3}) and (\ref{eq:FF5}) change over time, we rewrite the Equations as follows:
\begin{equation}
\begin{split}
 R_{i,t}-R_{f,t}=\alpha_{i,t}+\beta^{Mkt}_{i,t}(R_{m,t}-R_{f,t})+\beta^{SMB}_{i,t}SMB_t+\beta^{HML}_{i,t}HML_t+\varepsilon_{i,t},\label{eq:TVFF3}
\end{split}
\end{equation}
\vspace{-5mm}
\begin{equation}
\begin{split}
 R_{i,t}-R_{f,t}=\alpha_{i,t}+\beta^{Mkt}_{i,t}(R_{m,t}-R_{f,t})+\beta^{SMB}_{i,t}SMB_t+\beta^{HML}_{i,t}HML_t\\
+\beta^{RMW}_{i,t}RMW_t+\beta^{CMA}_{i,t}CMA_t+\varepsilon_{i,t}.\label{eq:TVFF5}
\end{split}
\end{equation}

\noindent  We first estimate the time-varying coefficients of Equations (\ref{eq:TVFF3}) and (\ref{eq:TVFF5}) for each window. In the second step, we estimate the time-varying risk premiums for each factor, using the time-varying estimates as independent variables as follows:
\begin{equation}
\begin{split}
 \ex[R_{i,t}]-\ex[R_{f,t}]=\lambda^{Mkt}_t\hat{\beta}^{Mkt}_{i,t}+\lambda^{SMB}_t\hat{\beta}^{SMB}_{i,t}+\lambda^{HML}_t\hat{\beta}^{HML}_{i,t}+v_{i,t},\label{eq:RPFF3}
\end{split}
\end{equation}
\vspace{-5mm}
\begin{equation}
\begin{split}
 \ex[R_{i,t}]-\ex[R_{f,t}]=\lambda^{Mkt}_t\hat{\beta}^{Mkt}_{i,t}+\lambda^{SMB}_t\hat{\beta}^{SMB}_{i,t}+\lambda^{HML}_t\hat{\beta}^{HML}_{i,t}\\
+\lambda^{RMW}_t\hat{\beta}^{RMW}_{i,t}+\lambda^{CMA}_t\hat{\beta}^{CMA}_{i,t}+v_{i,t},\label{eq:RPFF5}
\end{split} 
\end{equation}

\noindent where each $v_{i}$ is the $i$-th error term and follows an independent and identical distributed process. 

\subsection{Statistical Inference for Testing the Validity of the FF Models}\label{subsec:GGRS}
Here, we describe how to assess the validity of the Fama--French multi--factor models. We can understand that the models correctly capture the behavior of stock returns if all constant terms $\alpha_i$ for each portfolio in Equations (\ref{eq:FF3}) and (\ref{eq:FF5}) are not significantly different from zero. Many previous studies employ \citetapos{gibbons1989teg} test (hereafter called the GRS test) to confirm whether the above constant terms are zero or not, but the GRS test has an over-rejection problem as shown in \citet{kamstra2024tra}. Then we use the generalized GRS test proposed by \citet{kamstra2024tra} to examine the validity of the Fama--French multi--factor models. In the generalized GRS test, the null hypothesis is as follows:
\[
 H_0: \ve{\alpha}=\ve{0},\ \ H_1: {\rm{not}} \ H_0
\]
\noindent where $\ve{\alpha}=\left\{\alpha_i^\prime\right\}_{i=1}^n$ in each of the Equations (\ref{eq:FF3}) and (\ref{eq:FF5}). We assume that the number of portfolios is $n$, the number of risk factors is $m$, and the sample size is $T$. Then the generalized GRS test statistics is defined as follows:
\begin{equation}
 \frac{T(T-n-m)}{n(T-m-1)}(1+\ve{\bar{F}}^\prime\ve{\hat{\Omega}}^{-1}\ve{\bar{F}})^{-1}\ve{\hat{\alpha}}^\prime\ve{\hat{\Sigma}}^{-1}\ve{\hat{\alpha}}\sim F_{n,T-n-m},\label{eq:GGRS}
\end{equation}

\noindent The statistics follows the $F$-distribution with degrees of freedom $n$ and $T-n-m$ under the null hypothesis. Let us $\tilde{\ve{F}}_t$ be the vector of risk factors at time $t$, then each variable is also defined as follows:
\[
 \tilde{\ve{{F}}_t}=\left(\begin{array}{c}
		f_{1,t}\\
		f_{2,t}\\
		\vdots\\
		f_{m,t}
		     \end{array}
\right), \ \ 
\ve{\bar{F}}=\frac{1}{T}\sum_{t=1}^T\ve{\tilde{F}}_t,\ \
\ve{\hat{\Omega}}=\frac{1}{T}\sum_{t=1}^T(\ve{\tilde{F}}_t-\ve{\bar{F}})(\ve{\tilde{F}}_t-\ve{\bar{F}})^\prime, 
\]
\[
\ve{\hat{\alpha}}=\left(\begin{array}{c}
		 \hat{\alpha}_{1} \\
		 \hat{\alpha}_{2} \\
		 \vdots\\
		 \hat{\alpha}_{n}\\
		       \end{array}\right),\ \
 \ve{\hat{\varepsilon}}_t=\left(\begin{array}{c}
		 \hat{\varepsilon}_{1,t} \\
		 \hat{\varepsilon}_{2,t} \\
		 \vdots\\
		 \hat{\varepsilon}_{n,t}\\
		       \end{array}\right),\ \ 
\ve{\hat{\Sigma}}=\frac{1}{T-m-1}\sum_{t=1}^T\ve{\hat{\varepsilon}}_t\ve{\hat{\varepsilon}}_t^\prime,
\]

\noindent In particular, the vector of risk factors $\tilde{\ve{F}_t}$ in the FF3 and FF5 models is characterized as follows:

\[
\ve{\tilde{F}}^{FF3}_t=\left(\begin{array}{c}
		R_{m,t}-R_{f,t}\\
		SMB_t\\
		HML_t
		     \end{array}
\right), \ \ 
\ve{\tilde{F}}^{FF5}_t=\left(\begin{array}{c}
		R_{m,t}-R_{f,t}\\
		SMB_t\\
		HML_t\\
		RMW_t\\
		CMA_t
		     \end{array}
\right). \ \ 
\]
Moreover, we are not only concerned with the time-instability of the FF3 and FF5 models, but also with the factor redundancy of the models. We can see that which model is superior to other models from the perspective of the factor redundancy. However, we cannot use to evaluate the superiority of the models because the generalized GRS statistics tends to be large as the number of factors increases. In response to this problem, \citet{kamstra2024tra} propose the model ranking method based on the $p$-value of the generalized GRS statistic, which internalizes the difference in the degree of freedom with different number of factors. Therefore, we employ the generalized GRS statistics-based model ranking method for the FF3, FF5, and their nested models to discuss the factor redundancy and superiority.

Now, we examine whether the time-varying $\alpha_{i,t}$ of Equations (\ref{eq:TVFF3}) and (\ref{eq:TVFF5}) for each time period based on the rolling window method is not significantly different from zero to confirm the time-stability of the validity of the FF models.

%% file: ti_ff_data.tex
\section{Data}\label{sec:tv_cb_dat}
This study aims to explore the time-varying structure of the FF3 and FF5 models for the U.S., Japan, and Europe. We obtained the risk-free rate and returns on all risk factors from \href{http://mba.tuck.dartmouth.edu/pages/faculty/ken.french/data_library.html}{Professor Kenneth French's website}. While the starting sample periods differ by region, the ending sample periods are the same--June 30, 2023. For the U.S., the relevant risk factors for the FF3 model are available from July 1, 1926, but since the FF3 model can be interpreted as a nested model of the FF5 model, we use data from July 1, 1963, when the relevant risk factors for the FF5 model are available. For Japan and Europe, the FF3 and FF5 risk factors have been available since July 2, 1990. In addition, we use three types of returns for the 25 benchmark portfolios sorted by ``Size and Book-to-Market (MB25),'' ``Size and Operating Profitability (MO25),'' and ``Size and Investment (MI25)'' obtained from the above website.
\begin{center}
(Table \ref{ff_tv_table1} around here)
\end{center}

Table \ref{ff_tv_table1} shows the results of our descriptive analysis for returns on all risk factors. With the exception of RMW, the average returns on all risk factors are greater for the United States than for Japan and Europe. However, the standard deviations of returns on all risk factors are similar between locations. For the estimations, each variable that appeared in the moment conditions should be stationary. We apply the augmented Dickey--Fuller (ADF) test to confirm whether the stationarity condition is satisfied. We also employ \citetapos{schwarz1978edm} Bayesian information criterion to select the optimal lag length. Table \ref{ff_tv_table1} shows the results of the unit root test with descriptive statistics for the data. The ADF test rejects the null hypothesis that each variable contains a unit root at the 1\% significance level.\footnote{The ADF test also rejects the null hypothesis that the returns on the 25 benchmark portfolios contain a unit root at the 1\% significance level.}

%% file: ti_ff_empirical.tex
\section{Empirical Results}\label{sec:tv_cb_emp}

This section provides us the empirical results. First, we estimate the time-varying risk premiums in the FF3 and FF5 models using \citetapos{fama1973rre} two-step regression with rolling windows. In particular, we can confirm the time instability of the validity of the FF3 and FF5 models. Then we apply \citetapos{kamstra2024tra} model ranking method to examine the time instability of the factor redundancy.\footnote{For more detailed estimation results, please see the Online Appendix.}

\subsection{The Time-Varying Estimates of the Risk Premiums}
We apply the rolling window method to \citetapos{fama1973rre} two-step regression for examining the time instability of the risk premiums.

\begin{center}
(Figures \ref{ff_tv_fig1} to \ref{ff_tv_fig6} around here)
\end{center}

Figures \ref{ff_tv_fig1} to \ref{ff_tv_fig6} show the time-varying risk premiums in the FF3 and FF5 models for the U.S., Japan, Europe, which based on Equations (\ref{eq:TVFF3}) and (\ref{eq:TVFF5}). We can see that the statistical significance of the estimates of risk premiums are not stable through time, regardless of the manner of the portfolio sorting. This implies that the models are not stable over time, and may have some redundant factors. To explore these questions further, we focus on the time variation of the statistical significance of risk premiums by decade.

\begin{center}
(Table \ref{ff_tv_table2} around here)
\end{center}

Table \ref{ff_tv_table2} indicates the percentage of periods where the risk premiums are not statistically significantly different from zero by decade for the U.S. stock market. The intercepts almost are not statistically significantly different from zero in any decade or portfolio. In the FF3 and FF5 models, if the factors correctly explain stock returns, the intercept should not be statistically significantly different from zero. This result is partially consistent with \citet{ang2006csv}, who show that the risk premium is not statistically significantly different from zero when controlling for exposure to the VIX, using a sample period from January 1986 to December 2000. This suggests that if we include or control for exposure to the VIX, the percentage of periods for which the intercept is not statistically significantly different from zero will increase. The $MKT$ factor is not statistically significantly different from zero in most of the sample periods. This is consistent with \citet{ang2007clr}, who find that the validity of the CAPM changes over time when using a rolling window analysis with a sample from July 1924 to December 2001. Furtheremore, our result is in line with \citet{jagannathan1996ccc} who show that the risk premium for the $MKT$ factor is not statistically significantly different from zero in the FF3 model using a sample period from July 1963 to December 1990. They suggest that the CAPM fails to accurately explain the variation of the stock returns after 1963. The percentage of periods in which the risk premiums for the $SMB$ factor are not statistically significantly different from zero is low in each decade and portfolio, consistent with \citet{ang2006csv} and \citet{bartholody2005eer}. Next, the percentage of periods in which the risk premiums for $HML$ factor are not statistically significantly different from zero varies over time, and is affected by the method of portfolio sorting. Specifically, the percentage of periods where the risk premiums are not statistically significantly different from zero for $HML$ is low in MB25 portfolios are used but high in other portfolios are used. This suggests that the $HML$ factor is effective in explaining U.S. stock returns, although its effectiveness varies over time. This result is consistent with \citet{fama2006vpc,fama2021vp}, who show that the effectiveness of the $HML$ factor varies over time. For the specific factors of FF5 model, the effectiveness of the $RMW$ and $CMA$ factors is not stable over time, even though \citet{fama2015ffa,fama2016daf} conclude that these factors contribute to explaining stock returns.

\begin{center}
(Table \ref{ff_tv_table3} around here)
\end{center}

Table \ref{ff_tv_table3} show the percentage of periods where the risk premiums are not statistically significantly different from zero by decade for the Japanese stock market. The intercepts almost remain close to zero across any decade and portfolio. This suggests that the FF3 and FF5 models are well-suited to the Japanese stock market, aligning with the insights of \citet{kubota2003fsr,kubota2018dff}, and \citet{chou2023ccf}. However, our result are not consistent with \citet{chan1991fsr} and \citet{jagannathan1998rlr}, who show that the intercept is not statistically significantly from zero. We believe that this inconsistency may stem from the instability of the FF3 and FF5 models. For the $MKT$ factor, the percentage of periods in which the risk premiums are not statistically significantly different from zero is high in any decade and portfolio. This suggest that the other factors substitute for the $MKT$ factor, consistent with \citet{chan1991fsr}, \citet{kubota2018dff}, and \citet{chou2023ccf}. For the $SMB$ factor, the percentage of periods in which the risk premiums are not statistically significantly different from zero, is consistently low across almost all sample periods and portfolios, but these values are not stable over time. This result is consistent with some previous studies such as \citet{jagannathan1998rlr}, \citet{kubota2003fsr,kubota2018dff}, and \citet{chou2023ccf}. For the $HML$ factor, the percentage of periods in which the risk premiums are not statistically significantly different from zero varies over time across all portfolios, aligning with \citet{jagannathan1998rlr} and \citet{chou2023ccf}. However, \citet{chan1991fsr} and \citet{kubota2003fsr,kubota2018dff} indicate that the risk premium for $HML$ factor is significant. This implies that the effectiveness of $HML$ factor fluctuates over time. For the $RMW$ and $CMA$ factors, the percentage of periods in which the risk premiums are not statistically significantly different from zero is high in many decades across portfolios. This suggests that these factors are not necessary in explaining the Japanese stock returns, consistent with \citet{kubota2018dff}.

\begin{center}
(Table \ref{ff_tv_table4} around here)
\end{center}

Table \ref{ff_tv_table4} displays the percentage of periods during which the risk premiums were not statistically significantly different from zero, organized by decade for the European stock market. We can see that the intercepts are not  statistically significantly different from zero across all decades and portfolios. This finding aligns with the research of \citet{gregory2013cta} and \citet{foye2018tav}, who investigate the validity of the FF3 and FF5 models in the U.K., respectively. The risk premiums for the $MKT$ factor are not statistically significantly different from zero across all decades and portfolios. Conversely, the risk premiums for the $SMB$ factor are statistically significantly different from zero across almost all decades and portfolios. This suggests that the $SMB$ factor substitutes for the explanatory power of stock returns typically attributed to the $MKT$ in Europe, consistent with \citet{foye2018tav}. Then the statistical significance of the risk premiums for the $HML$ factor depends on the manner of portfolio sorting. In particular, the risk premiums are not statistically significantly different from zero in the MB25 portfolio, which is consistent with \citet{bauer2010cap}, who examine the validity of the FF3 model in the European stock market. However, the risk premiums for the $HML$ factor are statistically significantly different from zero over many decades in the MO25 and MI25 portfolios, aligning with the findings of \citet{foye2018tav}. Furtheremore, he concludes that the $CMA$ factor is not effective in the U.K. stock market, which is consistent with our result. Finally, we find that the $RMW$ factor lacks explanatory power for European stock returns, with the exception of the MO25 portfolio.

In summary, the effectiveness of many factors in the FF3 and FF5 models is not stable over time and is influenced by the manner of the portfolio sorting. This suggests that there are some redundant factors in the multi-factor models and that the redundancy varies and depends on how the portfolios are sorted. Moreover, the intercepts are almost statistically significantly different from zero across many decades in all countries, which implies that the FF3 and FF5 models are valid across many periods and regions. However, we cannot evaluate the factor redundancy and the validity of the FF3 and FF5 models based on \citetapos{fama1973rre} regression. Therefore, we employ \citetapos{kamstra2024tra} GRS-based model ranking method for the FF3, FF5, and their nested models to examine the factor redundancy. Moreover, we implement the model ranking using the rolling windows and the three types of portfolios to explore the time instablitily of the factor redundancy and the influence of portfolio sorting methods.

\subsection{Model Ranking based on the Generalized GRS Test Statistics}

We apply rolling window method to \citetapos{kamstra2024tra} model ranking to examine the factor reduandancy of the FF3 and FF5 models, taking into account the difference in the manner the portfolios are sorted.

\begin{center}
(Figures \ref{ff_tv_fig7} to \ref{ff_tv_fig12} around here)
\end{center}

Figures \ref{ff_tv_fig7} to \ref{ff_tv_fig12} show the time-varying generalized GRS test statistics of the FF3, FF5, and their nested models for the U.S. in the MB25, M025, and MI25 portfolios. We recognize that the validity of these models is not stable over time in any portfolio, which is consistent with some previous studies. In particular, \citet{griffin2002aff}, \citet{ang2006csv}, and \cite{fama2006vpc} show that these models ara not valid using the GRS test statistics. In addition, \citet{fama2015ffa,fama1996mea} show that the validity of these models depends on the manner of portfolio sorting. Moreover, we recognize that the CAPM outperform the other models since the late 2010s. This implies that the CAPM is not dead, and the validity of the CAPM evolves over time in the U.S. stock market. In the following, we focus on the validity of each model at the decadal scale to capture the time-instability of the models and the factor redundancy in more detail.

\begin{center}
(Table \ref{ff_tv_table5} around here)
\end{center}

Table \ref{ff_tv_table5} indicates the percentage of each decade for which each model is valid in the U.S. stock market. The most striking feature is that there are periods where any factor model is valid (or not) for more than half of one decade. In particular, any factor model, including the CAPM, is valid for more than half of the 2010's and 2020's, while no factor model is valid for more than half of the 1980's. This is consistent with \citet{griffin2002aff}, who examine the validity of the FF3 model using a sample period from January 1981 to Feburary 1995, and conclude that the FF3 model is not valid based on the GRS test. This result indicates that the validity of the models is heavily dependent on the sample periods in the U.S. stock market. However, the time instability of the factor redundancy and the superioriy among the models is still not clear. Therefore, we present the model ranking results by decadal scale.

\begin{center}
(Table \ref{ff_tv_table6} around here)
\end{center}

Table \ref{ff_tv_table6} shows the percentage of the period for which the nested models of the FF3 or FF5 models outperform the benchmark models by decade in the U.S. stock market. First, we recognize that the way of portfolio sorting affects the validity and the ranking of the asset pricing models. In particular, the FF3 (FF5) model outpeforms all of the nested models on the MB25 (MI25) portfolio until the 2000's, but not on the other portfolios. This implies that the factor redundancy is also affected by the way of portfolio sorting. Second, the CAPM outperforms the FF3 and FF5 models in the 2010's. This is surprising because \citet{fama1993crf,fama2015ffa} develop the FF3 and FF5 models to compensate for the lack of explanatory power of the CAPM. However, this result is consistent with \citet{ito2016eme}, who examine the time-instability of the market efficiency in the weak sense of \citet{fama1970ecm} for the U.S. stock market and conclude that the U.S. stock market is efficient in the 2010s. Therefore, we believe that the U.S. stock market is efficient the 2010's since the validity of the CAPM is a necessary condition for the efficient market.

\begin{center}
(Figures \ref{ff_tv_fig13} to \ref{ff_tv_fig18} around here)
\end{center}

Figures \ref{ff_tv_fig13} to \ref{ff_tv_fig18} show the time-varying generalized GRS test statistics of the FF3, FF5, and their nested models for the Japanese stock market. In contrast to the results for the U.S., each model, including the CAPM, is valid in most of the sample periods. This result is in line with the results of \citet{griffin2002aff}, who show that the FF3 model is valid in Japan based on the GRS test using monthly data from January 1981 to December 1995. Furthermore, our result is consistent with \citet{fama2012svm}, who examine the validity of the CAPM and the FF3 model in the Japanese stock market and conclude that these models are valid using the GRS test on a dataset covering November 1990 to March 2010. The manner of portfolio sorting does not seem to affect the behavior of the generalized GRS test statistics. This align with the findings of \citet{fama2015ffa}, who examine the effectiveness of the FF5 and its nested models in the Japanese stock market considering, and explore whether the manner of portfolio sorting affects their effectiveness. They find that all the models including the CAPM are valid in any portfolio based on the GRS test. In the following, we focus on the validity of each model at the decadal scale to capture the time-instability of the models and the factor redundancy in more detail.

\begin{center}
(Table \ref{ff_tv_table7} around here)
\end{center}

Table \ref{ff_tv_table7} shows that all models are valid in more than half of the sample periods in each decade. This means that the CAPM, FF3 and F55 models are indifferent in explaining Japanese stock returns over time. However, this result is inconsistent with \citet{chan1991fsr} and \citet{jagannathan1998rlr}, who examine whether the CAPM is supported in the Japanese stock market using monthly data from January 1971 to December 1988 and September 1981 to December 1993, respectively. They conclude that the CAPM is not supported in the Japanese stock market. We believe that the inconsistency is due to the differences in the criteria used to evaluate the models, in which they do not perform the GRS test. Moreover, the periods when the CAPM works seem to be consistent with \citet{noda2016amh}, who examines the time-instability of the market efficiency in the Japanese stock market. This implies that the Japanese stock market is efficient in most of the sample periods, since the CAPM must hold in the efficient market.

\begin{center}
(Table \ref{ff_tv_table8} around here)
\end{center}

Table \ref{ff_tv_table8} indicates the percentage of the period for which the nested models of the FF3 or FF5 outperform the benchmark model in the Japanese stock market on a decadal scale. We find that the FF3 model outperforms its nested models and the FF5 model in most of the decades. This is partially consistent with \citet{kubota2018dff}, who examine the validity of the CAPM and the FF3 and FF5 models in the Japanese stock market using the GRS test statistics over the sample period from January 1978 to December 2014, and conclude that the FF3 model is superior to the CAPM and the FF5 model. However, the GRS test statistics reported in their study differ significantly from ours, and all models are rejected in terms of mean-variance efficiency.

\begin{center}
(Figures \ref{ff_tv_fig19} to \ref{ff_tv_fig24} around here)
\end{center}

Figures \ref{ff_tv_fig7} to \ref{ff_tv_fig12} show the time-varying generalized GRS test statistics of the FF3, FF5, and their nested models for Europe. We recognize that these models are not valid in many periods, and the behavior of the GRS test statistics is affected by the portfolio used. This is consistent with \citet{griffin2002aff}, \citet{bauer2010cap}, \citet{gregory2013cta} and \citet{tauscher2016pob}, who find that the FF3 model are not valid based on the GRS test using the MB25 portfolio. In addition, \citet{fama2017itf} examine the validity of the FF3 and FF5 models using the GRS test and whether the result of the GRS test is affected by the way the portfolio is sorted. They show that both the FF3 and FF5 models fail to effectively explain the European stock returns, with the exception of the FF5 model in the MI25 portfolio. In order to gain a more detailed understanding of the time-instability of the models, we focus on the validity of each model at the decadal scale.

\begin{center}
(Table \ref{ff_tv_table9} around here)
\end{center}

Table \ref{ff_tv_table9} shows that the way of portfolio sorting affects the validity of the models, and the dependence between portfolio sorting and the validity of the model is not stable over time. In particular, all models in the MB25 and MI25 (MO25) portfolios are (not) valid for more than half of the 2010's. This is partially consistent with \citet{fama2017itf}. They find that the performance of the FF3 and FF5 models depends on the portfolio sorting. This implies that the factor redundancy is not irrelevant to the way of portfolio sorting in the European stock market. However, all the models in each portfolio are valid for more than half of the 2020's. 

\begin{center}
(Table \ref{ff_tv_table10} around here)
\end{center}

Table \ref{ff_tv_table10} indicates the percentage of the period for which the nested models of the FF3 or FF5 outperform the benchmark model in the European stock market on a decadal scale. We find that the CAPM does not outperform the FF3 model in many decades, consisting with \citet{fama2012svm}. Moreover, we recognize that the FF5 model outperforms the FF3 model in any decade except for the MB25 portfolio used. This implies that the way of portfolio sorting affects the superiority of the FF3 and FF5 models in the sense of effectiveness, which is partially consistent with the findings of \citet{fama2017itf}.

%% file: ti_ff_conclusion.tex
\section{Conclusion}\label{sec:tv_cb_cr}

In this study, we investigate the time-varying structure of the Fama--French multi-factor models (\citet{fama1993crf,fama2015ffa}). We first use the three different types of 25 benchmark portfolios for the U.S., Japan, and Europe to estimate the time-varying risk premium based on the Fama--French multi-factor models using \citetapos{fama1973rre} two-step estimation based on the rolling window method. We consider the risk factors in the model not to be effective for periods when each risk premium is not statistically significantly different from zero. If the effectiveness of the risk factor is not stable over time, this implies that the validity of the FF3 and FF5 models and factor redundancy should not also be stable either. And then we employ \citetapos{kamstra2024tra} generalized GRS statistics to not only examine the validity of the FF3 and FF5 models, but also explore which model is superior to other models from the perspective of the factor redundancy. In particular, we examine whether the time stability of the validity of the FF3 and FF5 models and factor redundancy depend on the manner in which the portfolios are sorted.

The main findings in this paper are as follows. First, the effectiveness of factors in the FF3 and FF5 models is not stable over time in all countries. This result is consistent with \citet{fama2006vpc,fama2016daf}, who show that the effectiveness of the $HML$ factor varies over time. Second, the manner of portfolio sorting affects the time stability of the effectiveness of the factors. This result supports the critique of \citet{lewellen2010asa}, who claim that the effectiveness of factors may depend on the way portfolios are sorted. Third, the validity of the FF3, FF5, and their nested models varies over time except for the Japanese stock market. In particular, all the models, including the CAPM, are valid in most of the sample periods in Japan. This insight is in line with \citet{griffin2002aff} and \citet{fama2012svm,fama2015ffa}, who conclude that the CAPM, FF3 and FF5 models are valid in the Japanese stock market. Moreover, our result implies that the efficient market hypothesis in the sense of \citet{fama1970ecm} holds, since the validity of the CAPM is a necessary condition for the efficient market. Finally, the factor redundancy changes over time and is also affected by the manner of portfolio sorting, except for Japan. In particular, we find that the CAPM outperforms the FF3 and FF5 models in the U.S. stock market in the 2010s, which means that the $SMB$, $HML$, $RMW$ and $CMA$ factors are redundant in the 2010s. In Europe, the FF5 model outperforms the FF3 model in every decade except for the MB25 portfolio used, which is partially consistent with the insights of \citet{fama2017itf}.

%% file: ti_ff_ack.tex
\section*{Acknowledgments}
The author would like to thank Takahiro Hoshino, Mikio Ito, Daisuke Nagakura, Tatsuyoshi Okimoto, Tatsuma Wada, and the conference participants at the 99th Annual Conference of the Western Economic Association International for their helpful comments and suggestions. The author (Noda) is also grateful for the financial assistance provided by the Japan Society for the Promotion of Science Grant in Aid for Scientific Research (grant number: 23H00838) and Japan Science and Technology Agency, Moonshot Research \& Development Program (grant number: JPMJMS2215). All data and programs used are available upon request.

%% file: ti_ff_table.tex
\bigskip

\bigskip

\bigskip

\setcounter{table}{0}
\renewcommand{\thetable}{\arabic{table}}

\clearpage

\begin{table}[p]
\caption{Descriptive Statistics and Unit Root Tests}\label{ff_tv_table1}
\begin{center}{\scriptsize
\begin{tabular}{cllcccccccccc}\hline\hline
 &  &  &  & Mean & SD & Min & Max &  & ADF & Lags & $\mathcal{N}$ & \\\cline{5-8}\cline{10-12}
 & FF3 Factors &  &  &  &  &  &  &  &  &  &  & \\
 & \ \ \ U.S. &  &  &  &  &  &  &  &  &  &  & \\
 &  & $R_{Mkt}$ &  & $0.0003$ & $0.0102$ & $-0.1744$ & $0.1135$ &  & $-20.4753$ & $35$ & $15229$ \\
 &  & $R_{SMB}$ &  & $0.0001$ & $0.0054$ & $-0.1163$ & $0.0624$ &  & $-23.5958$ & $19$ & $15229$ \\
 &  & $R_{HML}$ &  & $0.0001$ & $0.0058$ & $-0.0502$ & $0.0673$ &  & $-18.3823$ & $32$ & $15229$\\
 & \ \ \ Japan &  &  &  &  &  &  &  &  &  &  & \\
 &  & $R_{Mkt}$ &  & $0.0001$ & $0.0134$ & $-0.1085$ & $0.1302$ &  & $-20.5914$ & $19$ & $8740$ \\
 &  & $R_{SMB}$ &  & $0.0000$ & $0.0065$ & $-0.0918$ & $0.0459$ &  & $-19.8051$ & $19$ & $8740$ \\
 &  & $R_{HML}$ &  & $0.0002$ & $0.0059$ & $-0.0488$ & $0.0457$ &  & $-82.2832$ & $0$ & $8740$ \\
 & \ \ \ Europe &  &  &  &  &  &  &  &  &  &  & \\
 &  & $R_{Mkt}$ &  & $0.0002$ & $0.0114$ & $-0.1200$ & $0.1072$ &  & $-16.1878$ & $29$ & $8740$ \\
 &  & $R_{SMB}$ &  & $0.0000$ & $0.0057$ & $-0.0518$ & $0.0331$ &  & $-45.3183$ & $4$ & $8740$ \\
 &  & $R_{HML}$ &  & $0.0001$ & $0.0047$ & $-0.0414$ & $0.0438$ &  & $-15.0226$ & $26$ & $8740$ \\\hline
 & FF5 Factors &  &  &  &  &  &  &  &  &  &  & \\
 & \ \ \ U.S. &  &  &  &  &  &  &  &  &  &  & \\
 &  & $R_{Mkt}$ &  & $0.0003$ & $0.0102$ & $-0.1744$ & $0.1135$ &  & $-20.4753$ & $35$ & $15229$ \\
 &  & $R_{SMB}$ &  & $0.0001$ & $0.0054$ & $-0.1119$ & $0.0617$ &  & $-23.5356$ & $19$ & $15229$ \\
 &  & $R_{HML}$ &  & $0.0001$ & $0.0058$ & $-0.0502$ & $0.0673$ &  & $-18.3823$ & $32$ & $15229$ \\
 &  & $R_{RMW}$ &  & $0.0001$ & $0.0040$ & $-0.0301$ & $0.0451$ &  & $-22.6359$ & $23$ & $15229$ \\
 &  & $R_{CMA}$ &  & $0.0001$ & $0.0038$ & $-0.0587$ & $0.0253$ &  & $-31.2825$ & $12$ & $15229$ \\
 & \ \ \ Japan &  &  &  &  &  &  &  &  &  &  & \\
 &  & $R_{Mkt}$ &  & $0.0001$ & $0.0134$ & $-0.1085$ & $0.1302$ &  & $-20.5914$ & $19$ & $8740$ \\
 &  & $R_{SMB}$ &  & $0.0000$ & $0.0064$ & $-0.0905$ & $0.0463$ &  & $-21.1171$ & $16$ & $8740$ \\
 &  & $R_{HML}$ &  & $0.0002$ & $0.0059$ & $-0.0488$ & $0.0457$ &  & $-82.2832$ & $0$ & $8740$ \\
 &  & $R_{RMW}$ &  & $0.0000$ & $0.0043$ & $-0.0380$ & $0.0359$ &  & $-80.3342$ & $0$ & $8740$ \\
 &  & $R_{CMA}$ &  & $0.0000$ & $0.0044$ & $-0.0350$ & $0.0388$ &  & $-28.7767$ & $9$ & $8740$ \\
 & \ \ \ Europe &  &  &  &  &  &  &  &  &  &  & \\
 &  & $R_{Mkt}$ &  & $0.0002$ & $0.0114$ & $-0.1200$ & $0.1072$ &  & $-16.1878$ & $29$ & $8740$ \\
 &  & $R_{SMB}$ &  & $0.0000$ & $0.0057$ & $-0.0530$ & $0.0332$ &  & $-45.5505$ & $4$ & $8740$ \\
 &  & $R_{HML}$ &  & $0.0001$ & $0.0047$ & $-0.0414$ & $0.0438$ &  & $-15.0226$ & $26$ & $8740$ \\
 &  & $R_{RMW}$ &  & $0.0002$ & $0.0030$ & $-0.0218$ & $0.0418$ &  & $-15.5185$ & $28$ & $8740$ \\
 &  & $R_{CMA}$ &  & $0.0001$ & $0.0034$ & $-0.0254$ & $0.0216$ &  & $-16.8386$ & $20$ & $8740$ \\\hline\hline
\end{tabular}}
\vspace*{5pt}
{\scriptsize{\begin{minipage}{380pt}
{\underline{Notes:}}
 \begin{itemize}
  \item[(1)] ``$R_{Mkt}$,'' ``$R_{SMB}$,'' ``$R_{HML}$,'' ``$R_{RMW}$,'' ``$R_{CMA}$,'' and ``$R_{WML}$'' denote the returns on each risk factor, wich correspond to Fama-French multi-factor models.
  \item[(2)] ``ADF,'' ``Lag,'' and ``$\mathcal{N}$'' denote the augmented Dickey--Fuller test statistics, the lag order selected by the Bayesian information criterion, and the number of observations, respectively.
  \item[(3)] R version 4.4.0 was used to compute the statistics.
 \end{itemize}
\end{minipage}}}%
\end{center}
\end{table}

\clearpage

\begin{landscape}
\begin{figure}[p]
 \caption{Time-Varying Estimates of Risk Premiums in FF3 Model (U.S.)}\label{ff_tv_fig1}
 \begin{center}
 \includegraphics[scale=0.35]{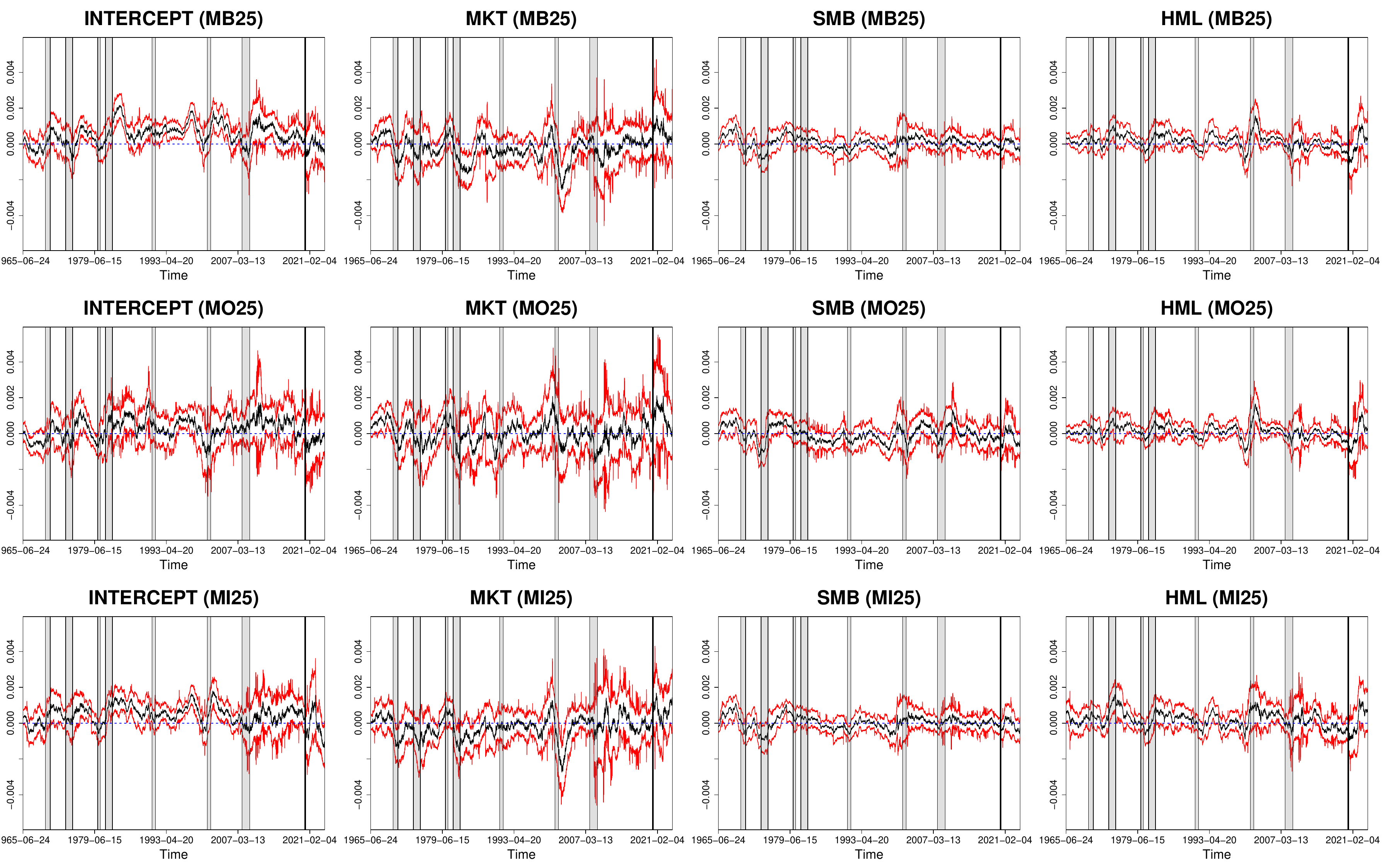}
\vspace*{5pt}
{
\begin{minipage}{420pt}
\scriptsize
\underline{Notes}:
\begin{itemize}
 \item[(1)] The dashed red and blue lines denote the critical value at the 5\% and 10\% significance levels for the generalized GRS test, respectively.
 \item[(2)]  The shade areas represent recessions reported by the NBER ``\href{https://www.nber.org/research/business-cycle-dating}{Business Cycle Dating}.''
 \item[(3)] R version 4.4.0 was used to compute the statistics.
\end{itemize}
\end{minipage}}%
\end{center}
\end{figure}
\end{landscape}

\clearpage

\begin{landscape}
\begin{figure}[p]
 \caption{Time-Varying Estimates of Risk Premiums in FF5 Model (U.S.)}\label{ff_tv_fig2}
 \begin{center}
 \includegraphics[scale=0.35]{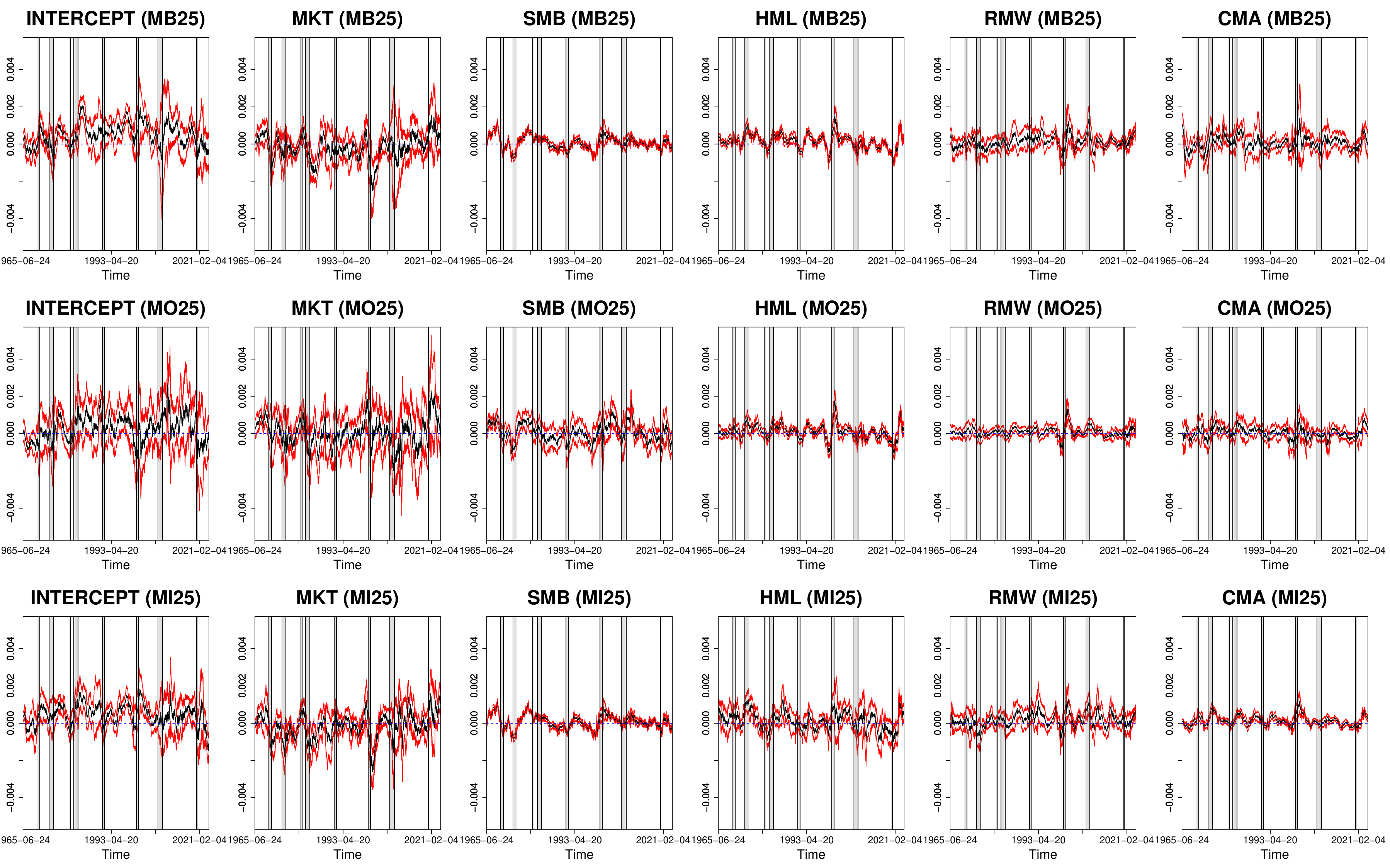}
\vspace*{5pt}
{
\begin{minipage}{420pt}
\scriptsize
\underline{Notes}:
\begin{itemize}
 \item[(1)] The dashed red and blue lines denote the critical value at the 5\% and 10\% significance levels for the generalized GRS test, respectively.
 \item[(2)]  The shade areas represent recessions reported by the NBER ``\href{https://www.nber.org/research/business-cycle-dating}{Business Cycle Dating}.''
 \item[(3)] R version 4.4.0 was used to compute the statistics.
\end{itemize}
\end{minipage}}%
\end{center}
\end{figure}
\end{landscape}

\clearpage

\begin{landscape}
\begin{figure}[p]
 \caption{Time-Varying Estimates of Risk Premiums in FF3 Model (Japan)}\label{ff_tv_fig3}
 \begin{center}
 \includegraphics[scale=0.35]{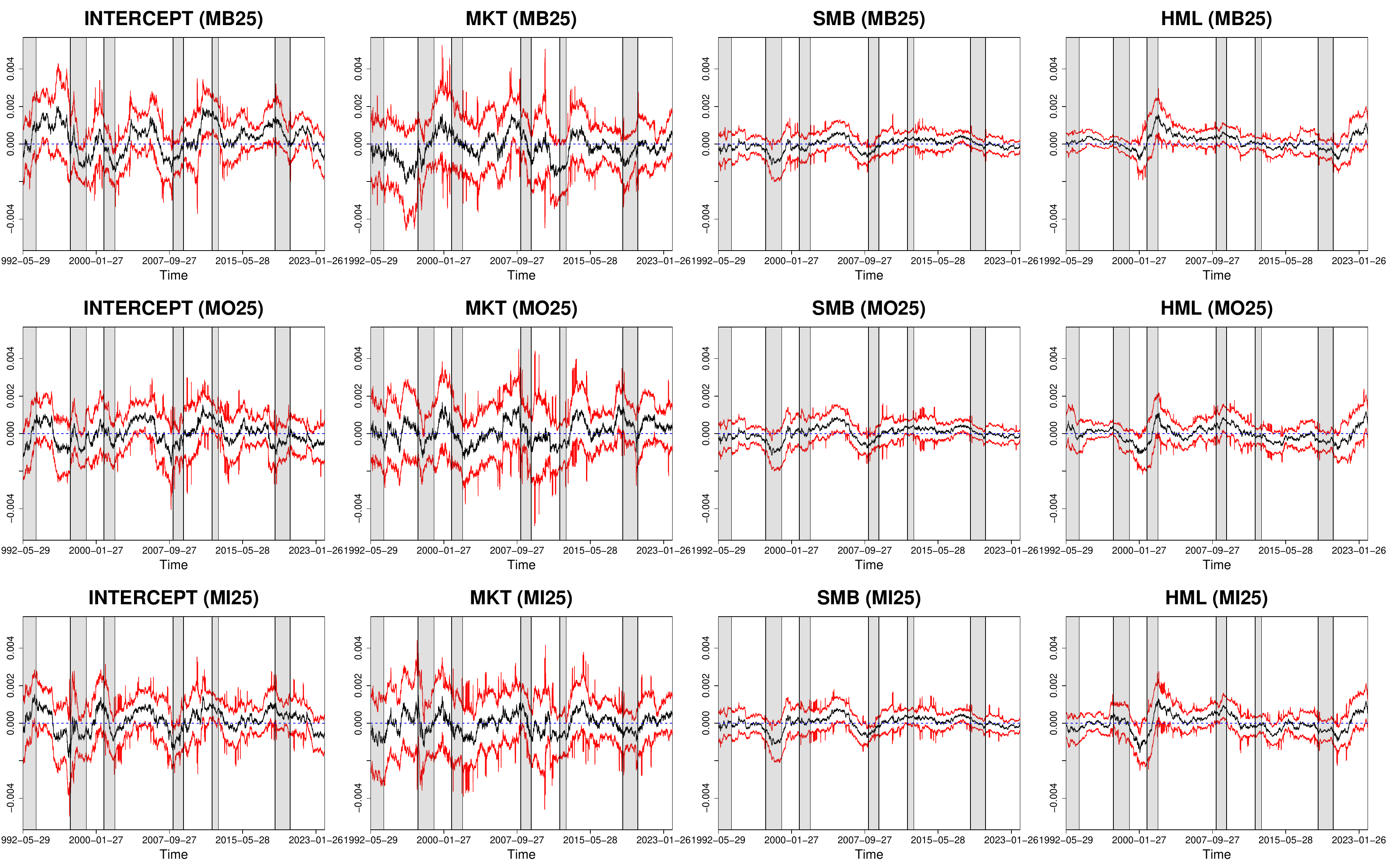}
\vspace*{5pt}
{
\begin{minipage}{420pt}
\scriptsize
\underline{Notes}:
\begin{itemize}
 \item[(1)] The dashed red and blue lines denote the critical value at the 5\% and 10\% significance levels for the generalized GRS test, respectively.
 \item[(2)]  The shade areas represent recessions reported by the Cabinet Office Japan ``\href{https://www.esri.cao.go.jp/en/stat/di/rdates.html}{Business Cycle Dating}.''
 \item[(3)] R version 4.4.0 was used to compute the statistics.
\end{itemize}
\end{minipage}}%
\end{center}
\end{figure}
\end{landscape}

\clearpage

\begin{landscape}
\begin{figure}[p]
 \caption{Time-Varying Estimates of Risk Premiums in FF5 Model (Japan)}\label{ff_tv_fig4}
 \begin{center}
 \includegraphics[scale=0.35]{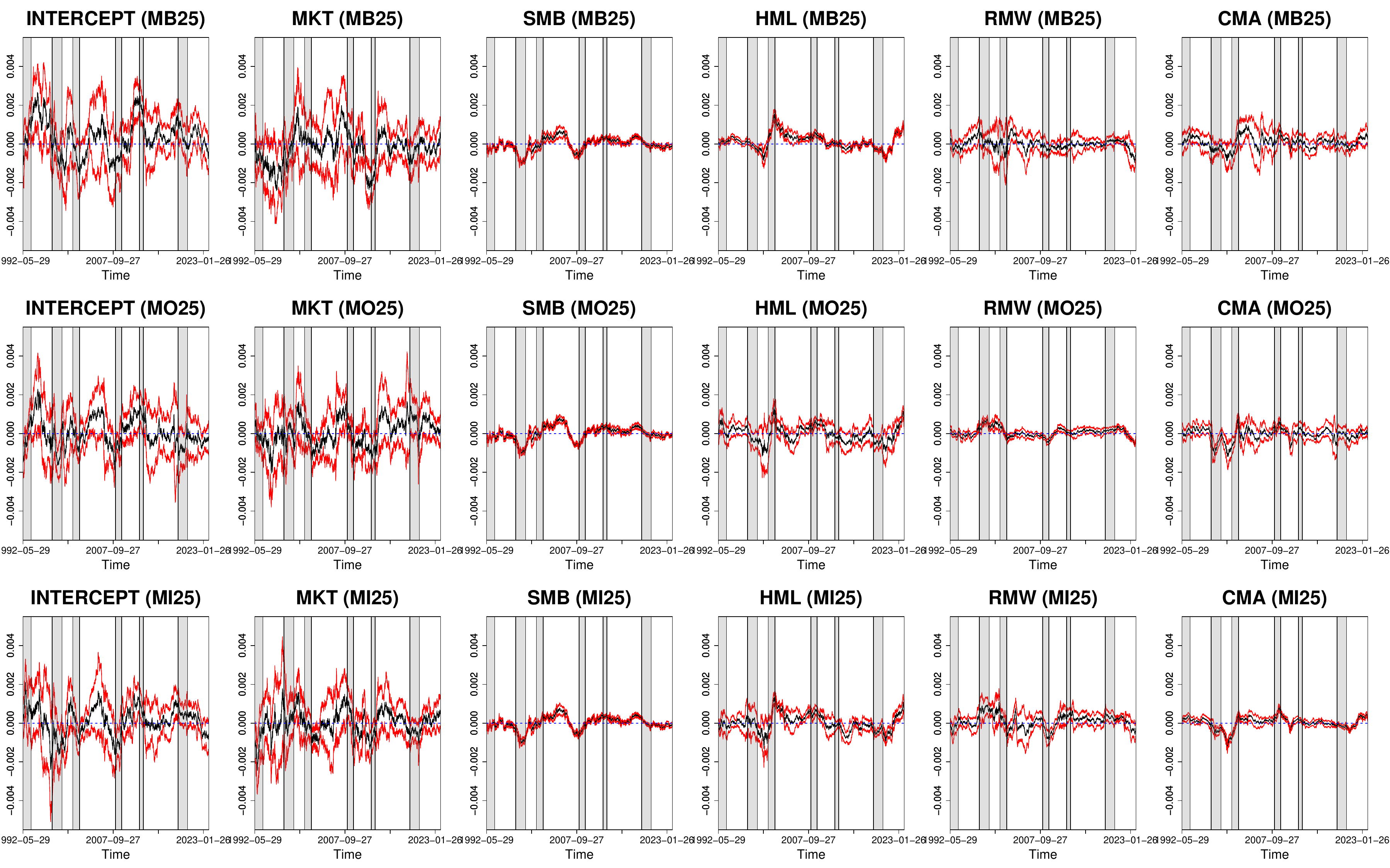}
\vspace*{5pt}
{
\begin{minipage}{420pt}
\scriptsize
\underline{Notes}:
\begin{itemize}
 \item[(1)] The dashed red and blue lines denote the critical value at the 5\% and 10\% significance levels for the generalized GRS test, respectively.
 \item[(2)]  The shade areas represent recessions reported by the Cabinet Office Japan ``\href{https://www.esri.cao.go.jp/en/stat/di/rdates.html}{Business Cycle Dating}.''
 \item[(3)] R version 4.4.0 was used to compute the statistics.
\end{itemize}
\end{minipage}}%
\end{center}
\end{figure}
\end{landscape}

\clearpage

\begin{landscape}
\begin{figure}[p]
 \caption{Time-Varying Estimates of Risk Premiums in FF3 Model (Europe)}\label{ff_tv_fig5}
 \begin{center}
 \includegraphics[scale=0.35]{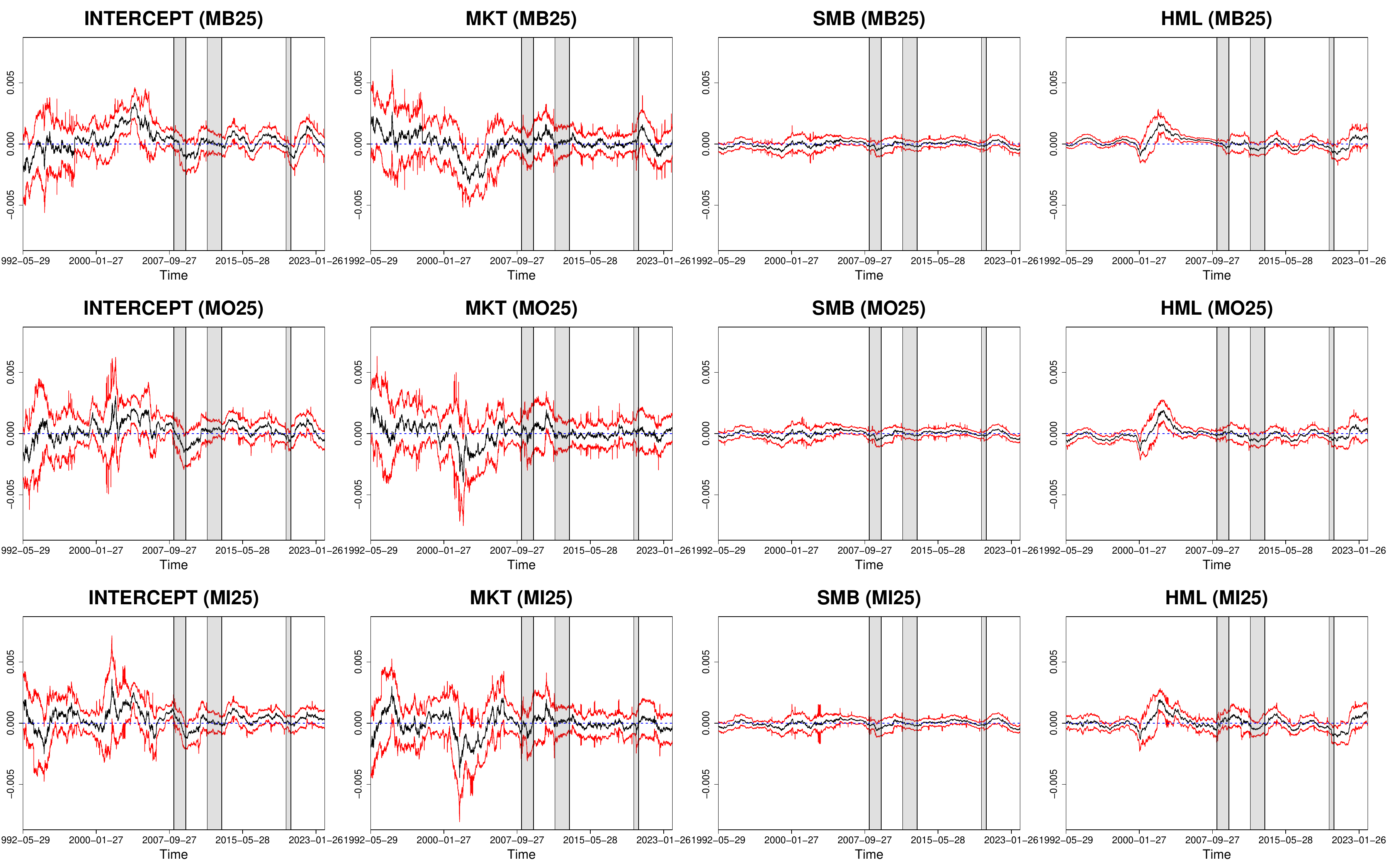}
\vspace*{5pt}
{
\begin{minipage}{420pt}
\scriptsize
\underline{Notes}:
\begin{itemize}
 \item[(1)] The dashed red and blue lines denote the critical value at the 5\% and 10\% significance levels for the generalized GRS test, respectively.
 \item[(2)]  The shade areas represent recessions reported by the Euro Area Business Cycle Network ``\href{https://eabcn.org/dc/chronology-euro-area-business-cycles}{Chronology of Euro Area Business Cycles}.''
 \item[(3)] R version 4.4.0 was used to compute the statistics.
\end{itemize}
\end{minipage}}%
\end{center}
\end{figure}
\end{landscape}

\clearpage

\begin{landscape}
\begin{figure}[p]
 \caption{Time-Varying Estimates of Risk Premiums in FF5 Model (Europe)}\label{ff_tv_fig6}
 \begin{center}
 \includegraphics[scale=0.35]{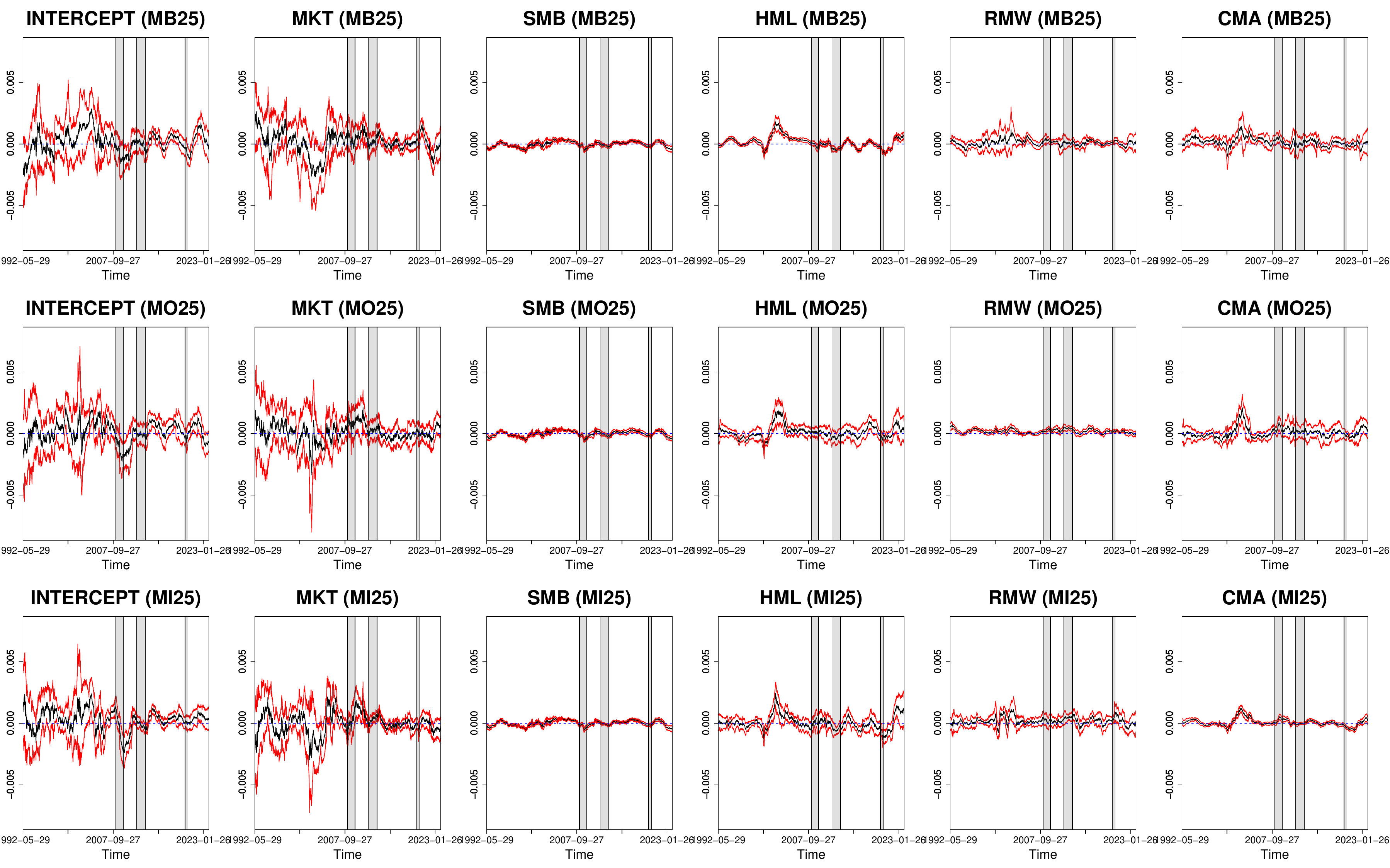}
\vspace*{5pt}
{
\begin{minipage}{420pt}
\scriptsize
\underline{Notes}:
\begin{itemize}
 \item[(1)] The dashed red and blue lines denote the critical value at the 5\% and 10\% significance levels for the generalized GRS test, respectively.
 \item[(2)]  The shade areas represent recessions reported by the Euro Area Business Cycle Network ``\href{https://eabcn.org/dc/chronology-euro-area-business-cycles}{Chronology of Euro Area Business Cycles}.''
 \item[(3)] R version 4.4.0 was used to compute the statistics.
\end{itemize}
\end{minipage}}%
\end{center}
\end{figure}
\end{landscape}

\clearpage

\begin{table}[p]
\caption{Percentage of the periods of $\hat{\ve{\lambda}}_t=\ve{0}$ (U.S.)}\label{ff_tv_table2}
\centering
\begin{tabular}{cc}
  \begin{minipage}[t]{0.45\textwidth}
    \begin{center}
\scalebox{0.55}{\begin{tabular}{cccccccccc} \hline\hline
In the 1960's & & \multicolumn{3}{c}{FF3} & & \multicolumn{3}{c}{FF5} &\\\cline{3-5}\cline{7-9}
 & & MB25 & MO25 & MI25 & & MB25 & MO25 & MI25 &\\\hline
$\alpha$ & & ${\color{blue}{0.9532}}$  & $0.4284$  & ${\color{blue}{0.7912}}$ & & ${\color{blue}{0.8785}}$  & ${\color{blue}{0.5455}}$  & ${\color{blue}{0.7282}}$ &\\
MKT & & ${\color{red}{0.6994}}$  & $0.1413$  & ${\color{red}{0.8218}}$  & & ${\color{red}{0.8650}}$  & $0.1773$  & ${\color{red}{0.7183}}$ &\\
SMB & & $0.0774$  & $0.0702$  & $0.0729$  & & $0.0666$  & $0.0747$  & $0.0666$ &\\
HML & & $0.4689$  & ${\color{red}{0.9217}}$  & ${\color{red}{0.7678}}$  & & $0.4707$  & ${\color{red}{0.7966}}$  & ${\color{red}{0.6436}}$ &\\
RMW & & $-$ & $-$ & $-$ & & ${\color{red}{0.9001}}$  & ${\color{red}{0.8173}}$  & ${\color{red}{0.9730}}$ &\\
CMA & & $-$ & $-$ & $-$ & & ${\color{red}{0.7678}}$  & ${\color{red}{0.8398}}$  & ${\color{red}{0.6121}}$ &\\\hline\hline
  \end{tabular}}
 \end{center}
\end{minipage}
 &
\begin{minipage}[t]{0.45\textwidth}
 \begin{center}
  \scalebox{0.55}{\begin{tabular}{cccccccccc} \hline\hline
In the 1970's & & \multicolumn{3}{c}{FF3} & & \multicolumn{3}{c}{FF5} &\\\cline{3-5}\cline{7-9}
 & & MB25 & MO25 & MI25 & & MB25 & MO25 & MI25 &\\\hline
$\alpha$ & & ${\color{blue}{0.6584}}$  & ${\color{blue}{0.6271}}$  & $0.4683$  & & ${\color{blue}{0.6219}}$  & ${\color{blue}{0.7704}}$  & ${\color{blue}{0.5024}}$ &\\
MKT & & ${\color{red}{0.6223}}$  & ${\color{red}{0.8155}}$  & ${\color{red}{0.5503}}$  & & ${\color{red}{0.6382}}$  & ${\color{red}{0.8361}}$  & ${\color{red}{0.5483}}$ &\\
SMB & & $0.1473$  & $0.1607$  & $0.1647$  & & $0.1239$  & $0.1599$  & $0.1584$ &\\
HML & & $0.2589$  & $0.3994$  & $0.3963$  & & $0.2593$  & ${\color{red}{0.8282}}$  & ${\color{red}{0.5416}}$ &\\
RMW & & $-$ & $-$ & $-$ & & ${\color{red}{0.8226}}$  & ${\color{red}{0.5150}}$  & ${\color{red}{0.9181}}$ &\\
CMA & & $-$ & $-$ & $-$ & & ${\color{red}{0.7838}}$  & ${\color{red}{0.8147}}$  & $0.2704$ &\\\hline\hline
  \end{tabular}}
 \end{center}
\end{minipage}\\ 

\vspace{-3mm}\\

\begin{minipage}[h]{0.45\textwidth}
 \begin{center}
  \scalebox{0.55}{\begin{tabular}{cccccccccc} \hline\hline
In the 1980's & & \multicolumn{3}{c}{FF3} & & \multicolumn{3}{c}{FF5} &\\\cline{3-5}\cline{7-9}
 & & MB25 & MO25 & MI25 & & MB25 & MO25 & MI25 &\\\hline
$\alpha$ & & ${\color{blue}{0.5245}}$  & $0.4110$  & $0.4201$ & & ${\color{blue}{0.6064}}$  & ${\color{blue}{0.7191}}$  & ${\color{blue}{0.5752}}$ &\\
MKT & & $0.4759$  & ${\color{red}{0.6309}}$  & ${\color{red}{0.5878}}$  & & ${\color{red}{0.6139}}$  & ${\color{red}{0.8050}}$  & $0.4822$ &\\
SMB & & $0.2690$  & $0.2563$  & $0.3196$  & & $0.2583$  & $0.1685$  & $0.1954$ &\\
HML & & $0.3212$  & ${\color{red}{0.7500}}$  & $0.3920$  & & $0.2789$  & ${\color{red}{0.8248}}$  & ${\color{red}{0.6697}}$ &\\
RMW & & $-$ & $-$ & $-$ & & ${\color{red}{1.0000}}$  & $0.4996$  & ${\color{red}{0.7706}}$ &\\
CMA & & $-$ & $-$ & $-$ & & ${\color{red}{0.7599}}$  & ${\color{red}{0.5756}}$  & $0.4221$ &\\\hline\hline
  \end{tabular}}
 \end{center}
\end{minipage}
  &
\begin{minipage}[h]{0.45\textwidth}
 \begin{center}
  \scalebox{0.55}{\begin{tabular}{cccccccccc} \hline\hline
In the 1990's & & \multicolumn{3}{c}{FF3} & & \multicolumn{3}{c}{FF5} &\\\cline{3-5}\cline{7-9}
 & & MB25 & MO25 & MI25 & & MB25 & MO25 & MI25 &\\\hline
$\alpha$ & & $0.3449$  & ${\color{blue}{0.6214}}$  & ${\color{blue}{0.5922}}$ & & ${\color{blue}{0.5123}}$  & ${\color{blue}{0.5831}}$  & ${\color{blue}{0.5767}}$ &\\
MKT & & ${\color{red}{0.8184}}$  & ${\color{red}{0.7468}}$  & ${\color{red}{0.9015}}$ & & ${\color{red}{0.8869}}$  & ${\color{red}{0.9684}}$  & ${\color{red}{0.8612}}$ &\\
SMB & & $0.4059$  & $0.4415$  & $0.3877$ & & $0.2690$  & $0.3149$  & $0.2923$ &\\
HML & & $0.3097$  & ${\color{red}{0.7496}}$  & ${\color{red}{0.6171}}$ & & $0.3394$  & ${\color{red}{0.6756}}$  & ${\color{red}{0.8125}}$ &\\
RMW & & $-$ & $-$ & $-$ & & ${\color{red}{0.6072}}$  & $0.3062$  & ${\color{red}{0.6503}}$ &\\
CMA & & $-$ & $-$ & $-$ & & ${\color{red}{0.7971}}$  & ${\color{red}{0.8188}}$  & ${\color{red}{0.5629}}$ &\\\hline\hline
 \end{tabular}}
 \end{center}
\end{minipage}\\ 

\vspace{-3mm}\\

\begin{minipage}[h]{0.45\textwidth}
 \begin{center}
  \scalebox{0.55}{\begin{tabular}{cccccccccc} \hline\hline
In the 2000's & & \multicolumn{3}{c}{FF3} & & \multicolumn{3}{c}{FF5} &\\\cline{3-5}\cline{7-9}
 & & MB25 & MO25 & MI25 & & MB25 & MO25 & MI25 &\\\hline
$\alpha$ & & ${\color{blue}{0.6795}}$  & ${\color{blue}{0.6243}}$  & ${\color{blue}{0.6767}}$ & & ${\color{blue}{0.6950}}$  & ${\color{blue}{0.6843}}$  & ${\color{blue}{0.7364}}$ &\\
MKT & & ${\color{red}{0.6131}}$  & ${\color{red}{0.5718}}$  & ${\color{red}{0.7034}}$ & & ${\color{red}{0.6930}}$  & ${\color{red}{0.6358}}$  & ${\color{red}{0.5960}}$ &\\
SMB & & $0.4827$  & $0.3841$  & $0.3889$ & & $0.4358$  & $0.2767$  & $0.3869$ &\\
HML & & $0.3256$  & $0.2879$  & $0.4433$ & & $0.2934$  & $0.4123$  & ${\color{red}{0.5670}}$ &\\
RMW & & $-$ & $-$ & $-$ & & $0.4895$  & $0.3185$  & $0.4807$ &\\
CMA & & $-$ & $-$ & $-$ & & ${\color{red}{0.8577}}$  & ${\color{red}{0.8183}}$  & ${\color{red}{0.5563}}$ &\\\hline\hline
  \end{tabular}}
 \end{center}
\end{minipage}
  &
\begin{minipage}[h]{0.45\textwidth}
 \begin{center}
  \scalebox{0.55}{\begin{tabular}{cccccccccc} \hline\hline
In the 2010's & & \multicolumn{3}{c}{FF3} & & \multicolumn{3}{c}{FF5} &\\\cline{3-5}\cline{7-9}
 & & MB25 & MO25 & MI25 & & MB25 & MO25 & MI25 &\\\hline
$\alpha$ & & ${\color{blue}{0.5994}}$  & ${\color{blue}{0.5525}}$  & ${\color{blue}{0.7186}}$ & & ${\color{blue}{0.7993}}$  & ${\color{blue}{0.6057}}$  & ${\color{blue}{0.8080}}$ &\\
MKT & & ${\color{red}{0.8533}}$  & ${\color{red}{0.6133}}$  & ${\color{red}{0.7496}}$ & & ${\color{red}{0.8998}}$  & ${\color{red}{0.7158}}$  & ${\color{red}{0.8716}}$ &\\
SMB & & $0.4706$  & $0.4074$  & ${\color{red}{0.5016}}$ & & $0.4948$  & $0.3998$  & ${\color{red}{0.5974}}$ &\\
HML & & $0.2985$  & ${\color{red}{0.5250}}$  & ${\color{red}{0.6204}}$ & & $0.2854$  & ${\color{red}{0.6379}}$  & ${\color{red}{0.8271}}$ &\\
RMW & & $-$ & $-$ & $-$ & & ${\color{red}{0.6188}}$  & ${\color{red}{0.5393}}$  & ${\color{red}{0.6411}}$ &\\
CMA & & $-$ & $-$ & $-$ & & ${\color{red}{0.7651}}$  & ${\color{red}{0.6252}}$  & ${\color{red}{0.5024}}$ &\\\hline\hline
  \end{tabular}}
 \end{center}
\end{minipage}\\ 

\vspace{-3mm}\\

\begin{minipage}[h]{0.45\textwidth}
 \begin{center}
  \scalebox{0.55}{\begin{tabular}{cccccccccc} \hline\hline
In the 2020's & & \multicolumn{3}{c}{FF3} & & \multicolumn{3}{c}{FF5} &\\\cline{3-5}\cline{7-9}
 & & MB25 & MO25 & MI25 & & MB25 & MO25 & MI25 &\\\hline
$\alpha$ & & ${\color{blue}{0.7273}}$  & $0.4323$  & ${\color{blue}{0.6102}}$ & & ${\color{blue}{0.9920}}$  & ${\color{blue}{0.7591}}$  & ${\color{blue}{0.7136}}$ &\\
MKT & & ${\color{red}{0.6625}}$  & $0.4107$  & ${\color{red}{0.7193}}$ & & $0.4239$  & ${\color{red}{0.7295}}$  & ${\color{red}{0.9102}}$ &\\
SMB & & $0.1273$  & $0.1604$  & $0.1489$ & & $0.2409$  & $0.1636$  & $0.1670$ &\\
HML & & $0.0841$  & $0.2059$  & $0.1080$ & & $0.0409$  & $0.2739$  & $0.2386$ &\\
RMW & & $-$ & $-$ & $-$ & & $0.4455$  & ${\color{red}{0.5318}}$  & ${\color{red}{0.8307}}$ &\\
CMA & & $-$ & $-$ & $-$ & & $0.4114$  & $0.4443$  & $0.3216$ &\\\hline\hline
  \end{tabular}}
 \end{center}
\end{minipage}
  &
\begin{minipage}[h]{0.45\textwidth}
 \begin{center}
  \scalebox{0.55}{\begin{tabular}{cccccccccc} \hline\hline
Whole period & & \multicolumn{3}{c}{FF3} & & \multicolumn{3}{c}{FF5} &\\\cline{3-5}\cline{7-9}
 & & MB25 & MO25 & MI25 & & MB25 & MO25 & MI25 &\\\hline
$\alpha$ & & ${\color{blue}{0.6010}}$  & ${\color{blue}{0.5485}}$  & ${\color{blue}{0.5935}}$  & & ${\color{blue}{0.6852}}$  & ${\color{blue}{0.6681}}$  & ${\color{blue}{0.6507}}$ &\\
MKT & & ${\color{red}{0.6774}}$  & ${\color{red}{0.6192}}$  & ${\color{red}{0.7091}}$  & & ${\color{red}{0.7359}}$  & ${\color{red}{0.7419}}$  & ${\color{red}{0.6897}}$ &\\
SMB & & $0.3200$  & $0.2999$  & $0.3188$  & & $0.2926$  & $0.2434$  & $0.2965$ &\\
HML & & $0.3022$  & ${\color{red}{0.5513}}$  & $0.4914$  & & $0.2899$  & ${\color{red}{0.6610}}$  & ${\color{red}{0.6537}}$ &\\
RMW & & $-$ & $-$ & $-$ & & ${\color{red}{0.7067}}$  & $0.4706$  & ${\color{red}{0.7221}}$ &\\
CMA & & $-$ & $-$ & $-$ & & ${\color{red}{0.7678}}$  & ${\color{red}{0.7216}}$  & $0.4656$ &\\\hline\hline
  \end{tabular}}
 \end{center}
\end{minipage}
\end{tabular}
\vspace*{5pt}
{
\begin{minipage}{420pt}
\scriptsize
\underline{Notes}:
\begin{itemize}
 \item[(1)] The highlighted values indicate the percentage of periods where $\hat{\ve{\lambda}}_t=\ve{0}$ is over half the period.
 \item[(2)] R version 4.4.0 was used to compute the statistics.
\end{itemize}
\end{minipage}}
\end{table}

\clearpage

\begin{table}[p]
\caption{Percentage of the periods of $\hat{\ve{\lambda}}_t=\ve{0}$ (Japan)}\label{ff_tv_table3}
\centering
\begin{tabular}{cc}
 \begin{minipage}[t]{0.45\textwidth}
  \begin{center}
   \scalebox{0.55}{\begin{tabular}{cccccccccc} \hline\hline
In the 1990's & & \multicolumn{3}{c}{FF3} & & \multicolumn{3}{c}{FF5} &\\\cline{3-5}\cline{7-9}
 & & MB25 & MO25 & MI25 & & MB25 & MO25 & MI25 &\\\hline
$\alpha$ & & ${\color{blue}{0.5346}}$  & $0.4811$  & ${\color{blue}{0.6996}}$  & & ${\color{blue}{0.5583}}$  & ${\color{blue}{0.5159}}$  & ${\color{blue}{0.7996}}$ &\\
MKT & & ${\color{red}{0.6295}}$  & ${\color{red}{0.6966}}$  & ${\color{red}{0.8016}}$ & & ${\color{red}{0.5321}}$  & ${\color{red}{0.8203}}$  & ${\color{red}{0.8677}}$ &\\
SMB & & $0.3614$  & $0.4255$  & $0.3725$ & & $0.2751$  & $0.3327$  & $0.3700$ &\\
HML & & $0.4558$  & ${\color{red}{0.6557}}$  & ${\color{red}{0.6608}}$ & & $0.4160$  & $0.4836$  & ${\color{red}{0.8157}}$ &\\
RMW & & $-$ & $-$ & $-$ & & ${\color{red}{0.8087}}$  & $0.4791$  & ${\color{red}{0.6073}}$ &\\
CMA & & $-$ & $-$ & $-$ & & ${\color{red}{0.5694}}$  & $0.3640$  & $0.3211$ &\\\hline\hline
  \end{tabular}}
 \end{center}
\end{minipage}
  &
\begin{minipage}[t]{0.45\textwidth}
 \begin{center}
  \scalebox{0.55}{\begin{tabular}{cccccccccc} \hline\hline
In the 2000's & & \multicolumn{3}{c}{FF3} & & \multicolumn{3}{c}{FF5} &\\\cline{3-5}\cline{7-9}
 & & MB25 & MO25 & MI25 & & MB25 & MO25 & MI25 &\\\hline
$\alpha$ & & ${\color{blue}{0.7582}}$  & ${\color{blue}{0.8429}}$  & ${\color{blue}{0.6501}}$ & & ${\color{blue}{0.7923}}$  & ${\color{blue}{0.6236}}$  & ${\color{blue}{0.7267}}$ &\\
MKT & & ${\color{red}{0.8088}}$  & ${\color{red}{0.6251}}$  & ${\color{red}{0.7014}}$ & & ${\color{red}{0.9187}}$  & ${\color{red}{0.7083}}$  & ${\color{red}{0.6934}}$ &\\
SMB & & $0.3142$  & $0.2281$  & $0.1840$ & & $0.2641$  & $0.2411$  & $0.1844$ &\\
HML & & $0.1050$  & $0.3714$  & $0.2741$ & & $0.1115$  & $0.3925$  & $0.3890$ &\\
RMW & & $-$ & $-$ & $-$ & & ${\color{red}{0.9222}}$  & $0.5109$  & ${\color{red}{0.7290}}$ &\\
CMA & & $-$ & $-$ & $-$ & & ${\color{red}{0.6481}}$  & ${\color{red}{0.5868}}$  & $0.3488$ &\\\hline\hline
  \end{tabular}}
 \end{center}
\end{minipage}\\ 

\vspace{-3mm}\\

\begin{minipage}[h]{0.45\textwidth}
 \begin{center}
  \scalebox{0.55}{\begin{tabular}{cccccccccc} \hline\hline
In the 2010's & & \multicolumn{3}{c}{FF3} & & \multicolumn{3}{c}{FF5} &\\\cline{3-5}\cline{7-9}
 & & MB25 & MO25 & MI25 & & MB25 & MO25 & MI25 &\\\hline
$\alpha$ & & $0.2519$  & ${\color{blue}{0.7964}}$  & ${\color{blue}{0.7350}}$ & & ${\color{blue}{0.6097}}$  & ${\color{blue}{0.6917}}$  & ${\color{blue}{0.7002}}$ &\\
MKT & & $0.4820$  & ${\color{red}{0.7558}}$  & ${\color{red}{0.7979}}$ & & ${\color{red}{0.7481}}$  & ${\color{red}{0.7722}}$  & ${\color{red}{0.8102}}$ &\\
SMB & & $0.3259$  & $0.2910$  & $0.2634$ & & $0.2634$  & $0.2745$  & $0.2396$ &\\
HML & & $0.4724$  & $0.4739$  & ${\color{red}{0.5813}}$ & & ${\color{red}{0.5510}}$  & ${\color{red}{0.5844}}$  & ${\color{red}{0.6196}}$ &\\
RMW & & $-$ & $-$ & $-$ & & ${\color{red}{0.8689}}$  & ${\color{red}{0.6438}}$  & ${\color{red}{0.6495}}$ &\\
CMA & & $-$ & $-$ & $-$ & & ${\color{red}{0.7086}}$  & ${\color{red}{0.8405}}$  & ${\color{red}{0.5230}}$ &\\\hline\hline
  \end{tabular}}
 \end{center}
\end{minipage}
  &
\begin{minipage}[h]{0.45\textwidth}
 \begin{center}
  \scalebox{0.55}{\begin{tabular}{cccccccccc} \hline\hline
In the 2020's & & \multicolumn{3}{c}{FF3} & & \multicolumn{3}{c}{FF5} &\\\cline{3-5}\cline{7-9}
 & & MB25 & MO25 & MI25 & & MB25 & MO25 & MI25 &\\\hline
$\alpha$& & ${\color{blue}{0.5323}}$  & ${\color{blue}{0.9901}}$  & ${\color{blue}{0.8938}}$ & & ${\color{blue}{0.7514}}$  & ${\color{blue}{0.9989}}$  & ${\color{blue}{0.9124}}$ &\\
MKT & & ${\color{red}{0.8335}}$  & ${\color{red}{1.0000}}$  & ${\color{red}{0.9847}}$ & & ${\color{red}{0.9858}}$  & ${\color{red}{0.9989}}$  & ${\color{red}{0.9781}}$ &\\
SMB & & $0.4392$  & ${\color{red}{0.6462}}$  & $0.3647$ & & $0.4436$  & $0.6824$  & $0.3406$ &\\
HML & & $0.0449$  & $0.2081$  & $0.1051$ & & $0.0329$  & ${\color{red}{0.8105}}$  & $0.2760$ &\\
RMW & & $-$ & $-$ & $-$ & & ${\color{red}{0.8751}}$  & $0.2903$  & ${\color{red}{0.6221}}$ &\\
CMA & & $-$ & $-$ & $-$ & & ${\color{red}{0.9726}}$  & ${\color{red}{0.9518}}$  & $0.1424$ &\\\hline\hline
  \end{tabular}}
 \end{center}
\end{minipage}\\ 

\vspace{-3mm}\\

\begin{minipage}[h]{0.45\textwidth}
 \begin{center}
  \scalebox{0.55}{\begin{tabular}{cccccccccc} \hline\hline
Whole period & & \multicolumn{3}{c}{FF3} & & \multicolumn{3}{c}{FF5} &\\\cline{3-5}\cline{7-9}
 & & MB25 & MO25 & MI25 & & MB25 & MO25 & MI25 &\\\hline
$\alpha$ & & ${\color{blue}{0.5153}}$  & ${\color{blue}{0.7561}}$  & ${\color{blue}{0.7169}}$  & & ${\color{blue}{0.6718}}$  & ${\color{blue}{0.6614}}$  & ${\color{blue}{0.7569}}$ &\\
MKT & & ${\color{red}{0.6627}}$  & ${\color{red}{0.7268}}$  & ${\color{red}{0.7888}}$  & & ${\color{red}{0.7770}}$  & ${\color{red}{0.7889}}$  & ${\color{red}{0.8056}}$ &\\
SMB & & $0.3436$  & $0.3436$  & $0.2759$  & & $0.2868$  & $0.3239$  & $0.2651$ &\\
HML & & $0.3021$  & $0.4554$  & $0.4483$  & & $0.3183$  & ${\color{red}{0.5235}}$  & ${\color{red}{0.5547}}$ &\\
RMW & & $-$ & $-$ & $-$ & & ${\color{red}{0.8720}}$  & ${\color{red}{0.5210}}$  & ${\color{red}{0.6617}}$ &\\
CMA & & $-$ & $-$ & $-$ & & ${\color{red}{0.6849}}$  & ${\color{red}{0.6550}}$  & $0.3748$ &\\\hline\hline
  \end{tabular}}
 \end{center}
\end{minipage}
  &
\end{tabular}
\vspace*{5pt}
{
\begin{minipage}{420pt}
\scriptsize
\underline{Notes}:
\begin{itemize}
 \item[(1)] The highlighted values indicate the percentage of periods where $\hat{\ve{\lambda}}_t=\ve{0}$ is over half the period.
 \item[(2)] R version 4.4.0 was used to compute the statistics.
\end{itemize}
\end{minipage}}
\end{table}

\clearpage

\begin{table}[p]
\caption{Percentage of the periods of $\hat{\ve{\lambda}}_t=\ve{0}$ (Europe)}\label{ff_tv_table4}
\centering
\begin{tabular}{cc}
 \begin{minipage}[t]{0.45\textwidth}
  \begin{center}
   \scalebox{0.55}{\begin{tabular}{cccccccccc} \hline\hline
In the 1990's & & \multicolumn{3}{c}{FF3} & & \multicolumn{3}{c}{FF5} &\\\cline{3-5}\cline{7-9}
 & & MB25 & MO25 & MI25 & & MB25 & MO25 & MI25 &\\\hline
$\alpha$ & & ${\color{blue}{0.9586}}$  & ${\color{blue}{0.9682}}$  & ${\color{blue}{0.7945}}$ & & ${\color{blue}{0.9561}}$  & ${\color{blue}{0.9515}}$  & ${\color{blue}{0.8869}}$ &\\
MKT & & ${\color{red}{0.8985}}$  & ${\color{red}{0.9435}}$  & ${\color{red}{0.7653}}$ & & ${\color{red}{0.8985}}$  & ${\color{red}{0.9722}}$  & ${\color{red}{0.9470}}$ &\\
SMB & & $0.2297$  & $0.1217$  & $0.2892$ & & $0.2448$  & $0.1161$  & $0.2347$ &\\
HML & & $0.3534$  & $0.4210$  & ${\color{red}{0.7203}}$ & & $0.3650$  & ${\color{red}{0.7168}}$  & ${\color{red}{0.9339}}$ &\\
RMW & & $-$ & $-$ & $-$ & & ${\color{red}{0.8859}}$  & $0.4003$  & ${\color{red}{0.9566}}$ &\\
CMA & & $-$ & $-$ & $-$ & & ${\color{red}{0.8682}}$  & ${\color{red}{0.8703}}$  & ${\color{red}{0.5593}}$ &\\\hline\hline
  \end{tabular}}
 \end{center}
\end{minipage}
  &
\begin{minipage}[t]{0.45\textwidth}
 \begin{center}
  \scalebox{0.55}{\begin{tabular}{cccccccccc} \hline\hline
In the 2000's & & \multicolumn{3}{c}{FF3} & & \multicolumn{3}{c}{FF5} &\\\cline{3-5}\cline{7-9}
 & & MB25 & MO25 & MI25 & & MB25 & MO25 & MI25 &\\\hline
$\alpha$ & & ${\color{blue}{0.5349}}$  & ${\color{blue}{0.5565}}$  & ${\color{blue}{0.6715}}$ & & ${\color{blue}{0.6612}}$  & ${\color{blue}{0.7321}}$  & ${\color{blue}{0.7221}}$ &\\
MKT & & ${\color{red}{0.6295}}$  & ${\color{red}{0.6271}}$  & ${\color{red}{0.6366}}$ & & ${\color{red}{0.6596}}$  & ${\color{red}{0.8356}}$  & ${\color{red}{0.7601}}$ &\\
SMB & & $0.4019$  & $0.3453$  & $0.3848$ & & $0.3894$  & $0.3392$  & $0.3695$ &\\
HML & & $0.1770$  & ${\color{red}{0.6044}}$  & ${\color{red}{0.6029}}$ & & $0.1817$  & ${\color{red}{0.5324}}$  & ${\color{red}{0.5830}}$ &\\
RMW & & $-$ & $-$ & $-$ & & ${\color{red}{0.7551}}$  & $0.4691$  & ${\color{red}{0.8160}}$ &\\
CMA & & $-$ & $-$ & $-$ & & ${\color{red}{0.5232}}$  & ${\color{red}{0.7382}}$  & $0.4327$ &\\\hline\hline
  \end{tabular}}
 \end{center}
\end{minipage}\\ 

\vspace{-3mm}\\

\begin{minipage}[h]{0.45\textwidth}
 \begin{center}
\scalebox{0.55}{\begin{tabular}{cccccccccc} \hline\hline
In the 2010's & & \multicolumn{3}{c}{FF3} & & \multicolumn{3}{c}{FF5} &\\\cline{3-5}\cline{7-9}
 & & MB25 & MO25 & MI25 & & MB25 & MO25 & MI25 &\\\hline
$\alpha$ & & ${\color{blue}{0.6298}}$  & ${\color{blue}{0.6898}}$  & ${\color{blue}{0.6541}}$ &  & ${\color{blue}{0.6844}}$  & ${\color{blue}{0.6867}}$  & ${\color{blue}{0.5878}}$ &\\
MKT & & ${\color{red}{0.6437}}$  & ${\color{red}{0.8884}}$  & ${\color{red}{0.7301}}$ & & ${\color{red}{0.6756}}$  & ${\color{red}{0.8386}}$  & ${\color{red}{0.6411}}$ &\\
SMB & & $0.4423$  & $0.4839$  & $0.4260$ & & $0.4287$  & ${\color{red}{0.5169}}$  & $0.4141$ &\\
HML & & $0.3237$  & ${\color{red}{0.5993}}$  & ${\color{red}{0.5801}}$ & & $0.3566$  & ${\color{red}{0.8600}}$  & ${\color{red}{0.6929}}$ &\\
RMW & & $-$ & $-$ & $-$ & & ${\color{red}{0.7258}}$  & $0.4387$  & ${\color{red}{0.7523}}$ &\\
CMA & & $-$ & $-$ & $-$ & & ${\color{red}{0.8206}}$  & ${\color{red}{0.9425}}$  & ${\color{red}{0.6710}}$ &\\\hline\hline
\end{tabular}}
    \end{center}
  \end{minipage}
  &
\begin{minipage}[h]{0.45\textwidth}
 \begin{center}
  \scalebox{0.55}{\begin{tabular}{cccccccccc} \hline\hline
In the 2020's & & \multicolumn{3}{c}{FF3} & & \multicolumn{3}{c}{FF5} &\\\cline{3-5}\cline{7-9}
 & & MB25 & MO25 & MI25 & & MB25 & MO25 & MI25 &\\\hline
$\alpha$ & & ${\color{blue}{0.6495}}$  & ${\color{blue}{0.5016}}$  & ${\color{blue}{0.6068}}$ & & ${\color{blue}{0.5060}}$  & ${\color{blue}{0.5455}}$  & ${\color{blue}{0.5915}}$ &\\
MKT & & ${\color{red}{0.7284}}$  & ${\color{red}{0.7820}}$  & ${\color{red}{0.8938}}$ & & ${\color{red}{0.5696}}$  & ${\color{red}{0.9529}}$  & ${\color{red}{0.8894}}$ &\\
SMB & & $0.2442$  & $0.1906$  & $0.2410$ & & $0.2760$  & $0.2092$  & $0.2848$ &\\
HML & & $0.0208$  & $0.4710$  & $0.2410$ & & $0.0197$  & ${\color{red}{0.9069}}$  & $0.1993$ &\\
RMW & & $-$ & $-$ & $-$ & & ${\color{red}{0.7393}}$  & $0.4984$  & ${\color{red}{0.7141}}$ &\\
CMA & & $-$ & $-$ & $-$ & & ${\color{red}{0.9934}}$  & ${\color{red}{0.9168}}$  & $0.3636$ &\\\hline\hline
  \end{tabular}}
 \end{center}
\end{minipage}\\ 

\vspace{-3mm}\\

\begin{minipage}[h]{0.45\textwidth}
 \begin{center}
\scalebox{0.55}{\begin{tabular}{cccccccccc} \hline\hline
Whole period & & \multicolumn{3}{c}{FF3} & & \multicolumn{3}{c}{FF5} &\\\cline{3-5}\cline{7-9}
 & & MB25 & MO25 & MI25 & & MB25 & MO25 & MI25 &\\\hline
$\alpha$ & & ${\color{blue}{0.6818}}$  & ${\color{blue}{0.6937}}$  & ${\color{blue}{0.6887}}$  & & ${\color{blue}{0.7232}}$  & ${\color{blue}{0.7501}}$  & ${\color{blue}{0.7045}}$ &\\
MKT & & ${\color{red}{0.7109}}$  & ${\color{red}{0.8058}}$  & ${\color{red}{0.7270}}$  & & ${\color{red}{0.7130}}$  & ${\color{red}{0.8831}}$  & ${\color{red}{0.7820}}$ &\\
SMB & & $0.3551$  & $0.3178$  & $0.3585$  & & $0.3540$  & $0.3272$  & $0.3414$ &\\
HML & & $0.2498$  & ${\color{red}{0.5430}}$  & ${\color{red}{0.5835}}$  & & $0.2645$  & ${\color{red}{0.7249}}$  & ${\color{red}{0.6608}}$ &\\
RMW & & $-$ & $-$ & $-$ & & ${\color{red}{0.7759}}$  & $0.4458$  & ${\color{red}{0.8184}}$ &\\
CMA & & $-$ & $-$ & $-$ & & ${\color{red}{0.7560}}$  & ${\color{red}{0.8562}}$  & ${\color{red}{0.5325}}$ &\\\hline\hline
  \end{tabular}}
 \end{center}
\end{minipage}
\end{tabular}
\vspace*{5pt}
{
\begin{minipage}{420pt}
\scriptsize
\underline{Notes}:
\begin{itemize}
 \item[(1)] The highlighted values indicate the percentage of periods where $\hat{\ve{\lambda}}_t=\ve{0}$ is over half the period.
 \item[(2)] R version 4.4.0 was used to compute the statistics.
\end{itemize}
\end{minipage}}
\end{table}

\clearpage

\begin{landscape}
\begin{figure}[p]
 \caption{Generalized GRS Statistics and Factor Redundancy sorted by Size--B/M Portfolios (U.S., FF3)}\label{ff_tv_fig7}
 \begin{center}
  \includegraphics[scale=0.35]{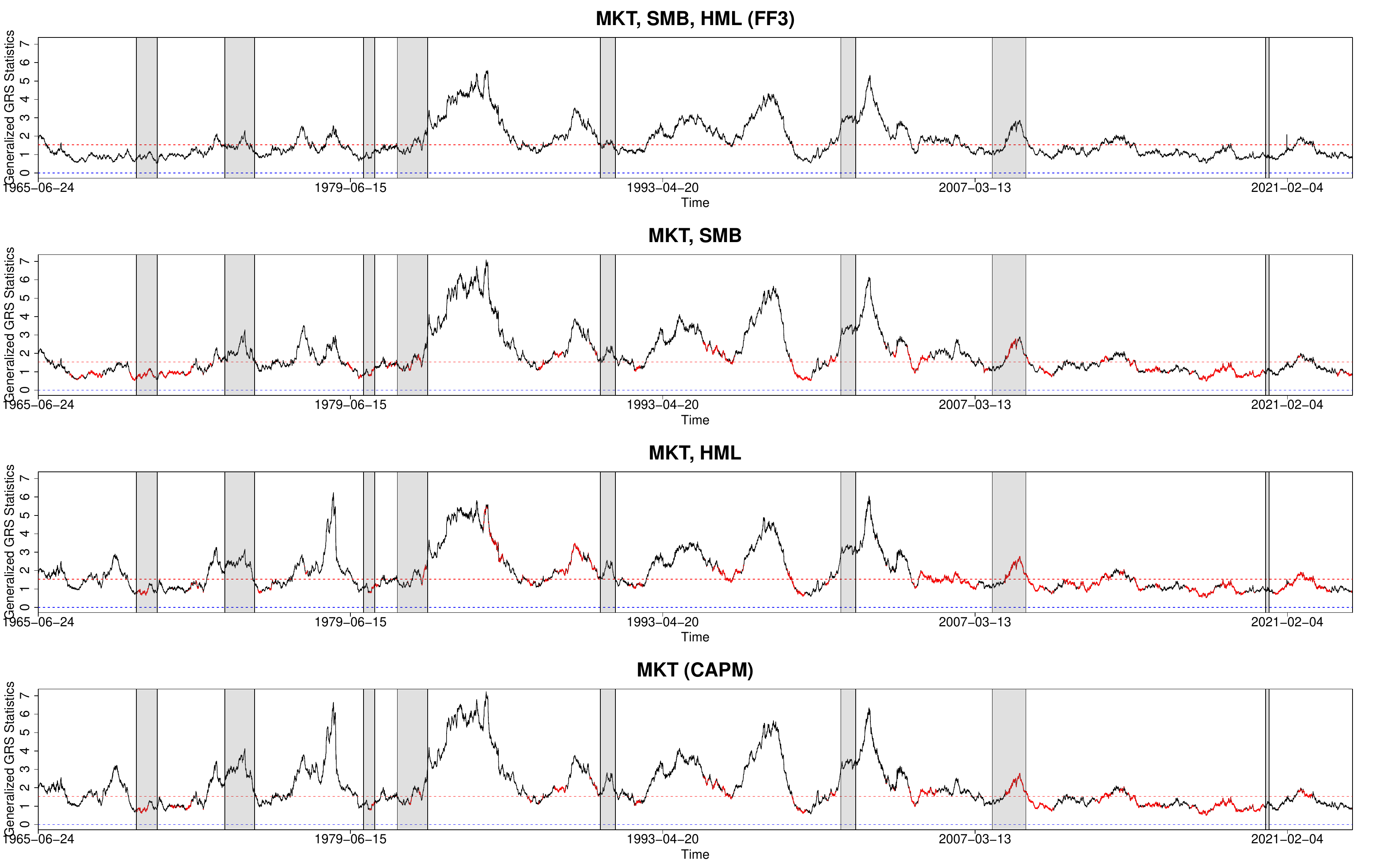}
\vspace*{5pt}
{
\begin{minipage}{600pt}
\scriptsize
\underline{Notes}:
\begin{itemize}
 \item[(1)] The dashed red lines denote the critical value at the 5\% significance levels for the generalized GRS test, and the dashed blue lines denote zero.
 \item[(2)] The red dots denote that the model outperforms the benchmark model at that time.
 \item[(3)] The shade areas represent recessions reported by the NBER ``\href{https://www.nber.org/research/business-cycle-dating}{Business Cycle Dating}.''
 \item[(4)] R version 4.4.0 was used to compute the statistics.
\end{itemize}
\end{minipage}}%
\end{center}
\end{figure}
\end{landscape}

\clearpage

\begin{landscape}
\begin{figure}[p]
 \caption{Generalized GRS Statistics and Factor Redundancy sorted by Size--OP Portfolios (U.S., FF3)}\label{ff_tv_fig8}
 \begin{center}
  \includegraphics[scale=0.35]{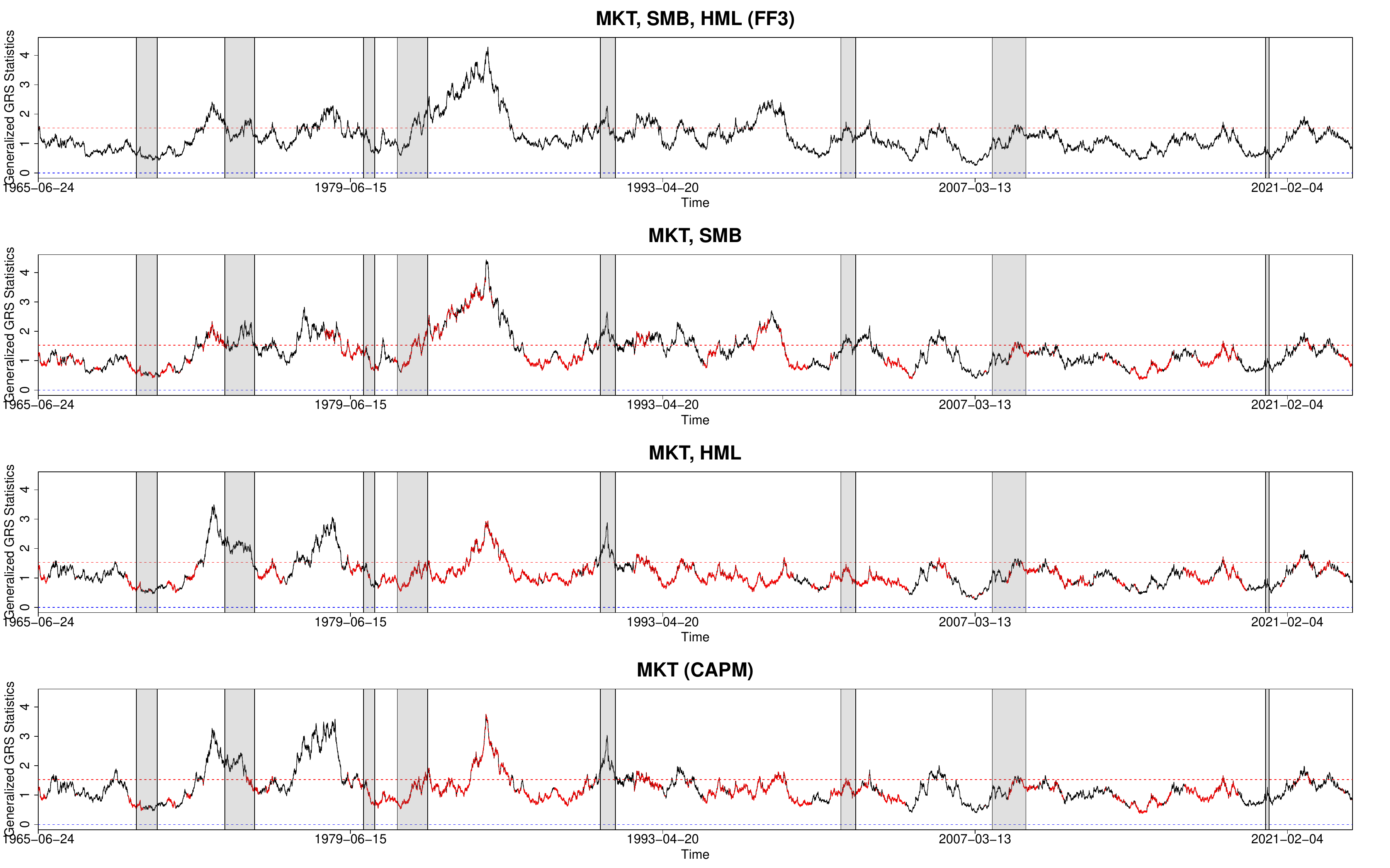}
\vspace*{5pt}
{
\begin{minipage}{600pt}
\scriptsize
\underline{Notes}:
\begin{itemize}
 \item[(1)] The dashed red lines denote the critical value at the 5\% significance levels for the generalized GRS test, and the dashed blue lines denote zero.
 \item[(2)] The red dots denote that the model outperforms the benchmark model at that time.
 \item[(3)] The shade areas represent recessions reported by the NBER ``\href{https://www.nber.org/research/business-cycle-dating}{Business Cycle Dating}.''
 \item[(4)] R version 4.4.0 was used to compute the statistics.
\end{itemize}
\end{minipage}}%
\end{center}
\end{figure}
\end{landscape}

\clearpage

\begin{landscape}
\begin{figure}[p]
 \caption{Generalized GRS Statistics and Factor Redundancy sorted by Size--Inv Portfolios (U.S., FF3)}\label{ff_tv_fig9}
 \begin{center}
  \includegraphics[scale=0.35]{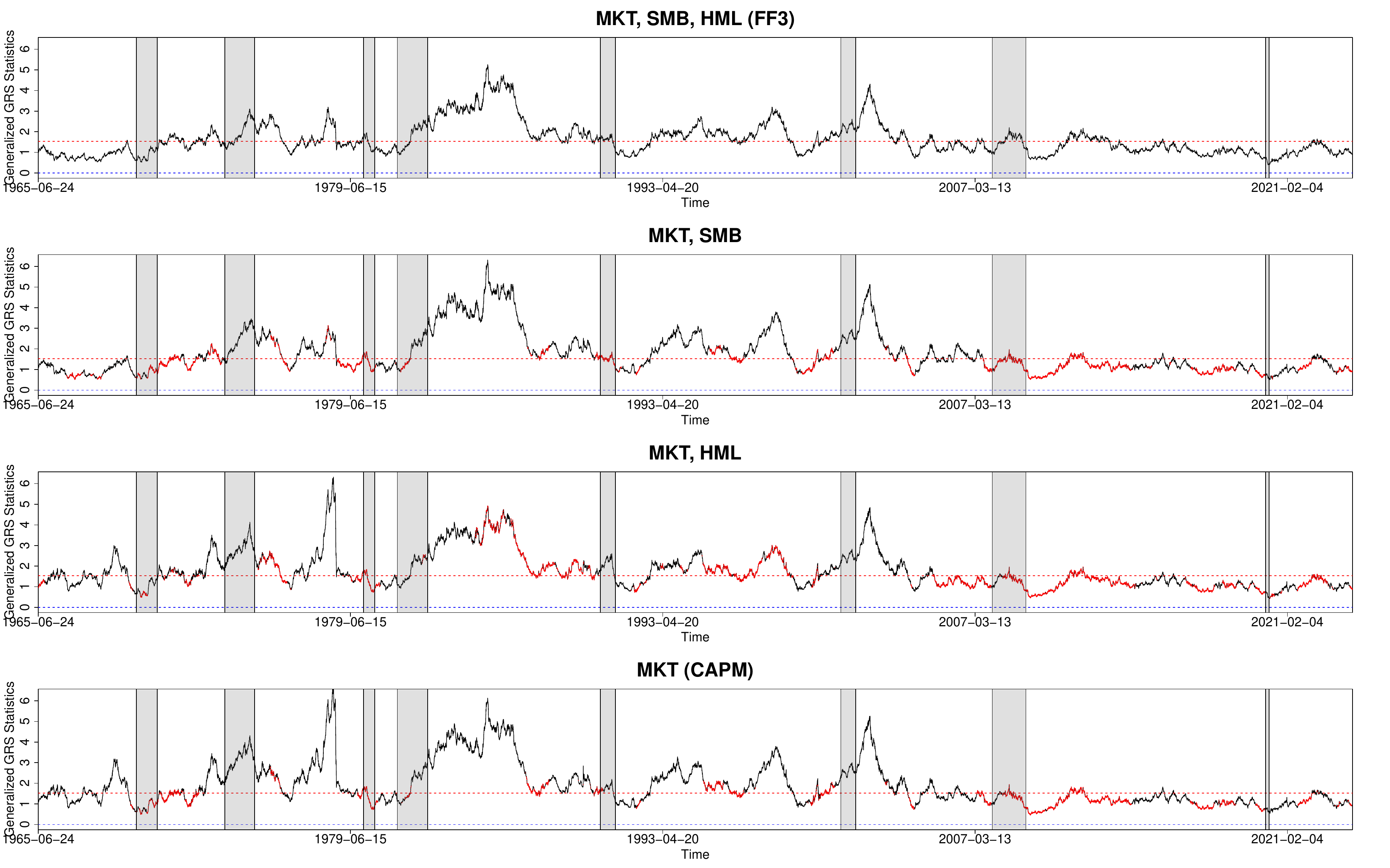}
\vspace*{5pt}
{
\begin{minipage}{600pt}
\scriptsize
\underline{Notes}:
\begin{itemize}
 \item[(1)] The dashed red lines denote the critical value at the 5\% significance levels for the generalized GRS test, and the dashed blue lines denote zero.
 \item[(2)] The red dots denote that the model outperforms the benchmark model at that time.
 \item[(3)] The shade areas represent recessions reported by the NBER ``\href{https://www.nber.org/research/business-cycle-dating}{Business Cycle Dating}.''
 \item[(4)] R version 4.4.0 was used to compute the statistics.
\end{itemize}
\end{minipage}}%
\end{center}
\end{figure}
\end{landscape}

\clearpage

\begin{landscape}
\begin{figure}[p]
 \caption{Generalized GRS Statistics and Factor Redundancy sorted by Size--B/M Portfolios (U.S., FF5)}\label{ff_tv_fig10}
 \begin{center}
  \includegraphics[scale=0.35]{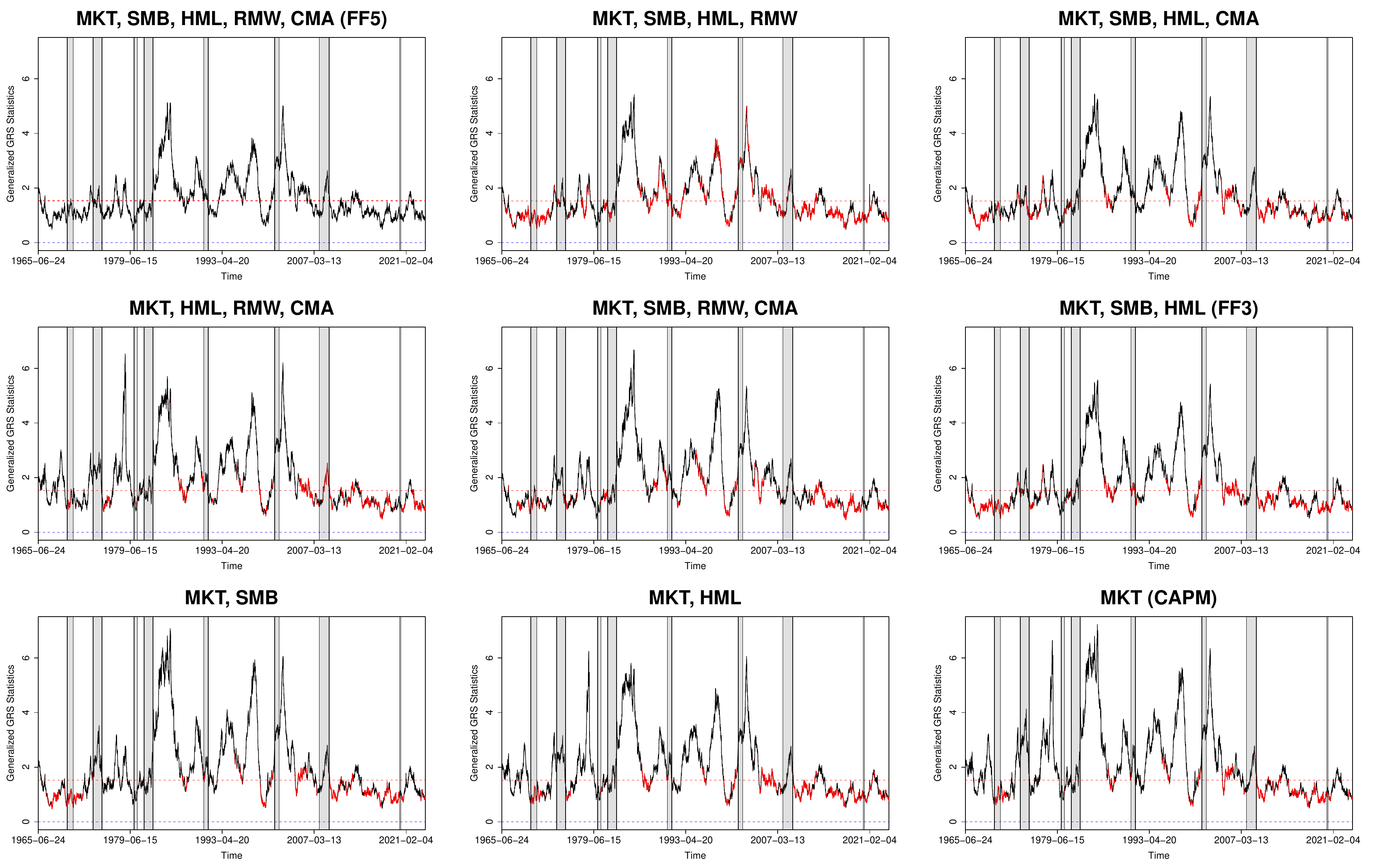}
\vspace*{5pt}
{
\begin{minipage}{600pt}
\scriptsize
\underline{Notes}:
\begin{itemize}
 \item[(1)] The dashed red lines denote the critical value at the 5\% significance levels for the generalized GRS test, and the dashed blue lines denote zero.
 \item[(2)] The red dots denote that the model outperforms the benchmark model at that time.
 \item[(3)] The shade areas represent recessions reported by the NBER ``\href{https://www.nber.org/research/business-cycle-dating}{Business Cycle Dating}.''
 \item[(4)] R version 4.4.0 was used to compute the statistics.
\end{itemize}
\end{minipage}}%
\end{center}
\end{figure}
\end{landscape}

\clearpage

\begin{landscape}
\begin{figure}[p]
 \caption{Generalized GRS Statistics and Factor Redundancy sorted by Size--OP Portfolios (U.S., FF5)}\label{ff_tv_fig11}
 \begin{center}
  \includegraphics[scale=0.35]{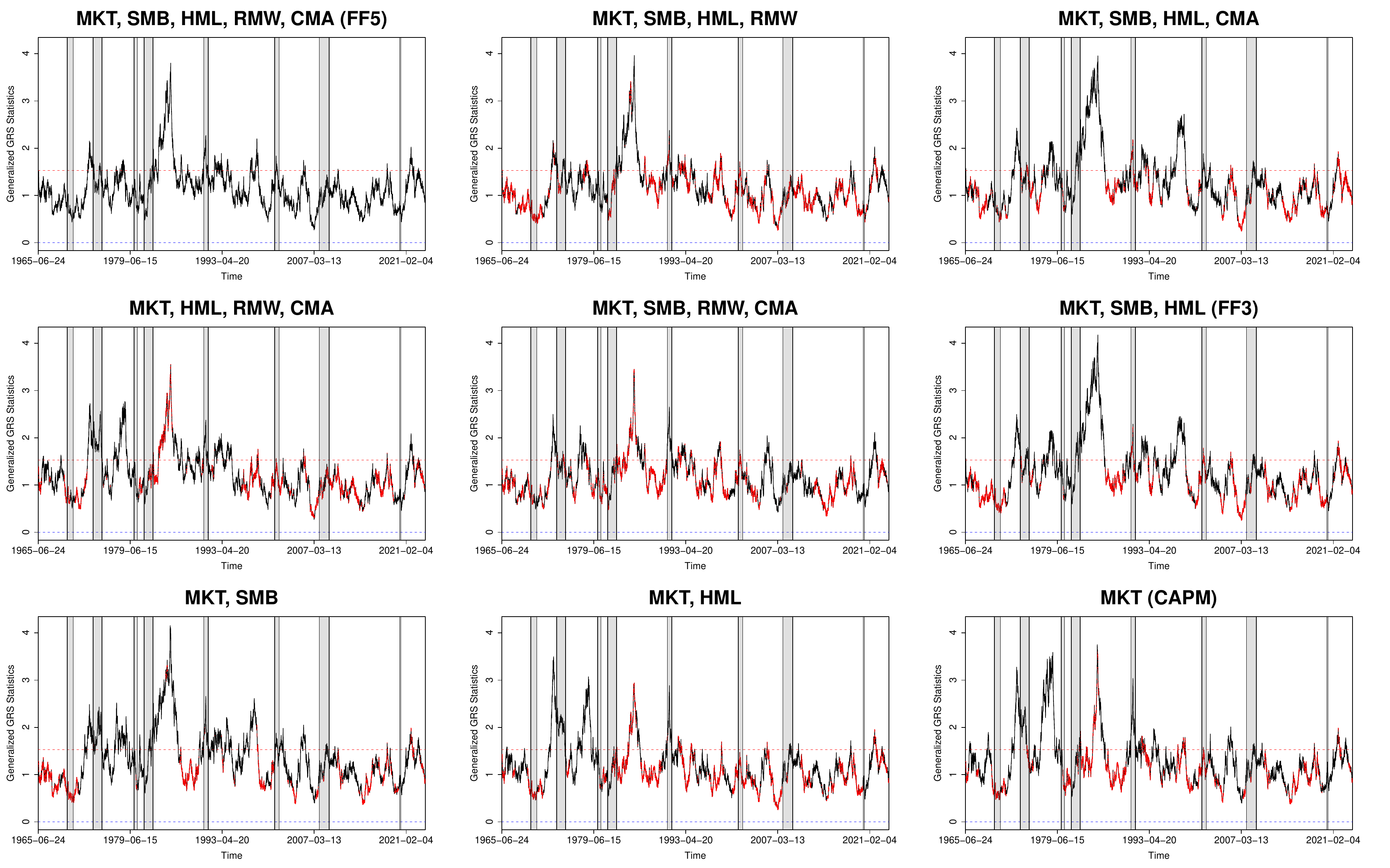}
\vspace*{5pt}
{
\begin{minipage}{600pt}
\scriptsize
\underline{Notes}:
\begin{itemize}
 \item[(1)] The dashed red lines denote the critical value at the 5\% significance levels for the generalized GRS test, and the dashed blue lines denote zero.
 \item[(2)] The red dots denote that the model outperforms the benchmark model at that time.
 \item[(3)] The shade areas represent recessions reported by the NBER ``\href{https://www.nber.org/research/business-cycle-dating}{Business Cycle Dating}.''
 \item[(4)] R version 4.4.0 was used to compute the statistics.
\end{itemize}
\end{minipage}}%
\end{center}
\end{figure}
\end{landscape}

\clearpage

\begin{landscape}
\begin{figure}[p]
 \caption{Generalized GRS Statistics and Factor Redundancy sorted by Size--Inv Portfolios (U.S., FF5)}\label{ff_tv_fig12}
 \begin{center}
  \includegraphics[scale=0.35]{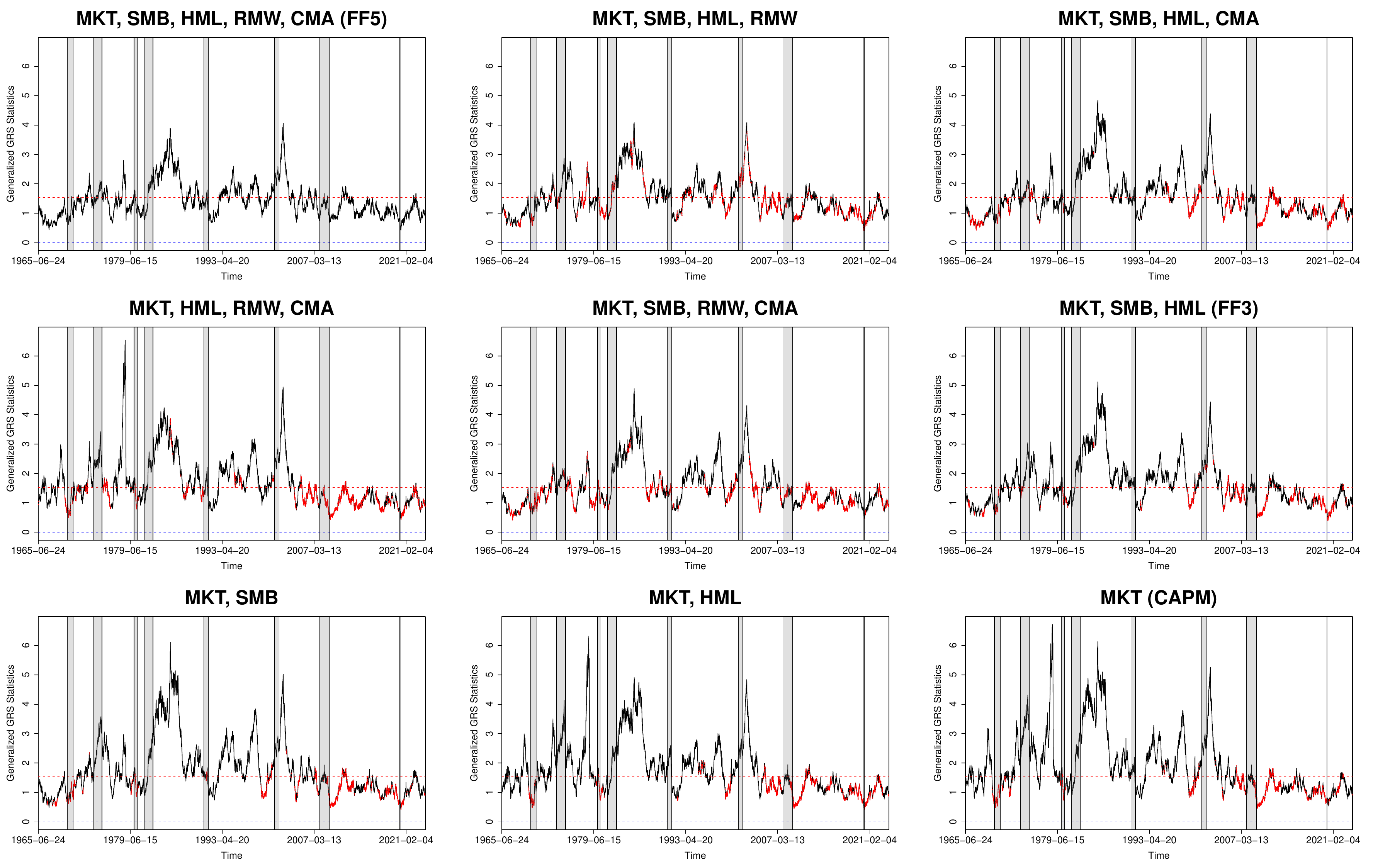}
\vspace*{5pt}
{
\begin{minipage}{600pt}
\scriptsize
\underline{Notes}:
\begin{itemize}
 \item[(1)] The dashed red lines denote the critical value at the 5\% significance levels for the generalized GRS test, and the dashed blue lines denote zero.
 \item[(2)] The red dots denote that the model outperforms the benchmark model at that time.
 \item[(3)] The shade areas represent recessions reported by the NBER ``\href{https://www.nber.org/research/business-cycle-dating}{Business Cycle Dating}.''
 \item[(4)] R version 4.4.0 was used to compute the statistics.
\end{itemize}
\end{minipage}}%
\end{center}
\end{figure}
\end{landscape}

\clearpage

\begin{table}[p]
\caption{Percentage of the periods of the model is valid (U.S.)}\label{ff_tv_table5}
\centering
\begin{tabular}{c}
 \begin{minipage}[h]{\textwidth}
  \begin{center}
   \scalebox{0.6}{\begin{tabular}{cccccccccccccccccc}\hline\hline
 & & \multicolumn{3}{c}{In the 1960's} & & \multicolumn{3}{c}{In the 1970's} & & \multicolumn{3}{c}{In the 1980's} & & \multicolumn{3}{c}{In the 1990's} & \\\cline{3-5}\cline{7-9}\cline{11-13}\cline{15-17}
 & & MB25 & MO25 & MI25 & & MB25 & MO25 & MI25 & & MB25 & MO25 & MI25 & & MB25 & MO25 & MI25 &\\\hline
MSHRC (FF5) & & ${\color{red}{0.9100}}$ & ${\color{red}{0.8560}}$ & ${\color{red}{0.9982}}$ & & ${\color{red}{0.7953}}$ & ${\color{red}{0.6291}}$ & ${\color{red}{0.6667}}$ & & $0.3501$ & $0.3865$ & $0.3382$ & & $0.2864$ & ${\color{red}{0.7354}}$ & ${\color{red}{0.5222}}$ & \\
MSHR & & ${\color{red}{0.9019}}$ & ${\color{red}{0.8938}}$ & ${\color{red}{0.9910}}$ & & ${\color{red}{0.7755}}$ & ${\color{red}{0.6675}}$ & ${\color{red}{0.5622}}$ & & $0.3552$ & $0.4407$ & $0.2824$ & & $0.3204$ & ${\color{red}{0.7658}}$ & $0.4205$ & \\
MSHC & & ${\color{red}{0.9073}}$ & ${\color{red}{0.8605}}$ & ${\color{red}{0.9964}}$ & & ${\color{red}{0.7906}}$ & $0.4565$ & ${\color{red}{0.6002}}$ & & $0.3121$ & $0.2239$ & $0.2520$ & & $0.2551$ & $0.3453$ & $0.3972$ & \\
MHRC & & $0.3186$ & $0.2934$ & ${\color{red}{0.5491}}$ & & ${\color{red}{0.5055}}$ & $0.4339$ & $0.4517$ & & $0.2678$ & $0.3291$ & $0.2801$ & & $0.2856$ & $0.4561$ & $0.4181$ &\\
MSRC & & ${\color{red}{0.8830}}$ & ${\color{red}{0.9478}}$ & ${\color{red}{0.9973}}$ & & ${\color{red}{0.6536}}$ & ${\color{red}{0.5962}}$ & ${\color{red}{0.5895}}$ & & $0.2951$ & $0.3481$ & $0.3477$ & & $0.2294$ & ${\color{red}{0.6618}}$ & $0.4434$ &\\
MSH (FF3) & & ${\color{red}{0.9001}}$ & ${\color{red}{0.8893}}$ & ${\color{red}{0.9883}}$ & & ${\color{red}{0.7700}}$ & ${\color{red}{0.5016}}$ & $0.4913$ & & $0.3236$ & $0.2987$ & $0.2017$ & & $0.2872$ & $0.4778$ & $0.3105$ & \\
MS & & ${\color{red}{0.8758}}$ & ${\color{red}{0.9685}}$ & ${\color{red}{0.9712}}$ & & ${\color{red}{0.6429}}$ & ${\color{red}{0.5736}}$ & $0.4386$ & & $0.2571$ & $0.3172$ & $0.1946$ & & $0.1938$ & $0.4767$ & $0.3125$ & \\
MH & & $0.3141$ & $0.3303$ & ${\color{red}{0.5095}}$ & & ${\color{red}{0.5020}}$ & $0.3203$ & $0.2866$ & & $0.3287$ & $0.3291$ & $0.2235$ & & $0.2595$ & $0.4130$ & $0.3568$ & \\
M (CAPM) & & $0.3069$ & $0.3411$ & $0.4923$ & & $0.4382$ & $0.3424$ & $0.2526$ & & $0.2820$ & $0.3398$ & $0.2077$ & & $0.1946$ & $0.4407$ & $0.3252$ & \\\hline\hline
\end{tabular}}
\end{center}
\end{minipage}\\ 

\vspace{-3mm}\\

\begin{minipage}[h]{\textwidth}
 \begin{center}
  \scalebox{0.6}{\begin{tabular}{cccccccccccccccccc}\hline\hline
 & & \multicolumn{3}{c}{In the 2000's} & & \multicolumn{3}{c}{In the 2010's} & & \multicolumn{3}{c}{In the 2020's} & & \multicolumn{3}{c}{Whole period} & \\\cline{3-5}\cline{7-9}\cline{11-13}\cline{15-17}
 & & MB25 & MO25 & MI25 & & MB25 & MO25 & MI25 & & MB25 & MO25 & MI25 & & MB25 & MO25 & MI25 &\\\hline
MSHRC (FF5) & & $0.3586$ & ${\color{red}{0.5499}}$ & ${\color{red}{0.5455}}$ & & ${\color{red}{0.8859}}$ & ${\color{red}{0.9344}}$ & ${\color{red}{0.9169}}$ & & ${\color{red}{0.8648}}$ & ${\color{red}{0.9977}}$ & ${\color{red}{0.9795}}$ & & ${\color{red}{0.5835}}$ & ${\color{red}{0.6839}}$ & ${\color{red}{0.6511}}$ & \\
MSHR & & $0.3646$ & ${\color{red}{0.5260}}$ & ${\color{red}{0.5221}}$ & & ${\color{red}{0.8724}}$ & ${\color{red}{0.9487}}$ & ${\color{red}{0.8565}}$ & & ${\color{red}{0.8409}}$ & ${\color{red}{0.9977}}$ & ${\color{red}{0.9511}}$ & & ${\color{red}{0.5835}}$ & ${\color{red}{0.7064}}$ & ${\color{red}{0.5891}}$ & \\
MSHC & & $0.3519$ & ${\color{red}{0.5026}}$ & ${\color{red}{0.5272}}$ & & ${\color{red}{0.8824}}$ & ${\color{red}{0.9145}}$ & ${\color{red}{0.9165}}$ & & ${\color{red}{0.7864}}$ & ${\color{red}{0.9966}}$ & ${\color{red}{0.9875}}$ & & ${\color{red}{0.5640}}$ & ${\color{red}{0.5471}}$ & ${\color{red}{0.6002}}$ & \\
MHRC & & $0.3964$ & $0.4740$ & ${\color{red}{0.5507}}$ & & ${\color{red}{0.8867}}$ & ${\color{red}{0.9388}}$ & ${\color{red}{0.9801}}$ & & ${\color{red}{0.8773}}$ & ${\color{red}{0.9989}}$ & ${\color{red}{0.9773}}$ & & ${\color{red}{0.4814}}$ & ${\color{red}{0.5368}}$ & ${\color{red}{0.5633}}$ & \\
MSRC & & $0.2954$ & $0.4477$ & $0.4588$ & & ${\color{red}{0.9185}}$ & ${\color{red}{0.9324}}$ & ${\color{red}{0.9718}}$ & & ${\color{red}{0.7955}}$ & ${\color{red}{0.9977}}$ & ${\color{red}{0.9852}}$ & & ${\color{red}{0.5281}}$ & ${\color{red}{0.6479}}$ & ${\color{red}{0.6206}}$ & \\
MSH (FF3) & & $0.4227$ & $0.4692$ & $0.4950$ & & ${\color{red}{0.8649}}$ & ${\color{red}{0.9344}}$ & ${\color{red}{0.8525}}$ & & ${\color{red}{0.7625}}$ & ${\color{red}{0.9932}}$ & ${\color{red}{0.9500}}$ & & ${\color{red}{0.5751}}$ & ${\color{red}{0.5904}}$ & ${\color{red}{0.5382}}$ & \\
MS & & $0.2815$ & $0.3706$ & $0.4457$ & & ${\color{red}{0.8792}}$ & ${\color{red}{0.9308}}$ & ${\color{red}{0.9201}}$ & & ${\color{red}{0.7489}}$ & ${\color{red}{0.9875}}$ & ${\color{red}{0.8841}}$ & & ${\color{red}{0.5010}}$ & ${\color{red}{0.5940}}$ & ${\color{red}{0.5261}}$ & \\
MH & & $0.4294$ & $0.4684$ & ${\color{red}{0.5316}}$ & & ${\color{red}{0.8680}}$ & ${\color{red}{0.9281}}$ & ${\color{red}{0.9300}}$ & & ${\color{red}{0.9705}}$ & ${\color{red}{0.9977}}$ & ${\color{red}{0.9955}}$ & & $0.4945$ & ${\color{red}{0.5097}}$ & ${\color{red}{0.5005}}$ & \\
M (CAPM) & & $0.2775$ & $0.4485$ & ${\color{red}{0.5121}}$ & & ${\color{red}{0.8943}}$ & ${\color{red}{0.9384}}$ & ${\color{red}{0.9424}}$ & & ${\color{red}{0.8352}}$ & ${\color{red}{0.9966}}$ & ${\color{red}{0.9466}}$ & & $0.4339$ & ${\color{red}{0.5192}}$ & $0.4810$ & \\\hline\hline
  \end{tabular}}
 \end{center}
\end{minipage}
\end{tabular}
\vspace*{5pt}
{
\begin{minipage}{420pt}
\scriptsize
\underline{Notes}:
\begin{itemize}
 \item[(1)] The highlighted values indicate that the model is valid over half the period.
 \item[(2)] R version 4.4.0 was used to compute the statistics.
\end{itemize}
\end{minipage}}
\end{table}

\clearpage

\begin{table}[p]
\caption{Model ranking results (U.S.)}\label{ff_tv_table6}
\centering
\begin{tabular}{cc}
  \begin{minipage}[t]{0.45\textwidth}
   \begin{center}
    \scalebox{0.5}{\begin{tabular}{cccccccccc} \hline\hline
In the 1960's & & \multicolumn{3}{c}{FF3} & & \multicolumn{3}{c}{FF5} &\\\cline{3-5}\cline{7-9}
 & & MB25 & MO25 & MI25 & & MB25 & MO25 & MI25 &\\\hline
MSHRC (FF5) & & $-$ &  $-$ &  $-$ & & $-$ &  $-$ &  $-$ & \\
MSHR & & $-$  & $-$ & $-$ & & ${\color{red}{0.5563}}$ &  ${\color{red}{0.6202}}$ & $0.1881$ &\\
MSHC & & $-$ & $-$ &  $-$ & & ${\color{red}{0.5122}}$ & ${\color{red}{0.5320}}$ & $0.3933$& \\
MHRC & & $-$ & $-$ &  $-$ & & $0.0153$  & $0.1053$  & $0.0459$&\\
MSRC & &$-$ & $-$ &  $-$ & & $0.0144$ &  $0.0706$  & $0.0270$&\\
MSH (FF3) & & $-$ &  $-$ &  $-$ & & ${\color{red}{0.6409}}$  & ${\color{red}{0.7759}}$  & $0.1899$&\\
MS & &$0.2529$ &  ${\color{red}{0.7678}}$ &  $0.3573$ & & $0.3843$ &  ${\color{red}{0.8749}}$  & $0.1341$&\\
MH & &$0.0153$ &  $0.0081$ &  $0.0639$ & & $0.0306$ & $0.1269$  & $0.0297$&\\
M (CAPM) & &$0.0180$ & $0.0092$ &  $0.0549$ & & $0.0324$ &  $0.1422$  & $0.0441$&\\\hline\hline
  \end{tabular}}
 \end{center}
\end{minipage}
 &
 \begin{minipage}[t]{0.45\textwidth}
  \begin{center}
   \scalebox{0.5}{\begin{tabular}{cccccccccc} \hline\hline
In the 1970's & & \multicolumn{3}{c}{FF3} & & \multicolumn{3}{c}{FF5} &\\\cline{3-5}\cline{7-9}
 & & MB25 & MO25 & MI25 & & MB25 & MO25 & MI25 &\\\hline
MSHRC (FF5) & & $-$ & $-$ & $-$ & & $-$ & $-$ & $-$ &\\
MSHR & & $-$ & $-$ & $-$ & & $0.4303$ & ${\color{red}{0.5158}}$ & $0.2035$&\\
MSHC & & $-$ & $-$ & $-$ & & $0.3056$ & $0.1849$ & $0.1489$&\\
MHRC & & $-$ & $-$ & $-$ & & $0.1132$ & $0.2130$ & $0.2585$ &\\
MSRC & & $-$ & $-$ & $-$ & & $0.0382$ & $0.0801$ & $0.0819$ &\\
MSH (FF3) & & $-$ & $-$ & $-$ & & $0.4173$ & $0.3199$ & $0.0736$ &\\
MS & & $0.3266$ & ${\color{red}{0.5669}}$ & $0.4462$ & & $0.2284$ & $0.3076$ & $0.1120$ &\\
MH & & $0.1833$ & $0.0359$ & $0.3116$ & & $0.2415$ & $0.1896$ & $0.0717$&\\
M (CAPM) & & $0.0990$ & $0.0536$ & $0.2831$ & & $0.1675$ & $0.2447$ & $0.1243$&\\\hline\hline
  \end{tabular}}
 \end{center}
\end{minipage}\\ 

\vspace{-3mm}\\

\begin{minipage}[h]{0.45\textwidth}
 \begin{center}
  \scalebox{0.5}{\begin{tabular}{cccccccccc} \hline\hline
In the 1980's & & \multicolumn{3}{c}{FF3} & & \multicolumn{3}{c}{FF5} &\\\cline{3-5}\cline{7-9}
 & & MB25 & MO25 & MI25 & & MB25 & MO25 & MI25 &\\\hline
MSHRC (FF5) & & $-$ & $-$ & $-$ & & $-$ & $-$ & $-$&\\
MSHR & & $-$ & $-$ & $-$ & & $0.2967$ & $0.4676$ & $0.1974$& \\
MSHC & & $-$ & $-$ & $-$ & & $0.1689$ & $0.0000$ & $0.0083$&\\
MHRC & & $-$ & $-$ & $-$ & &$0.1697$ & $0.1863$ & $0.2251$&\\
MSRC & & $-$ & $-$ & $-$ & &$0.0145$ & $0.0568$ & $0.0538$&\\
MSH (FF3) & & $-$ & $-$ & $-$ & & $0.1705$ & $0.0538$ & $0.0637$&\\
MS & & $0.0775$ & ${\color{red}{0.6562}}$ & $0.1705$ & & $0.0376$ & $0.1523$ & $0.0427$&\\
MH & & $0.4288$ & $0.0932$ & ${\color{red}{0.5079}}$ & & $0.1483$ & $0.1108$ & $0.0593$&\\
M (CAPM) & & $0.1938$ & $0.1144$ & $0.1808$ & & $0.0902$ & $0.2409$ & $0.0494$ &\\\hline\hline
  \end{tabular}}
 \end{center}
\end{minipage}
&
\begin{minipage}[h]{0.45\textwidth}
 \begin{center}
  \scalebox{0.5}{\begin{tabular}{cccccccccc} \hline\hline
In the 1990's & & \multicolumn{3}{c}{FF3} & & \multicolumn{3}{c}{FF5} &\\\cline{3-5}\cline{7-9}
 & & MB25 & MO25 & MI25 & & MB25 & MO25 & MI25 &\\\hline
MSHRC (FF5) & & $-$ & $-$ & $-$ & & $-$ & $-$ & $-$&\\
MSHR & & $-$ & $-$ & $-$ & & ${\color{red}{0.5063}}$ & ${\color{red}{0.6574}}$ & $0.2429$&\\
MSHC & & $-$ & $-$ & $-$ & & $0.1286$ & $0.0174$ & $0.1424$&\\
MHRC & & $-$ & $-$ & $-$ & & $0.2650$ & $0.2591$ & $0.2429$&\\
MSRC & & $-$ & $-$ & $-$ & & $0.0542$ & $0.0635$ & $0.0299$&\\
MSH (FF3) & & $-$ & $-$ & $-$ & & $0.2239$ & $0.2057$ & $0.1119$&\\
MS & & $0.2504$ & $0.4509$ & $0.2211$ & & $0.2124$ & $0.2591$ & $0.1005$&\\
MH & & $0.3228$ & $0.0638$ & ${\color{red}{0.5581}}$ & & $0.2029$ & $0.2239$ & $0.1970$&\\
M (CAPM) & &$0.1812$ & $0.0603$ & $0.3521$ & & $0.1531$ & $0.1930$ & $0.1721$&\\\hline\hline
  \end{tabular}}
 \end{center}
\end{minipage}\\ 

\vspace{-3mm}\\

\begin{minipage}[h]{0.45\textwidth}
 \begin{center}
  \scalebox{0.5}{\begin{tabular}{cccccccccc} \hline\hline
In the 2000's & & \multicolumn{3}{c}{FF3} & & \multicolumn{3}{c}{FF5} &\\\cline{3-5}\cline{7-9}
 & & MB25 & MO25 & MI25 & & MB25 & MO25 & MI25 &\\\hline
MSHRC (FF5) & & $-$ & $-$ & $-$ & & $-$ & $-$ & $-$ &\\
MSHR & &$-$ & $-$ & $-$ & & ${\color{red}{0.5336}}$ & $0.4930$ & $0.4648$&\\
MSHC & &$-$ & $-$ & $-$ & & $0.3241$ & $0.1356$ & $0.3980$&\\
MHRC & &$-$ & $-$ & $-$ & & $0.4286$ & $0.4131$ & $0.4604$&\\
MSRC & &$-$ & $-$ & $-$ & & $0.0221$ & $0.0626$ & $0.0527$&\\
MSH (FF3) & &$-$ & $-$ & $-$ & & $0.3865$ & $0.2016$ & $0.3268$&\\
MS & &$0.1889$ & $0.3547$ & $0.3590$ & & $0.2537$ & $0.1571$ & $0.2155$&\\
MH & &$0.4111$ & $0.0957$ & $0.4903$ &&  $0.3590$ & $0.3451$ & $0.3511$&\\
M (CAPM) & &$0.1984$ & $0.0683$ & $0.3738$ & & $0.2207$ & $0.2080$ & $0.3141$&\\\hline\hline
  \end{tabular}}
 \end{center}
\end{minipage}
&
\begin{minipage}[h]{0.45\textwidth}
 \begin{center}
  \scalebox{0.5}{\begin{tabular}{cccccccccc} \hline\hline
In the 2010's & & \multicolumn{3}{c}{FF3} & & \multicolumn{3}{c}{FF5} &\\\cline{3-5}\cline{7-9}
 & & MB25 & MO25 & MI25 & & MB25 & MO25 & MI25 &\\\hline
MSHRC (FF5) & & $-$ & $-$ & $-$ & & $-$ & $-$ & $-$\\
MSHR & & $-$ & $-$ & $-$ & & ${\color{red}{0.6729}}$ & $0.4348$ & $0.3732$\\
MSHC & & $-$ & $-$ & $-$ & & $0.4173$ & $0.4372$ & $0.4690$\\
MHRC & & $-$ & $-$ & $-$ & & $0.4249$ & ${\color{red}{0.6061}}$ & ${\color{red}{0.6705}}$\\
MSRC & & $-$ & $-$ & $-$ & & $0.1134$ & $0.0789$ & $0.1022$\\
MSH (FF3) & & $-$ & $-$ & $-$ & & ${\color{red}{0.5358}}$ & $0.4634$ & $0.3855$\\
MS & & ${\color{red}{0.5346}}$ & ${\color{red}{0.5628}}$ & ${\color{red}{0.5878}}$ & & ${\color{red}{0.5815}}$ & ${\color{red}{0.5012}}$ & ${\color{red}{0.5234}}$\\
MH & &${\color{red}{0.5723}}$ & $0.0925$ & ${\color{red}{0.6300}}$ & & ${\color{red}{0.5541}}$ & ${\color{red}{0.5314}}$ & ${\color{red}{0.5501}}$\\
M (CAPM) & &${\color{red}{0.6490}}$ & $0.0964$ & ${\color{red}{0.5763}}$ & & ${\color{red}{0.6701}}$ & ${\color{red}{0.5199}}$ & ${\color{red}{0.5700}}$\\\hline\hline
  \end{tabular}}
 \end{center}
\end{minipage}\\ 

\vspace{-3mm}\\

\begin{minipage}[h]{0.45\textwidth}
 \begin{center}
  \scalebox{0.5}{\begin{tabular}{cccccccccc} \hline\hline
In the 2020's & & \multicolumn{3}{c}{FF3} & & \multicolumn{3}{c}{FF5} &\\\cline{3-5}\cline{7-9}
 & & MB25 & MO25 & MI25 & & MB25 & MO25 & MI25 &\\\hline
MSHRC (FF5) & & $-$ & $-$ & $-$ & & $-$ & $-$ & $-$\\
MSHR & & $-$ & $-$ & $-$ & & ${\color{red}{0.5341}}$ & $0.4926$ & $0.4784$&\\
MSHC & & $-$ & $-$ & $-$ & & $0.3284$ & $0.3568$ & ${\color{red}{0.5625}}$&\\
MHRC & & $-$ & $-$ & $-$ & & ${\color{red}{0.5352}}$ & ${\color{red}{0.5455}}$ & $0.4432$&\\
MSRC & & $-$ & $-$ & $-$ & & $0.0127$ & $0.0145$ & $0.0175$&\\
MSH (FF3) & & $-$ & $-$ & $-$ & & $0.4239$ & $0.4676$ & $0.4216$&\\
MS & & $0.0761$ & $0.2765$ & $0.3773$ & & $0.1693$ & $0.1615$ & $0.4102$&\\
MH & & ${\color{red}{0.7489}}$ & $0.0386$ & ${\color{red}{0.7068}}$ & & ${\color{red}{0.7625}}$ & ${\color{red}{0.5670}}$ & ${\color{red}{0.6580}}$&\\
M (CAPM) & & $0.2932$ & $0.0203$ & $0.2648$ & & $0.2818$ & $0.3670$ & $0.2795$ &\\
\hline\hline
  \end{tabular}}
  \end{center}
 \end{minipage}
 &
 \begin{minipage}[h]{0.45\textwidth}
  \begin{center}
   \scalebox{0.5}{\begin{tabular}{cccccccccc} \hline\hline
Whole period & & \multicolumn{3}{c}{FF3} & & \multicolumn{3}{c}{FF5} &\\\cline{3-5}\cline{7-9}
 & & MB25 & MO25 & MI25 & & MB25 & MO25 & MI25 &\\\hline
MSHRC (FF5) & & $-$ & $-$ & $-$ & & $-$ & $-$ & $-$&\\
MSHR & & $-$ & $-$ & $-$ & & $0.4958$ & ${\color{red}{0.5206}}$ & $0.2989$&\\
MSHC & & $-$ & $-$ & $-$ & & $0.2908$ & $0.1956$ & $0.2650$&\\
MHRC & & $-$ & $-$ & $-$ & & $0.2753$ & $0.3304$ & $0.3507$& \\
MSRC & & $-$ & $-$ & $-$ & & $0.2696$ & $0.4270$ & $0.3650$&\\
MSH (FF3) & & $-$ & $-$ & $-$ & & $0.3736$ & $0.3020$ & $0.2057$& \\
MS & & $0.2617$ & ${\color{red}{0.5228}}$ & $0.3580$ & & $0.2661$ & $0.3141$ & $0.2064$&\\
MH & & $0.3775$ & $0.4277$ & $0.4788$ & & $0.3081$ & $0.2855$ & $0.2539$&\\
M (CAPM) & & $0.2471$ & $0.4227$ & $0.3250$ & & $0.2440$ & $0.2757$ & $0.2323$&\\\hline\hline
   \end{tabular}}
  \end{center}
 \end{minipage}
\end{tabular}
\vspace*{5pt}
{
\begin{minipage}{420pt}
\scriptsize
\underline{Notes}:
\begin{itemize}
 \item[(1)] The highlighted values indicate that the model outperforms the benchmark model over the half period.
 \item[(2)] R version 4.4.0 was used to compute the statistics.
\end{itemize}
\end{minipage}}
\end{table}

\clearpage

\begin{landscape}
\begin{figure}[p]
 \caption{Generalized GRS Statistics and Factor Redundancy sorted by Size--B/M Portfolios (Japan, FF3)}\label{ff_tv_fig13}
 \begin{center}
  \includegraphics[scale=0.35]{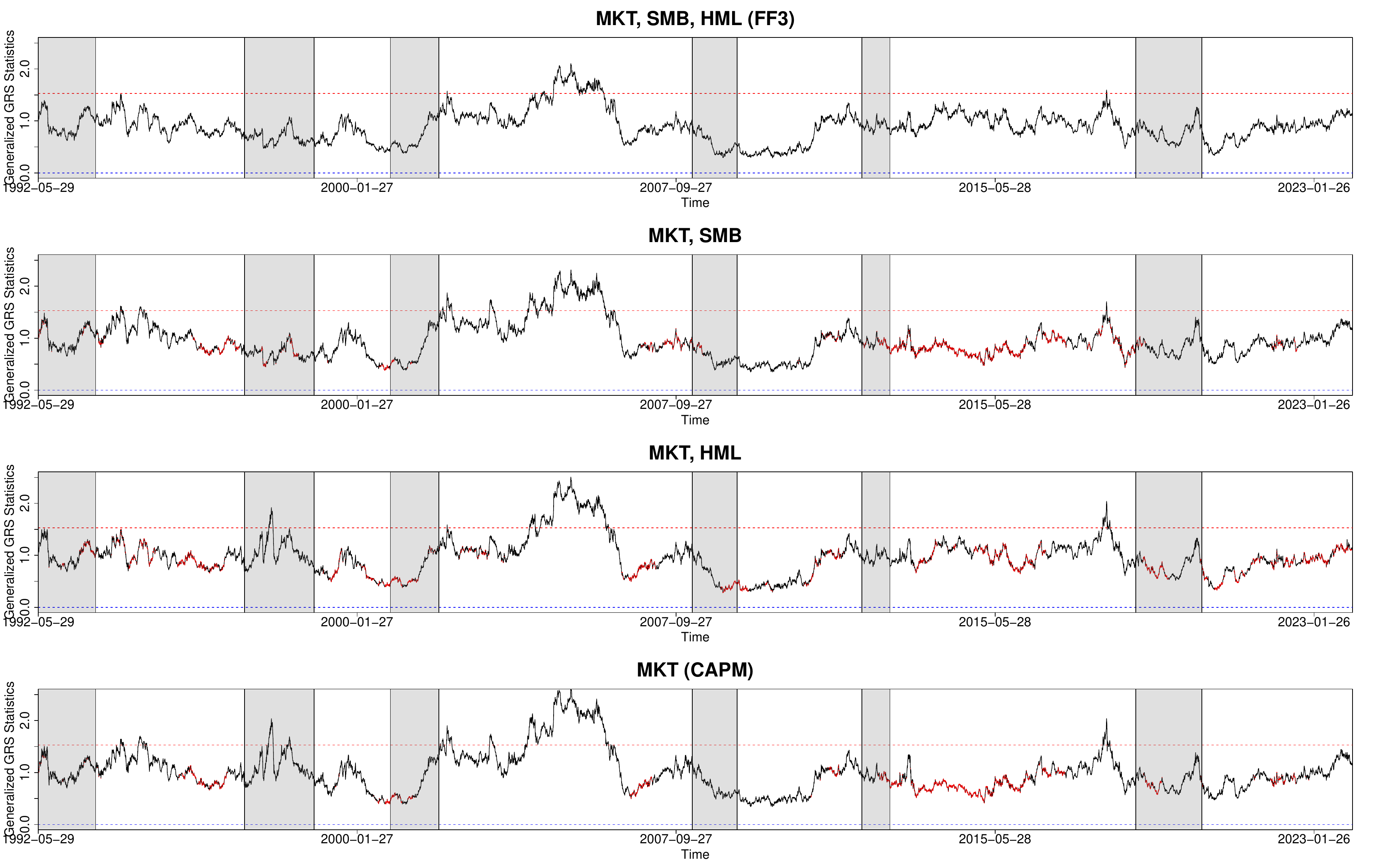}
\vspace*{5pt}
{
\begin{minipage}{600pt}
\scriptsize
\underline{Notes}:
\begin{itemize}
 \item[(1)] The dashed red lines denote the critical value at the 5\% significance levels for the generalized GRS test, and the dashed blue lines denote zero.
 \item[(2)] The red dots denote that the model outperforms the benchmark model at that time.
 \item[(3)] The shade areas represent recessions reported by the NBER ``\href{https://www.nber.org/research/business-cycle-dating}{Business Cycle Dating}.''
 \item[(4)] R version 4.4.0 was used to compute the statistics.
\end{itemize}
\end{minipage}}%
\end{center}
\end{figure}
\end{landscape}

\clearpage

\begin{landscape}
\begin{figure}[p]
 \caption{Generalized GRS Statistics and Factor Redundancy sorted by Size--OP Portfolios (Japan, FF3)}\label{ff_tv_fig14}
 \begin{center}
  \includegraphics[scale=0.35]{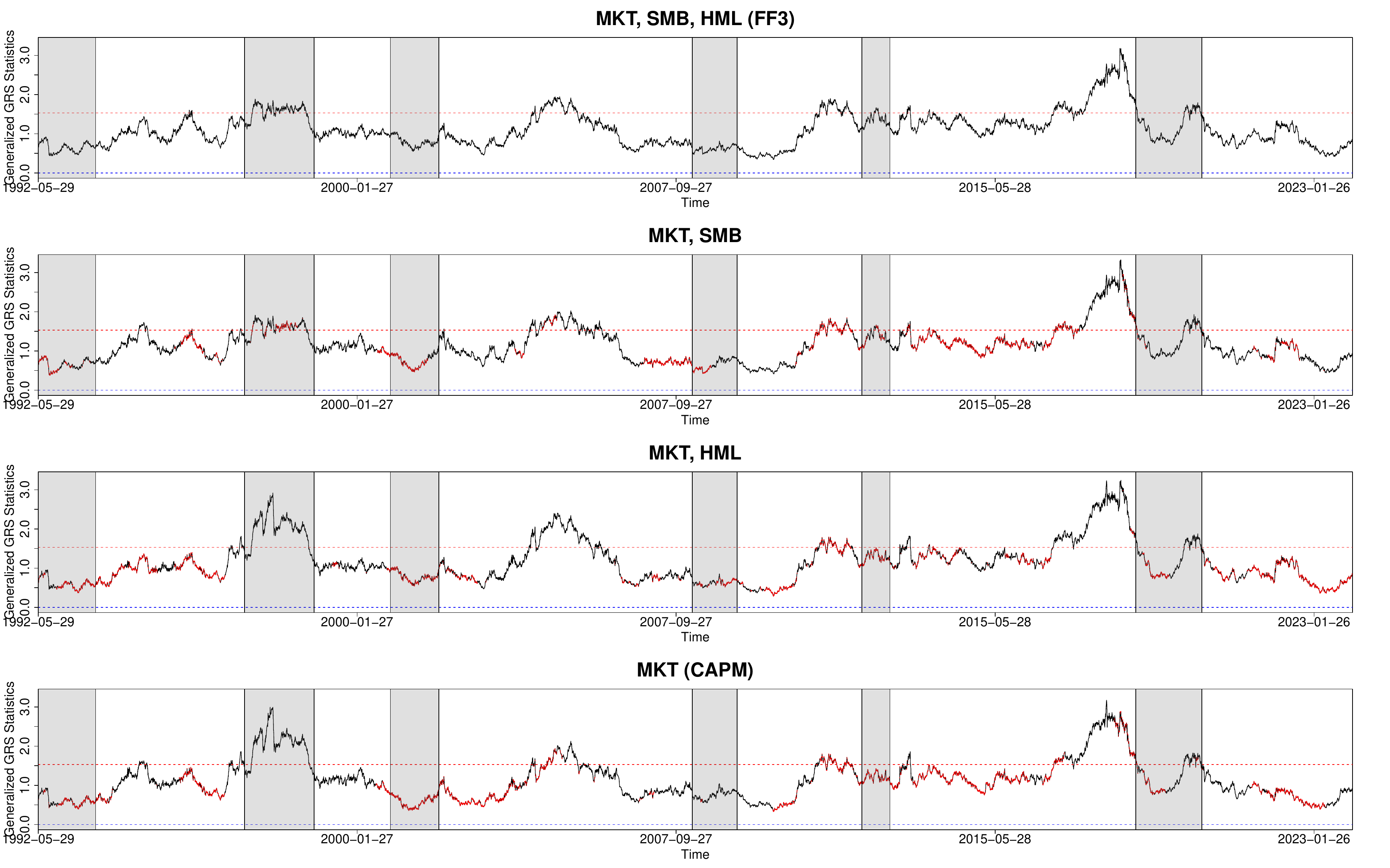}
\vspace*{5pt}
{
\begin{minipage}{600pt}
\scriptsize
\underline{Notes}:
\begin{itemize}
 \item[(1)] The dashed red lines denote the critical value at the 5\% significance levels for the generalized GRS test, and the dashed blue lines denote zero.
 \item[(2)] The red dots denote that the model outperforms the benchmark model at that time.
 \item[(3)] The shade areas represent recessions reported by the NBER ``\href{https://www.nber.org/research/business-cycle-dating}{Business Cycle Dating}.''
 \item[(4)] R version 4.4.0 was used to compute the statistics.
\end{itemize}
\end{minipage}}%
\end{center}
\end{figure}
\end{landscape}

\clearpage

\begin{landscape}
\begin{figure}[p]
 \caption{Generalized GRS Statistics and Factor Redundancy sorted by Size--Inv Portfolios (Japan, FF3)}\label{ff_tv_fig15}
 \begin{center}
  \includegraphics[scale=0.35]{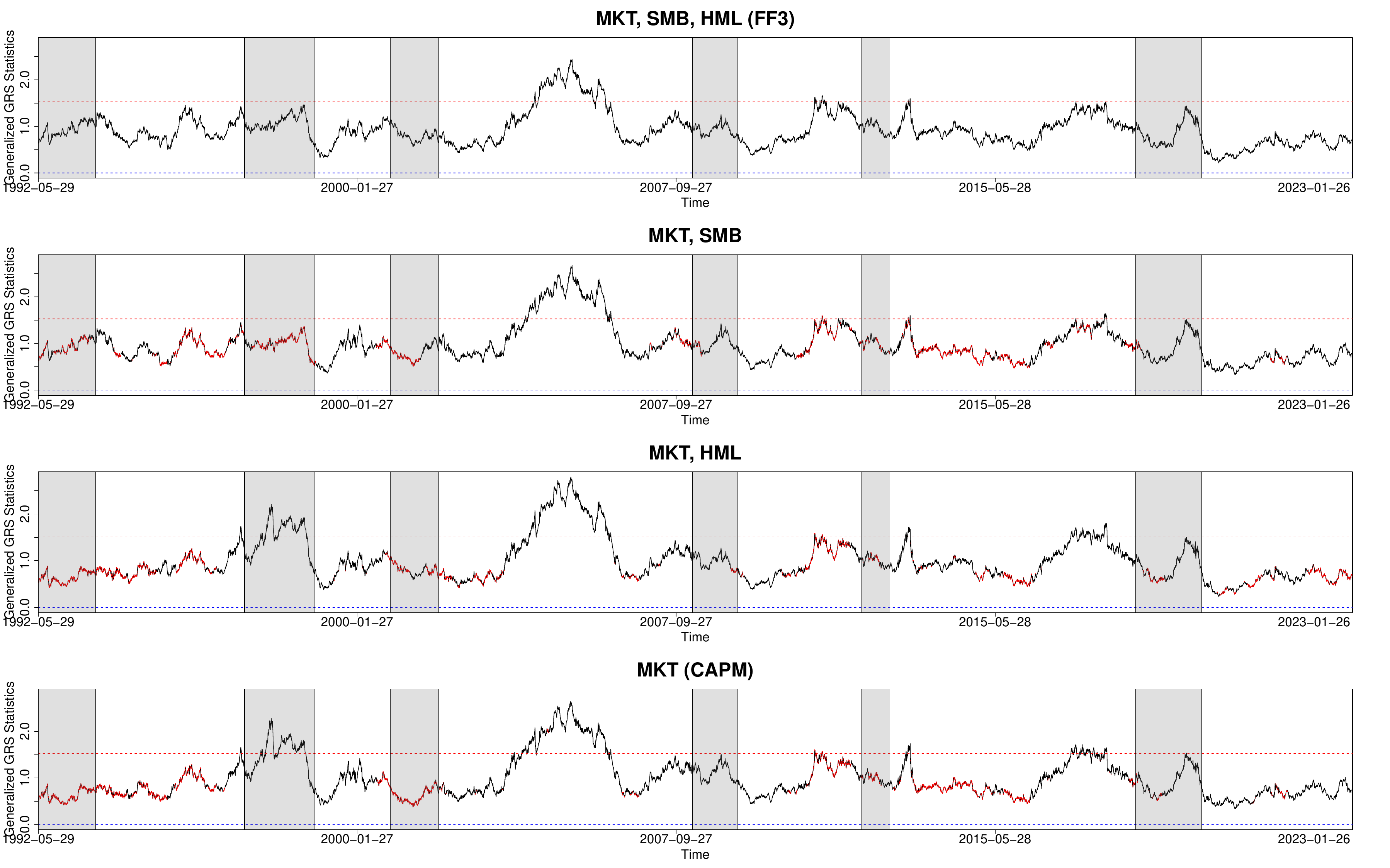}
\vspace*{5pt}
{
\begin{minipage}{600pt}
\scriptsize
\underline{Notes}:
\begin{itemize}
 \item[(1)] The dashed red lines denote the critical value at the 5\% significance levels for the generalized GRS test, and the dashed blue lines denote zero.
 \item[(2)] The red dots denote that the model outperforms the benchmark model at that time.
 \item[(3)] The shade areas represent recessions reported by the NBER ``\href{https://www.nber.org/research/business-cycle-dating}{Business Cycle Dating}.''
 \item[(4)] R version 4.4.0 was used to compute the statistics.
\end{itemize}
\end{minipage}}%
\end{center}
\end{figure}
\end{landscape}

\clearpage

\begin{landscape}
\begin{figure}[p]
 \caption{Generalized GRS Statistics and Factor Redundancy sorted by Size--B/M Portfolios (Japan, FF5)}\label{ff_tv_fig16}
 \begin{center}
  \includegraphics[scale=0.35]{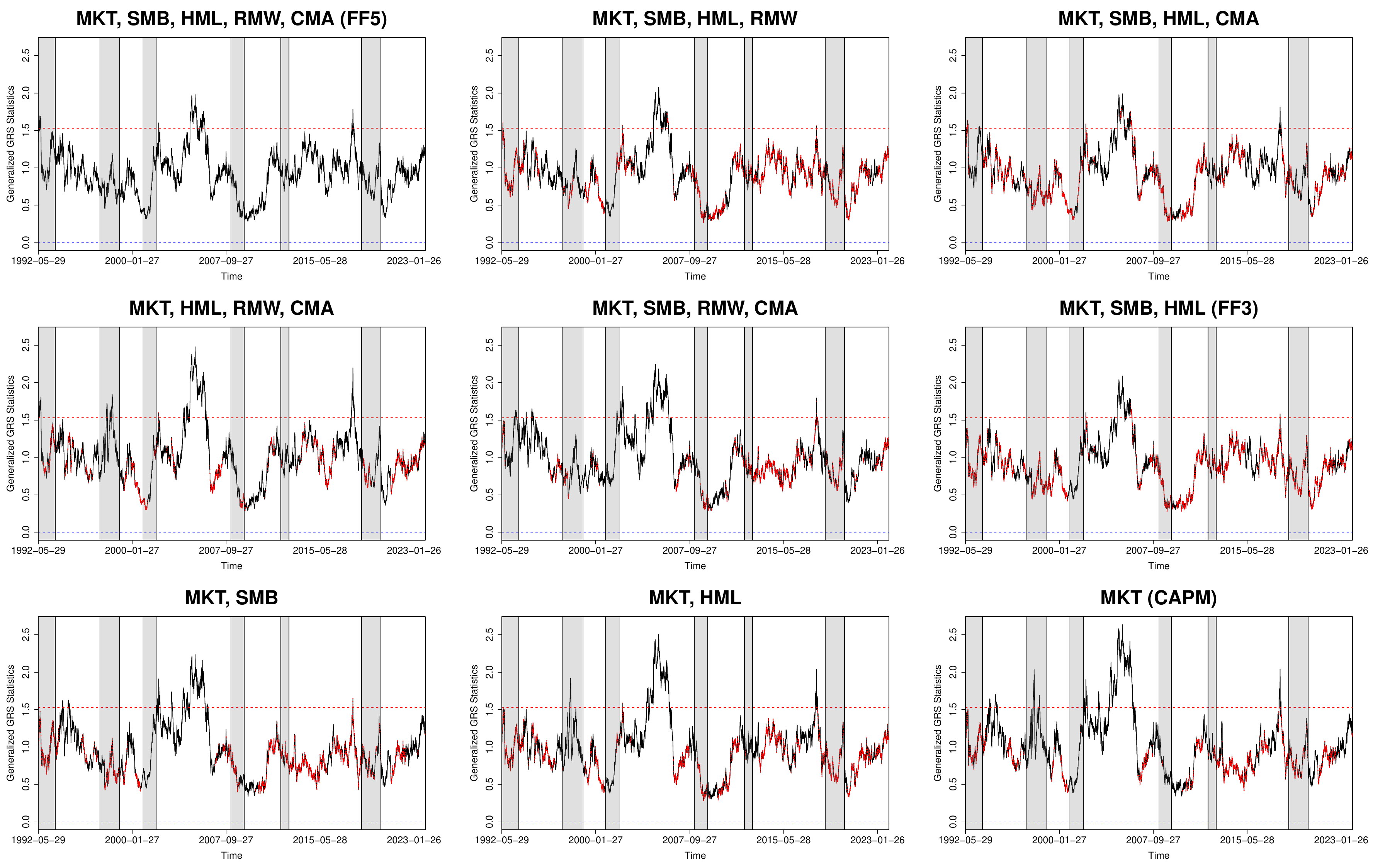}
\vspace*{5pt}
{
\begin{minipage}{600pt}
\scriptsize
\underline{Notes}:
\begin{itemize}
 \item[(1)] The dashed red lines denote the critical value at the 5\% significance levels for the generalized GRS test, and the dashed blue lines denote zero.
 \item[(2)] The red dots denote that the model outperforms the benchmark model at that time.
 \item[(3)] The shade areas represent recessions reported by the NBER ``\href{https://www.nber.org/research/business-cycle-dating}{Business Cycle Dating}.''
 \item[(4)] R version 4.4.0 was used to compute the statistics.
\end{itemize}
\end{minipage}}%
\end{center}
\end{figure}
\end{landscape}

\clearpage

\begin{landscape}
\begin{figure}[p]
 \caption{Generalized GRS Statistics and Factor Redundancy sorted by Size--OP Portfolios (Japan, FF5)}\label{ff_tv_fig17}
 \begin{center}
  \includegraphics[scale=0.35]{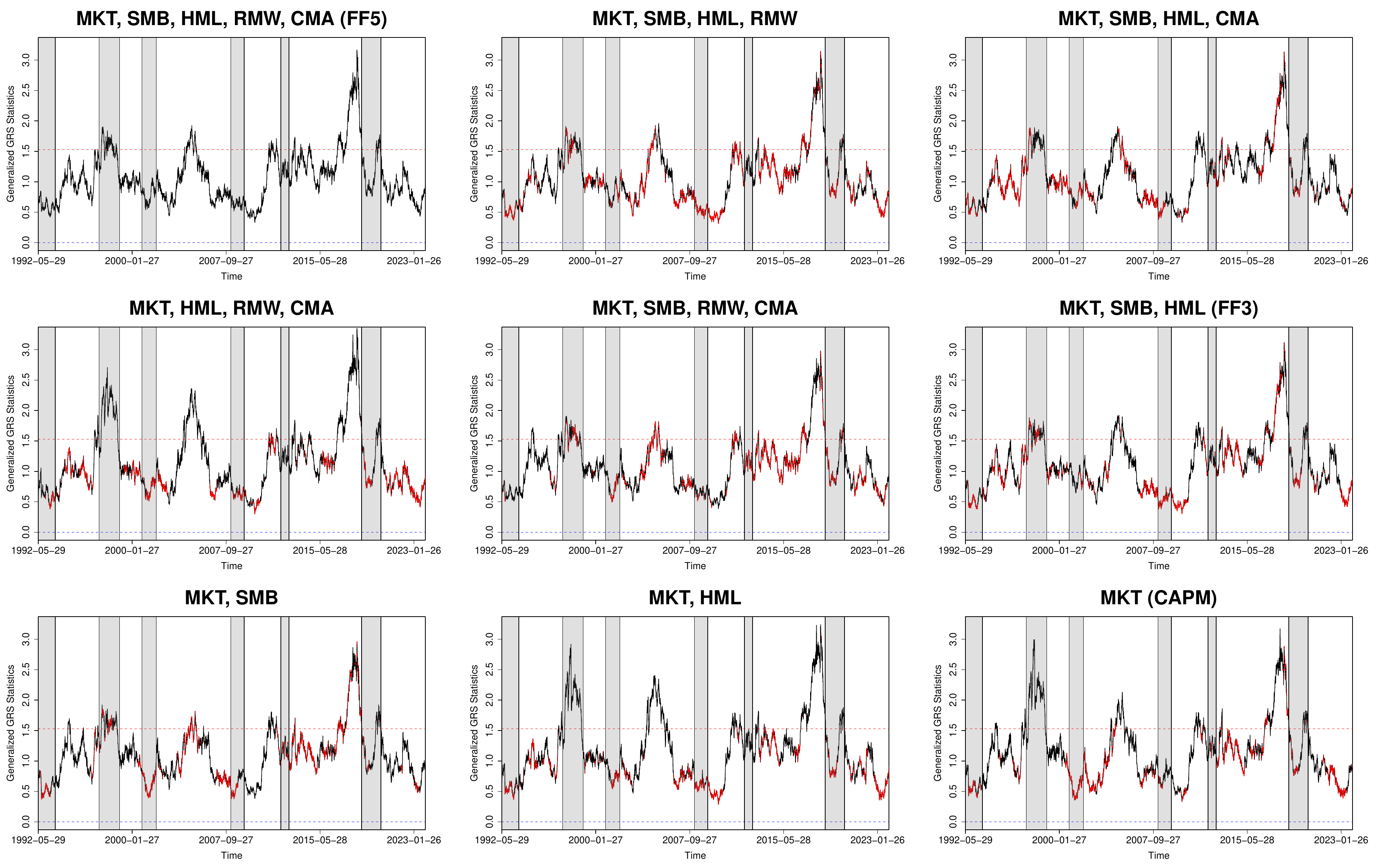}
\vspace*{5pt}
{
\begin{minipage}{600pt}
\scriptsize
\underline{Notes}:
\begin{itemize}
 \item[(1)] The dashed red lines denote the critical value at the 5\% significance levels for the generalized GRS test, and the dashed blue lines denote zero.
 \item[(2)] The red dots denote that the model outperforms the benchmark model at that time.
 \item[(3)] The shade areas represent recessions reported by the NBER ``\href{https://www.nber.org/research/business-cycle-dating}{Business Cycle Dating}.''
 \item[(4)] R version 4.4.0 was used to compute the statistics.
\end{itemize}
\end{minipage}}%
\end{center}
\end{figure}
\end{landscape}

\clearpage

\begin{landscape}
\begin{figure}[p]
 \caption{Generalized GRS Statistics and Factor Redundancy sorted by Size--Inv Portfolios (Japan, FF5)}\label{ff_tv_fig18}
 \begin{center}
  \includegraphics[scale=0.35]{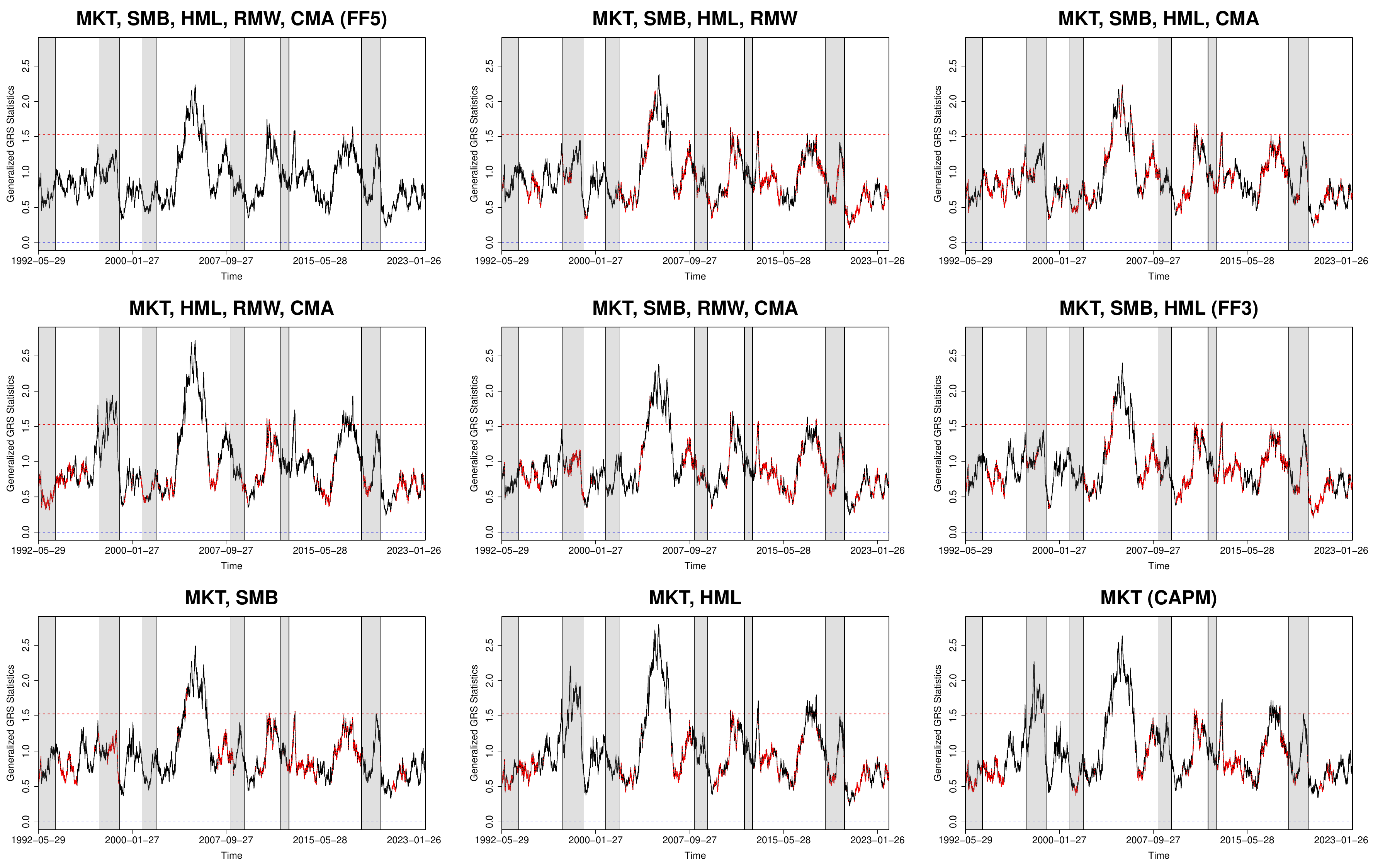}
\vspace*{5pt}
{
\begin{minipage}{600pt}
\scriptsize
\underline{Notes}:
\begin{itemize}
 \item[(1)] The dashed red lines denote the critical value at the 5\% significance levels for the generalized GRS test, and the dashed blue lines denote zero.
 \item[(2)] The red dots denote that the model outperforms the benchmark model at that time.
 \item[(3)] The shade areas represent recessions reported by the NBER ``\href{https://www.nber.org/research/business-cycle-dating}{Business Cycle Dating}.''
 \item[(4)] R version 4.4.0 was used to compute the statistics.
\end{itemize}
\end{minipage}}%
\end{center}
\end{figure}
\end{landscape}

\clearpage

\begin{table}[p]
\caption{Percentage of the periods of the model is valid (Japan)}\label{ff_tv_table7}
\centering
\begin{tabular}{c}
 \begin{minipage}[h]{\textwidth}
  \begin{center}
\scalebox{0.8}{\begin{tabular}{cccccccccccccc}\hline\hline
 & & \multicolumn{3}{c}{In the 1990's} & & \multicolumn{3}{c}{In the 2000's} & & \multicolumn{3}{c}{In the 2010's} & \\\cline{3-5}\cline{7-9}\cline{11-13}
 & & MB25 & MO25 & MI25 & & MB25 & MO25 & MI25 & & MB25 & MO25 & MI25 &\\\hline
MSHRC (FF5) & & ${\color{red}{0.9783}}$ & ${\color{red}{0.8844}}$ & ${\color{red}{1.0000}}$ & & ${\color{red}{0.9126}}$ & ${\color{red}{0.9322}}$ & ${\color{red}{0.8559}}$ & & ${\color{red}{0.9962}}$ & ${\color{red}{0.7220}}$ & ${\color{red}{0.9854}}$ &\\
MSHR & & ${\color{red}{0.9975}}$ & ${\color{red}{0.8698}}$ & ${\color{red}{1.0000}}$ & & ${\color{red}{0.8984}}$ & ${\color{red}{0.9214}}$ & ${\color{red}{0.8429}}$ & & ${\color{red}{1.0000}}$ & ${\color{red}{0.7546}}$ & ${\color{red}{0.9965}}$ &\\
MSHC & & ${\color{red}{0.9889}}$ & ${\color{red}{0.8486}}$ & ${\color{red}{1.0000}}$ & & ${\color{red}{0.9145}}$ & ${\color{red}{0.9268}}$ & ${\color{red}{0.8555}}$ & & ${\color{red}{0.9954}}$ & ${\color{red}{0.6840}}$ & ${\color{red}{0.9877}}$ &\\
MHRC & & ${\color{red}{0.9465}}$ & ${\color{red}{0.7895}}$ & ${\color{red}{0.8824}}$ & & ${\color{red}{0.8329}}$ & ${\color{red}{0.8521}}$ & ${\color{red}{0.8026}}$ & & ${\color{red}{0.9781}}$ & ${\color{red}{0.7209}}$ & ${\color{red}{0.9452}}$ &\\
MSRC & & ${\color{red}{0.9591}}$ & ${\color{red}{0.8460}}$ & ${\color{red}{1.0000}}$ & & ${\color{red}{0.8099}}$ & ${\color{red}{0.9621}}$ & ${\color{red}{0.8214}}$ & & ${\color{red}{0.9942}}$ & ${\color{red}{0.7435}}$ & ${\color{red}{0.9866}}$ &\\
MSH (FF3) & & ${\color{red}{0.9990}}$ & ${\color{red}{0.8571}}$ & ${\color{red}{1.0000}}$ & & ${\color{red}{0.8950}}$ & ${\color{red}{0.9172}}$ & ${\color{red}{0.8421}}$ & & ${\color{red}{1.0000}}$ & ${\color{red}{0.7297}}$ & ${\color{red}{0.9992}}$ &\\
MS & & ${\color{red}{0.9899}}$ & ${\color{red}{0.8334}}$ & ${\color{red}{1.0000}}$ & & ${\color{red}{0.8291}}$ & ${\color{red}{0.9686}}$ & ${\color{red}{0.8233}}$ & & ${\color{red}{1.0000}}$ & ${\color{red}{0.7880}}$ & ${\color{red}{1.0000}}$ &\\
MH & & ${\color{red}{0.9778}}$ & ${\color{red}{0.8067}}$ & ${\color{red}{0.8516}}$ & & ${\color{red}{0.8225}}$ & ${\color{red}{0.8421}}$ & ${\color{red}{0.8061}}$ & & ${\color{red}{0.9896}}$ & ${\color{red}{0.7051}}$ & ${\color{red}{0.9793}}$ &\\
M (CAPM) & & ${\color{red}{0.9647}}$ & ${\color{red}{0.8062}}$ & ${\color{red}{0.8556}}$ & & ${\color{red}{0.7735}}$ & ${\color{red}{0.9099}}$ & ${\color{red}{0.7919}}$ & & ${\color{red}{0.9942}}$ & ${\color{red}{0.7500}}$ & ${\color{red}{0.9927}}$ &\\\hline\hline
\end{tabular}}
\end{center}
\end{minipage}\\ 

\vspace{-3mm}\\

\begin{minipage}[h]{\textwidth}
 \begin{center}
  \scalebox{0.8}{\begin{tabular}{cccccccccccccc}\hline\hline
 & & \multicolumn{3}{c}{In the 2020's} & & \multicolumn{3}{c}{Whole period} & & \multicolumn{3}{c}{$-$} & \\\cline{3-5}\cline{7-9}\cline{11-13}
 & & MB25 & MO25 & MI25 & & MB25 & MO25 & MI25 & & MB25 & MO25 & MI25 & \\\hline
MSHRC (FF5) & & ${\color{red}{1.0000}}$ & ${\color{red}{0.9507}}$ & ${\color{red}{1.0000}}$ & & ${\color{red}{0.9654}}$ & ${\color{red}{0.8550}}$ & ${\color{red}{0.9490}}$ & & $-$ & $-$ & $-$ &\\
MSHR & & ${\color{red}{1.0000}}$ & ${\color{red}{0.9507}}$ & ${\color{red}{1.0000}}$ & & ${\color{red}{0.9667}}$ & ${\color{red}{0.8585}}$ & ${\color{red}{0.9483}}$ & & $-$ & $-$ & $-$ &\\
MSHC & & ${\color{red}{1.0000}}$ & ${\color{red}{0.9584}}$ & ${\color{red}{1.0000}}$ & & ${\color{red}{0.9683}}$ & ${\color{red}{0.8332}}$ & ${\color{red}{0.9496}}$ & & $-$ & $-$ & $-$ &\\
MHRC & & ${\color{red}{1.0000}}$ & ${\color{red}{0.9496}}$ & ${\color{red}{1.0000}}$ & & ${\color{red}{0.9261}}$ & ${\color{red}{0.8056}}$ & ${\color{red}{0.8901}}$ & & $-$ & $-$ & $-$ &\\
MSRC & & ${\color{red}{1.0000}}$ & ${\color{red}{0.9529}}$ & ${\color{red}{1.0000}}$ & & ${\color{red}{0.9270}}$ & ${\color{red}{0.8624}}$ & ${\color{red}{0.9382}}$ & & $-$ & $-$ & $-$ &\\
MSH (FF3) & & ${\color{red}{1.0000}}$ & ${\color{red}{0.9628}}$ & ${\color{red}{1.0000}}$ & & ${\color{red}{0.9660}}$ & ${\color{red}{0.8474}}$ & ${\color{red}{0.9490}}$ & & $-$ & $-$ & $-$ &\\
MS & & ${\color{red}{1.0000}}$ & ${\color{red}{0.9562}}$ & ${\color{red}{0.9989}}$ & & ${\color{red}{0.9425}}$ & ${\color{red}{0.8761}}$ & ${\color{red}{0.9430}}$ & & $-$ & $-$ & $-$ &\\
MH & & ${\color{red}{1.0000}}$ & ${\color{red}{0.9639}}$ & ${\color{red}{1.0000}}$ & & ${\color{red}{0.9342}}$ & ${\color{red}{0.8031}}$ & ${\color{red}{0.8947}}$ & & $-$ & $-$ & $-$ &\\
M (CAPM) & & ${\color{red}{1.0000}}$ & ${\color{red}{0.9814}}$ & ${\color{red}{0.9978}}$ & & ${\color{red}{0.9167}}$ & ${\color{red}{0.8412}}$ & ${\color{red}{0.8952}}$ & & $-$ & $-$ & $-$ &\\\hline\hline
\end{tabular}}
 \end{center}
\end{minipage}
\end{tabular}
\vspace*{5pt}
{
\begin{minipage}{420pt}
\scriptsize
\underline{Notes}:
\begin{itemize}
 \item[(1)] The highlighted values indicate that the model is valid over half the period.
 \item[(2)] R version 4.4.0 was used to compute the statistics.
\end{itemize}
\end{minipage}}
\end{table}

\clearpage

\begin{table}[p]
\caption{Model ranking results (Japan)}\label{ff_tv_table8}
\centering
\begin{tabular}{cc}
 \begin{minipage}[t]{0.45\textwidth}
  \begin{center}
   \scalebox{0.5}{\begin{tabular}{cccccccccc} \hline\hline
In the 1990's & & \multicolumn{3}{c}{FF3} & & \multicolumn{3}{c}{FF5} &\\\cline{3-5}\cline{7-9}
 & & MB25 & MO25 & MI25 & & MB25 & MO25 & MI25 &\\\hline
MSHRC (FF5) & & $-$ & $-$ & $-$ & & $-$ & $-$ & $-$ &\\
MSHR & & $-$ & $-$ & $-$ & & $0.3841$ & $0.3165$ & $0.2140$&\\
MSHC & & $-$ & $-$ & $-$ & & ${\color{red}{0.7042}}$ & ${\color{red}{0.6451}}$ & $0.4422$&\\
MHRC & & $-$ & $-$ & $-$ & & $0.4008$ & $0.3973$ & $0.4240$&\\
MSRC & & $-$ & $-$ & $-$ & & $0.0592$ & $0.0501$ & $0.1029$&\\
MSH (FF3) & & $-$ & $-$ & $-$ & & ${\color{red}{0.7259}}$ & ${\color{red}{0.5018}}$ & $0.2832$&\\
MS & & $0.3715$ & $0.3382$ & ${\color{red}{0.6572}}$ & & ${\color{red}{0.5391}}$ & $0.2867$ & $0.3705$&\\
MH & & $0.4881$ & $0.0949$ & ${\color{red}{0.5179}}$ & & ${\color{red}{0.5906}}$ & $0.3791$ & $0.3281$&\\
M (CAPM) & & $0.2948$ & $0.0660$ & ${\color{red}{0.5618}}$ & & $0.4321$ & $0.3786$ & $0.3831$&\\\hline\hline
  \end{tabular}}
 \end{center}
\end{minipage}
&
\begin{minipage}[t]{0.45\textwidth}
 \begin{center}
  \scalebox{0.5}{\begin{tabular}{cccccccccc} \hline\hline
In the 2000's & & \multicolumn{3}{c}{FF3} & & \multicolumn{3}{c}{FF5} &\\\cline{3-5}\cline{7-9}
 & & MB25 & MO25 & MI25 & & MB25 & MO25 & MI25 &\\\hline
MSHRC (FF5) & & $-$ & $-$ & $-$ & & $-$ & $-$ & $-$ &\\
MSHR & & $-$ & $-$ & $-$ & & $0.3687$ & ${\color{red}{0.6044}}$ & $0.4109$&\\
MSHC & & $-$ & $-$ & $-$ & & ${\color{red}{0.5332}}$ & ${\color{red}{0.5036}}$ & $0.4676$&\\
MHRC & & $-$ & $-$ & $-$ & & $0.3630$ & $0.4281$ & $0.3028$&\\
MSRC & & $-$ & $-$ & $-$ & & $0.0567$ & $0.1279$ & $0.0634$&\\
MSH (FF3) & & $-$ & $-$ & $-$ & & $0.4078$ & $0.4074$ & $0.3377$&\\
MS & & $0.1100$ & $0.4695$ & $0.2806$ & & $0.1867$ & ${\color{red}{0.5067}}$ & $0.1610$&\\
MH & & $0.4304$ & $0.1139$ & $0.3469$ & & $0.3197$ & $0.3772$ & $0.2844$&\\
M (CAPM) & &$0.1821$ & $0.1366$ & $0.2150$ & & $0.1234$ & $0.4377$ & $0.1142$&\\\hline\hline
  \end{tabular}}
 \end{center}
\end{minipage}\\ 

\vspace{-3mm}\\

\begin{minipage}[h]{0.45\textwidth}
 \begin{center}
  \scalebox{0.5}{\begin{tabular}{cccccccccc} \hline\hline
In the 2010's & & \multicolumn{3}{c}{FF3} & & \multicolumn{3}{c}{FF5} &\\\cline{3-5}\cline{7-9}
 & & MB25 & MO25 & MI25 & & MB25 & MO25 & MI25 &\\\hline
MSHRC (FF5) & &$-$ & $-$ & $-$ & & $-$ & $-$ & $-$\\
MSHR & &$-$ & $-$ & $-$ & & ${\color{red}{0.8202}}$ & ${\color{red}{0.6315}}$ & ${\color{red}{0.6656}}$&\\
MSHC & &$-$ & $-$ & $-$ & & $0.4317$ & $0.3566$ & $0.4847$&\\
MHRC & &$-$ & $-$ & $-$ & & $0.2584$ & ${\color{red}{0.5038}}$ & $0.3512$&\\
MSRC & &$-$ & $-$ & $-$ & & $0.2135$ & $0.2013$ & $0.1674$&\\
MSH (FF3) & &$-$ & $-$ & $-$ & & ${\color{red}{0.8163}}$ & ${\color{red}{0.5288}}$ & ${\color{red}{0.6883}}$&\\
MS & &${\color{red}{0.7170}}$ & ${\color{red}{0.6561}}$ & ${\color{red}{0.5817}}$ & & ${\color{red}{0.8554}}$ & ${\color{red}{0.5874}}$ & ${\color{red}{0.6116}}$&\\
MH & &$0.2703$ & $0.1551$ & $0.3171$ & & ${\color{red}{0.5583}}$ & $0.4087$ & $0.4620$&\\
M (CAPM) & &$0.4375$ & $0.1915$ & $0.3689$ & & ${\color{red}{0.6863}}$ & ${\color{red}{0.5564}}$ & ${\color{red}{0.5583}}$&\\\hline\hline
  \end{tabular}}
 \end{center}
\end{minipage}
&
\begin{minipage}[h]{0.45\textwidth}
 \begin{center}
  \scalebox{0.5}{\begin{tabular}{cccccccccc} \hline\hline
In the 2020's & & \multicolumn{3}{c}{FF3} & & \multicolumn{3}{c}{FF5} &\\\cline{3-5}\cline{7-9}
 & & MB25 & MO25 & MI25 & & MB25 & MO25 & MI25 &\\\hline
MSHRC (FF5) & & $-$ & $-$ & $-$ & & $-$ & $-$ & $-$&\\
MSHR & & $-$ & $-$ & $-$ & & ${\color{red}{0.7536}}$ & ${\color{red}{0.6637}}$ & ${\color{red}{0.5214}}$&\\
MSHC & & $-$ & $-$ & $-$ & & ${\color{red}{0.5663}}$ & $0.2990$ & $0.4600$&\\
MHRC & & $-$ & $-$ & $-$ & & ${\color{red}{0.6484}}$ & ${\color{red}{0.8007}}$ & $0.3702$&\\
MSRC & & $-$ & $-$ & $-$ & & $0.0150$ & $0.0432$ & $0.0460$&\\
MSH (FF3) & & $-$ & $-$ & $-$ & & ${\color{red}{0.7766}}$ & ${\color{red}{0.5301}}$ & ${\color{red}{0.5400}}$&\\
MS & & $0.1698$ & $0.2618$ & $0.0613$ & & $0.4359$ & $0.2267$ & $0.2311$&\\
MH & & ${\color{red}{0.5444}}$ & $0.0810$ & $0.2103$ & & ${\color{red}{0.8664}}$ & ${\color{red}{0.7185}}$ & $0.3691$&\\
M (CAPM) & & $0.3034$ & $0.0658$ & $0.1238$ & & $0.4688$ & ${\color{red}{0.5871}}$ & $0.2475$&\\\hline\hline
  \end{tabular}}
 \end{center}
\end{minipage}\\ 

\vspace{-3mm}\\

\begin{minipage}[h]{0.45\textwidth}
 \begin{center}
  \scalebox{0.5}{\begin{tabular}{cccccccccc} \hline\hline
Whole period & & \multicolumn{3}{c}{FF3} & & \multicolumn{3}{c}{FF5} &\\\cline{3-5}\cline{7-9}
 & & MB25 & MO25 & MI25 & & MB25 & MO25 & MI25 &\\\hline
MSHRC (FF5) & & $-$ & $-$ & $-$ & & $-$ & $-$ & $-$ &\\
MSHR & & $-$ & $-$ & $-$ & & ${\color{red}{0.5610}}$ & ${\color{red}{0.5495}}$ & $0.4572$&\\
MSHC & &$-$ & $-$ & $-$ & & ${\color{red}{0.5460}}$ & $0.4679$ & $0.4660$&\\
MHRC & &$-$ & $-$ & $-$ & & $0.3707$ & $0.4869$ & $0.3556$&\\
MSRC & &$-$ & $-$ & $-$ & & $0.3445$ & $0.4224$ & $0.3797$&\\
MSH (FF3) & &$-$ & $-$ & $-$ & & ${\color{red}{0.6584}}$ & $0.4833$ & $0.4599$&\\
MS & & $0.3758$ & $0.4740$ & $0.4447$ & & ${\color{red}{0.5158}}$ & $0.4474$ & $0.3649$& \\
MH & & $0.4059$ & $0.4450$ & $0.3637$ & & $0.5241$ & $0.4262$ & $0.3617$&\\
M (CAPM) & & $0.3054$ & $0.4599$ & $0.3389$ & & $0.4187$ & $0.4782$ & $0.3377$&\\\hline\hline
  \end{tabular}}
 \end{center}
\end{minipage}
&
\end{tabular}
\vspace*{5pt}
{
\begin{minipage}{420pt}
\scriptsize
\underline{Notes}:
\begin{itemize}
 \item[(1)] The highlighted values indicate that the model outperforms the benchmark model over the half period.
 \item[(2)] R version 4.4.0 was used to compute the statistics.
\end{itemize}
\end{minipage}}
\end{table}

\clearpage

\begin{landscape}
\begin{figure}[p]
 \caption{Generalized GRS Statistics and Factor Redundancy sorted by Size--B/M Portfolios (Europe, FF3)}\label{ff_tv_fig19}
 \begin{center}
  \includegraphics[scale=0.35]{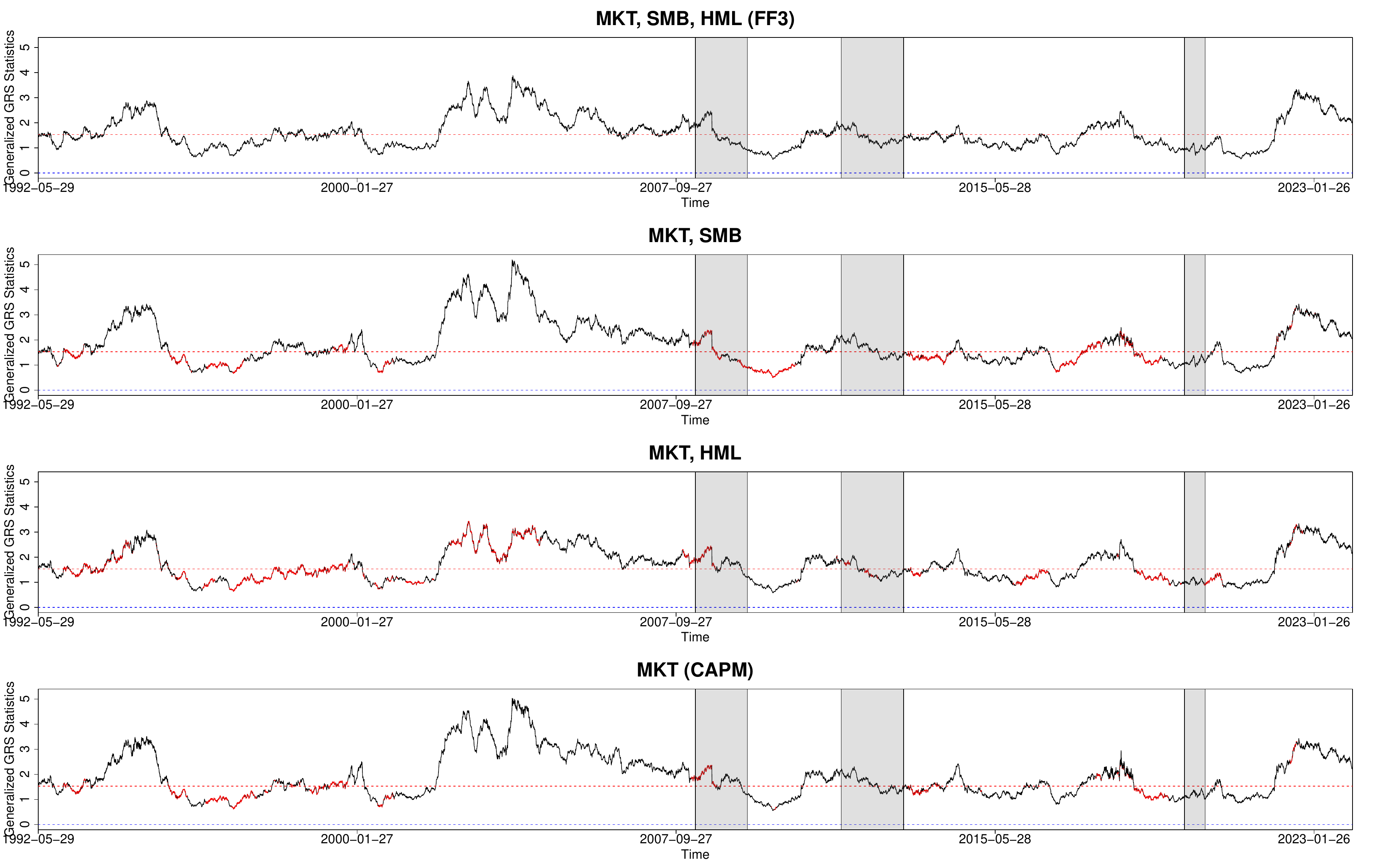}
\vspace*{5pt}
{
\begin{minipage}{600pt}
\scriptsize
\underline{Notes}:
\begin{itemize}
 \item[(1)] The dashed red lines denote the critical value at the 5\% significance levels for the generalized GRS test, and the dashed blue lines denote zero.
 \item[(2)] The red dots denote that the model outperforms the benchmark model at that time.
 \item[(3)] The shade areas represent recessions reported by the NBER ``\href{https://www.nber.org/research/business-cycle-dating}{Business Cycle Dating}.''
 \item[(4)] R version 4.4.0 was used to compute the statistics.
\end{itemize}
\end{minipage}}%
\end{center}
\end{figure}
\end{landscape}

\clearpage

\begin{landscape}
\begin{figure}[p]
 \caption{Generalized GRS Statistics and Factor Redundancy sorted by Size--OP Portfolios (Europe, FF3)}\label{ff_tv_fig20}
 \begin{center}
  \includegraphics[scale=0.35]{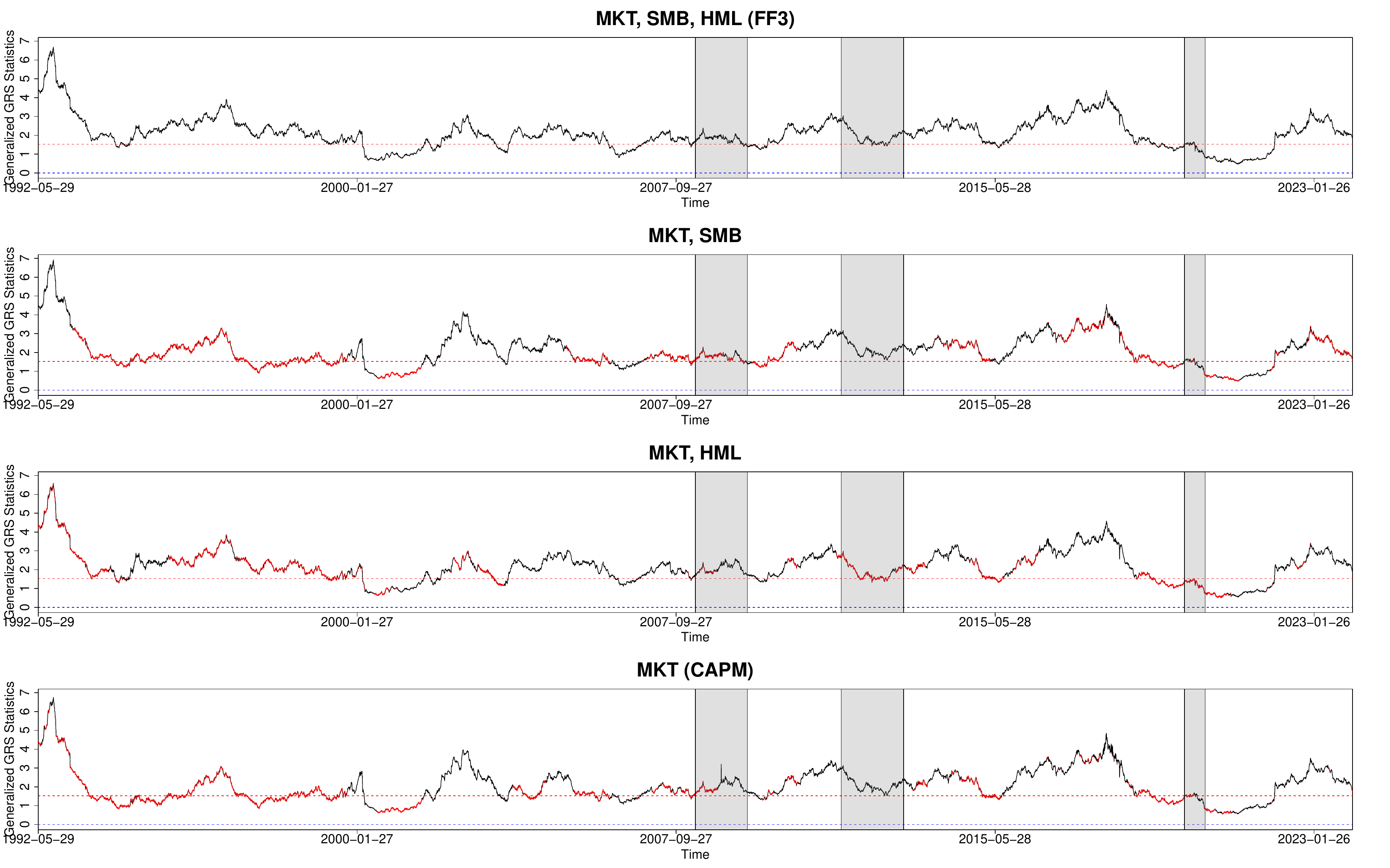}
\vspace*{5pt}
{
\begin{minipage}{600pt}
\scriptsize
\underline{Notes}:
\begin{itemize}
 \item[(1)] The dashed red lines denote the critical value at the 5\% significance levels for the generalized GRS test, and the dashed blue lines denote zero.
 \item[(2)] The red dots denote that the model outperforms the benchmark model at that time.
 \item[(3)] The shade areas represent recessions reported by the NBER ``\href{https://www.nber.org/research/business-cycle-dating}{Business Cycle Dating}.''
 \item[(4)] R version 4.4.0 was used to compute the statistics.
\end{itemize}
\end{minipage}}%
\end{center}
\end{figure}
\end{landscape}

\clearpage

\begin{landscape}
\begin{figure}[p]
 \caption{Generalized GRS Statistics and Factor Redundancy sorted by Size--Inv Portfolios (Europe, FF3)}\label{ff_tv_fig21}
 \begin{center}
  \includegraphics[scale=0.35]{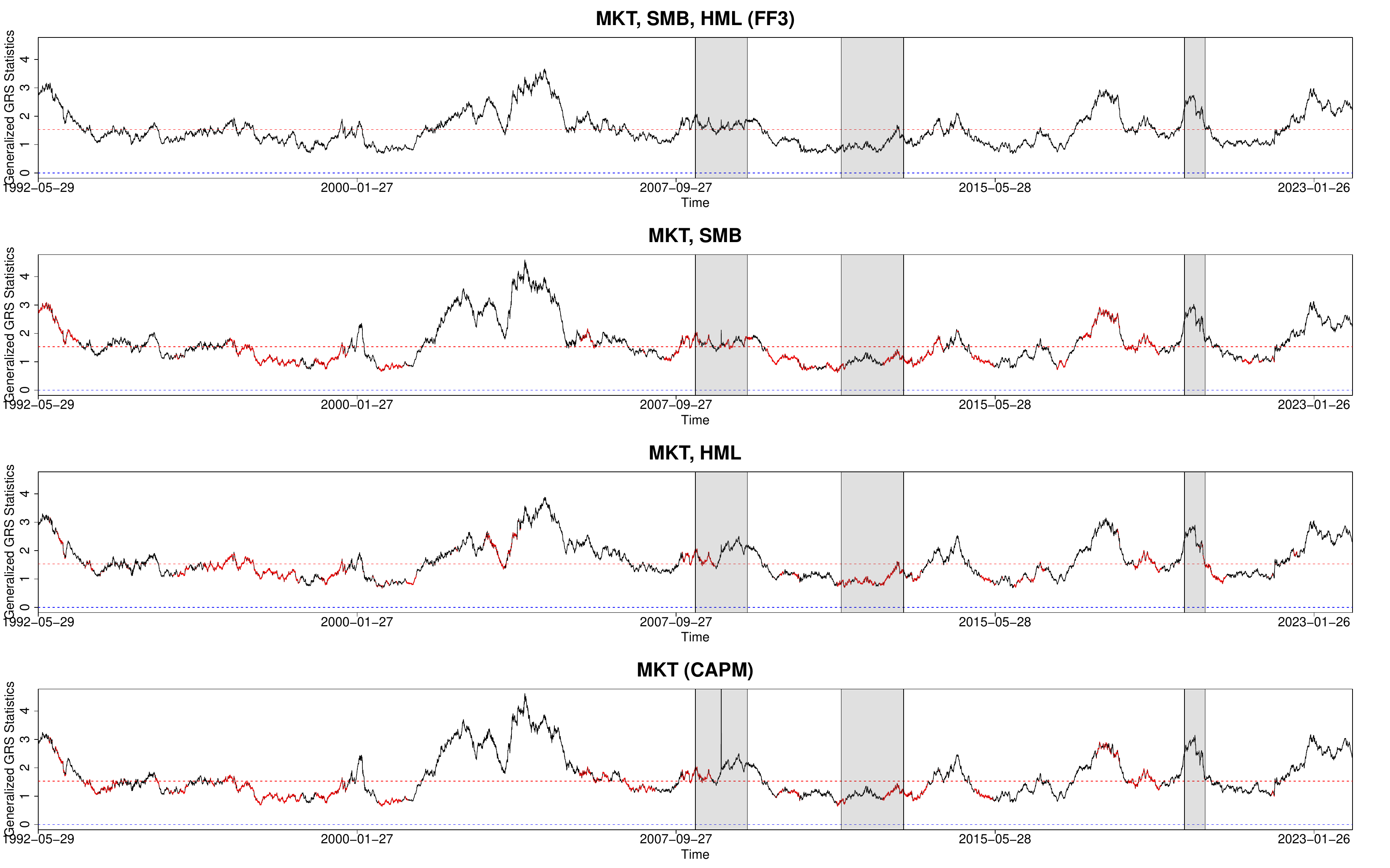}
\vspace*{5pt}
{
\begin{minipage}{600pt}
\scriptsize
\underline{Notes}:
\begin{itemize}
 \item[(1)] The dashed red lines denote the critical value at the 5\% significance levels for the generalized GRS test, and the dashed blue lines denote zero.
 \item[(2)] The red dots denote that the model outperforms the benchmark model at that time.
 \item[(3)] The shade areas represent recessions reported by the NBER ``\href{https://www.nber.org/research/business-cycle-dating}{Business Cycle Dating}.''
 \item[(4)] R version 4.4.0 was used to compute the statistics.
\end{itemize}
\end{minipage}}%
\end{center}
\end{figure}
\end{landscape}

\clearpage

\begin{landscape}
\begin{figure}[p]
 \caption{Generalized GRS Statistics and Factor Redundancy sorted by Size--B/M Portfolios (Europe, FF5)}\label{ff_tv_fig22}
 \begin{center}
  \includegraphics[scale=0.35]{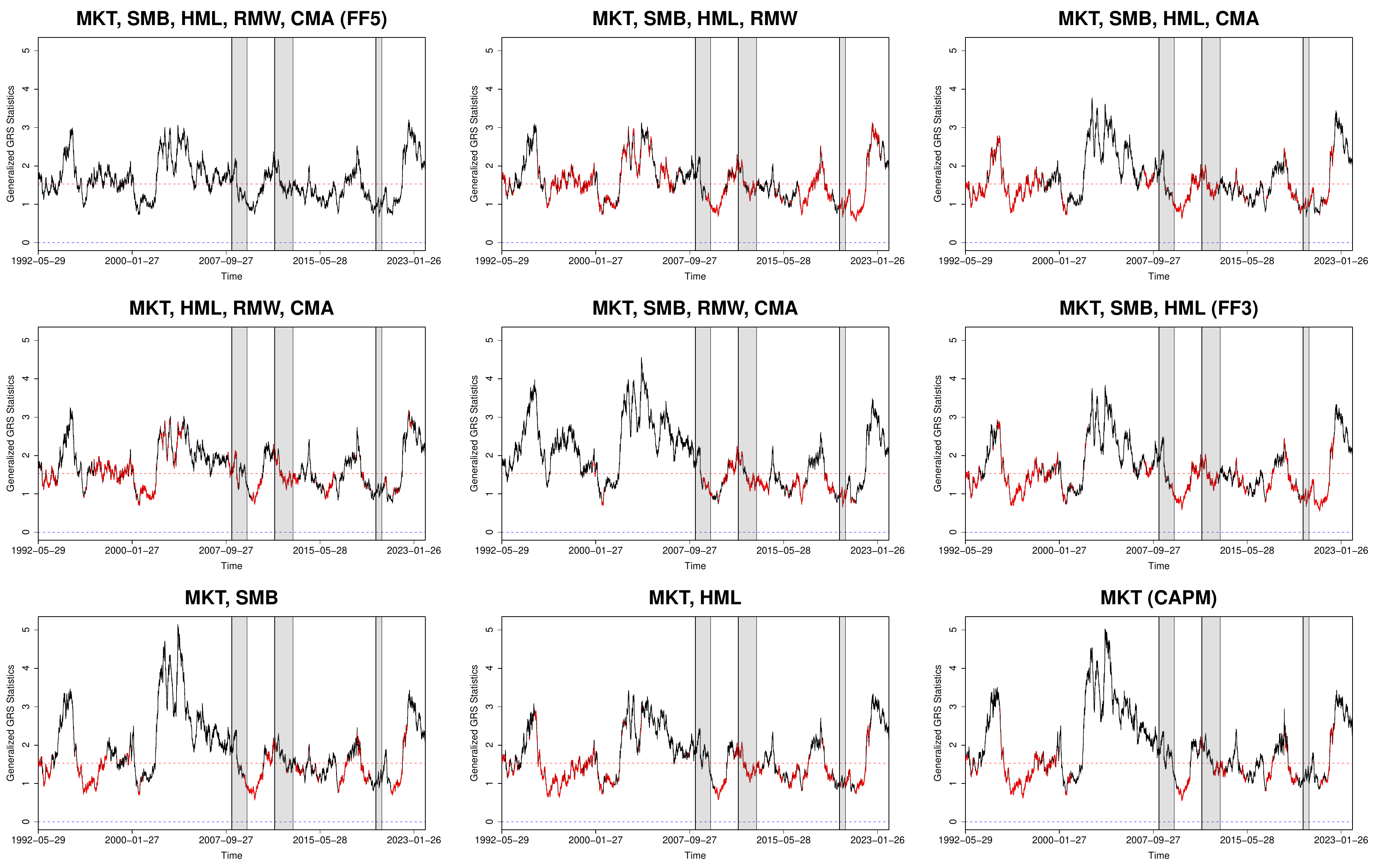}
\vspace*{5pt}
{
\begin{minipage}{600pt}
\scriptsize
\underline{Notes}:
\begin{itemize}
 \item[(1)] The dashed red lines denote the critical value at the 5\% significance levels for the generalized GRS test, and the dashed blue lines denote zero.
 \item[(2)] The red dots denote that the model outperforms the benchmark model at that time.
 \item[(3)] The shade areas represent recessions reported by the NBER ``\href{https://www.nber.org/research/business-cycle-dating}{Business Cycle Dating}.''
 \item[(4)] R version 4.4.0 was used to compute the statistics.
\end{itemize}
\end{minipage}}%
\end{center}
\end{figure}
\end{landscape}

\clearpage

\begin{landscape}
\begin{figure}[p]
 \caption{Generalized GRS Statistics and Factor Redundancy sorted by Size--OP Portfolios (Europe, FF5)}\label{ff_tv_fig23}
 \begin{center}
  \includegraphics[scale=0.35]{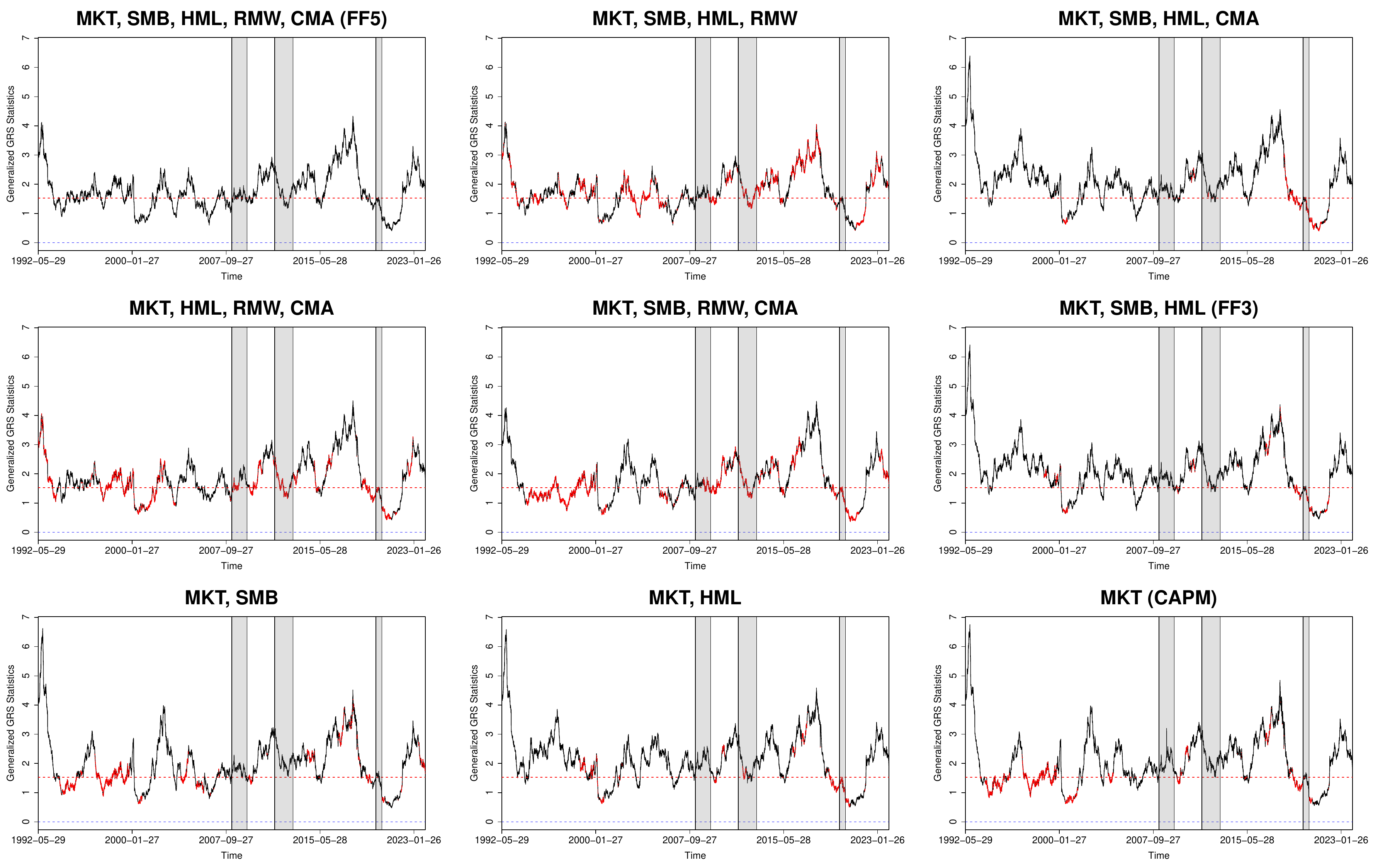}
\vspace*{5pt}
{
\begin{minipage}{600pt}
\scriptsize
\underline{Notes}:
\begin{itemize}
 \item[(1)] The dashed red lines denote the critical value at the 5\% significance levels for the generalized GRS test, and the dashed blue lines denote zero.
 \item[(2)] The red dots denote that the model outperforms the benchmark model at that time.
 \item[(3)] The shade areas represent recessions reported by the NBER ``\href{https://www.nber.org/research/business-cycle-dating}{Business Cycle Dating}.''
 \item[(4)] R version 4.4.0 was used to compute the statistics.
\end{itemize}
\end{minipage}}%
\end{center}
\end{figure}
\end{landscape}

\clearpage

\begin{landscape}
\begin{figure}[p]
 \caption{Generalized GRS Statistics and Factor Redundancy sorted by Size--Inv Portfolios (Europe, FF5)}\label{ff_tv_fig24}
 \begin{center}
  \includegraphics[scale=0.35]{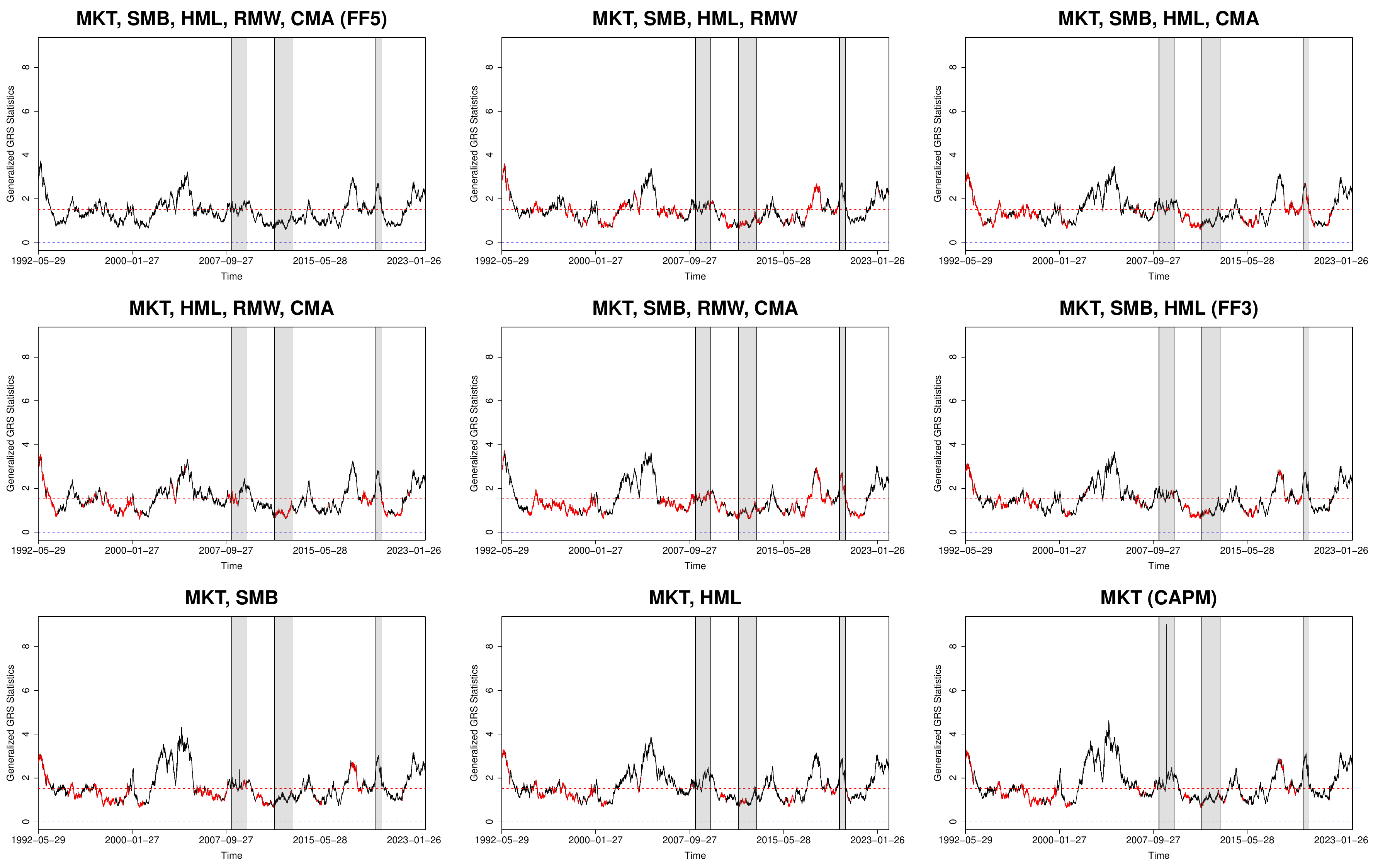}
\vspace*{5pt}
{
\begin{minipage}{600pt}
\scriptsize
\underline{Notes}:
\begin{itemize}
 \item[(1)] The dashed red lines denote the critical value at the 5\% significance levels for the generalized GRS test, and the dashed blue lines denote zero.
 \item[(2)] The red dots denote that the model outperforms the benchmark model at that time.
 \item[(3)] The shade areas represent recessions reported by the NBER ``\href{https://www.nber.org/research/business-cycle-dating}{Business Cycle Dating}.''
 \item[(4)] R version 4.4.0 was used to compute the statistics.
\end{itemize}
\end{minipage}}%
\end{center}
\end{figure}
\end{landscape}

\clearpage

\begin{table}[p]
\caption{Percentage of the periods of the model is valid (Europe)}\label{ff_tv_table9}
\centering
\begin{tabular}{c}
 \begin{minipage}[h]{\textwidth}
  \begin{center}
\scalebox{0.8}{\begin{tabular}{cccccccccccccc}\hline\hline
 & & \multicolumn{3}{c}{In the 1990's} & & \multicolumn{3}{c}{In the 2000's} & & \multicolumn{3}{c}{In the 2010's} & \\\cline{3-5}\cline{7-9}\cline{11-13}
 & & MB25 & MO25 & MI25 & & MB25 & MO25 & MI25 & & MB25 & MO25 & MI25 &\\\hline
MSHRC (FF5) & & $0.3695$ & $0.3084$ & ${\color{red}{0.6572}}$ & & $0.3933$ & ${\color{red}{0.5895}}$ & ${\color{red}{0.5063}}$ & & ${\color{red}{0.7124}}$ & $0.2166$ & ${\color{red}{0.8048}}$ &\\
MSHR & & $0.4119$ & $0.3024$ & ${\color{red}{0.5568}}$ & & $0.3913$ & ${\color{red}{0.5983}}$ & $0.4937$ & & ${\color{red}{0.7232}}$ & $0.2097$ & ${\color{red}{0.7803}}$ &\\
MSHC & & ${\color{red}{0.5689}}$ & $0.0515$ & ${\color{red}{0.7527}}$ & & $0.3607$ & $0.3637$ & $0.4017$ & & ${\color{red}{0.7182}}$ & $0.1603$ & ${\color{red}{0.8052}}$ &\\
MHRC & & $0.4260$ & $0.2923$ & ${\color{red}{0.6234}}$ & & $0.3043$ & ${\color{red}{0.5136}}$ & $0.4209$ & & ${\color{red}{0.6545}}$ & $0.2312$ & ${\color{red}{0.7634}}$ &\\
MSRC & & $0.0858$ & ${\color{red}{0.5583}}$ & ${\color{red}{0.7779}}$ & & $0.3128$ & ${\color{red}{0.5017}}$ & ${\color{red}{0.5010}}$ & & ${\color{red}{0.7063}}$ & $0.2189$ & ${\color{red}{0.7906}}$ &\\
MSH (FF3) & & ${\color{red}{0.5785}}$ & $0.0480$ & ${\color{red}{0.6734}}$ & & $0.3549$ & $0.3476$ & $0.3645$ & & ${\color{red}{0.7189}}$ & $0.1668$ & ${\color{red}{0.7788}}$ &\\
MS & & $0.4912$ & $0.4023$ & ${\color{red}{0.6623}}$ & & $0.3013$ & $0.4806$ & $0.4469$ & & ${\color{red}{0.6975}}$ & $0.1442$ & ${\color{red}{0.7630}}$ &\\
MH & & $0.6411$ & $0.0944$ & ${\color{red}{0.6986}}$ & & $0.2698$ & $0.2967$ & $0.3258$ & & ${\color{red}{0.6649}}$ & $0.1787$ & ${\color{red}{0.7285}}$ &\\
M (CAPM) & & ${\color{red}{0.5437}}$ & ${\color{red}{0.5250}}$ & ${\color{red}{0.7415}}$ & & $0.2545$ & $0.3634$ & $0.3292$ & & ${\color{red}{0.6745}}$ & $0.1572$ & ${\color{red}{0.7343}}$ &\\\hline\hline
\end{tabular}}
\end{center}
\end{minipage}\\ 

\vspace{-3mm}\\

\begin{minipage}[h]{\textwidth}
 \begin{center}
  \scalebox{0.8}{\begin{tabular}{cccccccccccccc}\hline\hline
 & & \multicolumn{3}{c}{In the 2020's} & & \multicolumn{3}{c}{Whole period} & & \multicolumn{3}{c}{$-$} & \\\cline{3-5}\cline{7-9}\cline{11-13}
 & & MB25 & MO25 & MI25 & & MB25 & MO25 & MI25 & & MB25 & MO25 & MI25 & \\\hline
MSHRC (FF5) & & ${\color{red}{0.6068}}$ & ${\color{red}{0.5433}}$ & ${\color{red}{0.5947}}$ & & ${\color{red}{0.5141}}$ & $0.3958$ & ${\color{red}{0.6491}}$ & & $-$ & $-$ & $-$ &\\
MSHR & & ${\color{red}{0.6068}}$ & ${\color{red}{0.5411}}$ & ${\color{red}{0.5761}}$ & & ${\color{red}{0.5273}}$ & $0.3946$ & ${\color{red}{0.6105}}$ & & $-$ & $-$ & $-$ &\\
MSHC & & ${\color{red}{0.6090}}$ & ${\color{red}{0.5487}}$ & ${\color{red}{0.5794}}$ & & ${\color{red}{0.5544}}$ & $0.2429$ & ${\color{red}{0.6372}}$ & & $-$ & $-$ & $-$ &\\
MHRC & & ${\color{red}{0.6057}}$ & ${\color{red}{0.5652}}$ & ${\color{red}{0.5564}}$ & & $0.4806$ & $0.3746$ & ${\color{red}{0.5957}}$ & & $-$ & $-$ & $-$ &\\
MSRC & & ${\color{red}{0.5904}}$ & ${\color{red}{0.5575}}$ & ${\color{red}{0.5257}}$ & & $0.4151$ & $0.4309$ & ${\color{red}{0.6645}}$ & & $-$ & $-$ & $-$ &\\
MSH (FF3)& & ${\color{red}{0.6177}}$ & ${\color{red}{0.5487}}$ & ${\color{red}{0.5706}}$ & & ${\color{red}{0.5562}}$ & $0.2389$ & ${\color{red}{0.5964}}$ & & $-$ & $-$ & $-$ &\\
MS & & ${\color{red}{0.5652}}$ & ${\color{red}{0.5334}}$ & ${\color{red}{0.5378}}$ & & ${\color{red}{0.5047}}$ & $0.3593$ & ${\color{red}{0.6114}}$ & & $-$ & $-$ & $-$ &\\
MH & & ${\color{red}{0.6057}}$ & ${\color{red}{0.6002}}$ & ${\color{red}{0.5214}}$ & & ${\color{red}{0.5253}}$ & $0.2435$ & ${\color{red}{0.5684}}$ & & $-$ & $-$ & $-$ &\\
M (CAPM) & & ${\color{red}{0.5794}}$ & ${\color{red}{0.5619}}$ & ${\color{red}{0.5345}}$ & & $0.4967$ & $0.3589$ & ${\color{red}{0.5833}}$ & & $-$ & $-$ & $-$ &\\\hline\hline
\end{tabular}}
 \end{center}
\end{minipage}
\end{tabular}
\vspace*{5pt}
{
\begin{minipage}{420pt}
\scriptsize
\underline{Notes}:
\begin{itemize}
 \item[(1)] The highlighted values indicate that the model is valid over half the period.
 \item[(2)] R version 4.4.0 was used to compute the statistics.
\end{itemize}
\end{minipage}}
\end{table}

\clearpage

\begin{table}[p]
\caption{Model ranking results (Europe)}\label{ff_tv_table10}
\centering
\begin{tabular}{cc}
 \begin{minipage}[t]{0.45\textwidth}
  \begin{center}
   \scalebox{0.5}{\begin{tabular}{cccccccccc} \hline\hline
In the 1990's & & \multicolumn{3}{c}{FF3} & & \multicolumn{3}{c}{FF5} &\\\cline{3-5}\cline{7-9}
 & & MB25 & MO25 & MI25 & & MB25 & MO25 & MI25 &\\\hline
MSHRC (FF5) & & $-$ & $-$ & $-$ & & $-$ & $-$ & $-$ &\\
MSHR & & $-$ & $-$ & $-$ & & ${\color{red}{0.6042}}$ & $0.4639$ & $0.3362$&\\
MSHC & & $-$ & $-$ & $-$ & & ${\color{red}{0.7708}}$ & $0.0091$ & ${\color{red}{0.6345}}$&\\
MHRC & & $-$ & $-$ & $-$ & & ${\color{red}{0.7062}}$ & ${\color{red}{0.6451}}$ & ${\color{red}{0.5482}}$&\\
MSRC & & $-$ & $-$ & $-$ & & $0.0021$ & $0.1782$ & $0.1636$&\\
MSH (FF3) & & $-$ & $-$ & $-$ & & ${\color{red}{0.7057}}$ & $0.0636$ & $0.4119$&\\
MS & & $0.2857$ & ${\color{red}{0.6586}}$ & $0.4955$ & & ${\color{red}{0.5194}}$ & ${\color{red}{0.5411}}$ & ${\color{red}{0.5659}}$&\\
MH & & ${\color{red}{0.5232}}$ & $0.4553$ & $0.4273$ & & ${\color{red}{0.8198}}$ & $0.1439$ & ${\color{red}{0.5114}}$&\\
M (CAPM) & & $0.3575$ & ${\color{red}{0.5544}}$ & $0.4019$ & & ${\color{red}{0.6668}}$ & ${\color{red}{0.6522}}$ & ${\color{red}{0.5664}}$&\\\hline\hline
  \end{tabular}}
 \end{center}
\end{minipage}
&
\begin{minipage}[t]{0.45\textwidth}
 \begin{center}
  \scalebox{0.5}{\begin{tabular}{cccccccccc} \hline\hline
In the 2000's & & \multicolumn{3}{c}{FF3} & & \multicolumn{3}{c}{FF5} &\\\cline{3-5}\cline{7-9}
 & & MB25 & MO25 & MI25 & & MB25 & MO25 & MI25 &\\\hline
MSHRC (FF5) & & $-$ & $-$ & $-$ & & $-$ & $-$ & $-$\\
MSHR & & $-$ & $-$ & $-$ & & ${\color{red}{0.5125}}$ & $0.3883$ & $0.4120$&\\
MSHC & & $-$ & $-$ & $-$ & & $0.1855$ & $0.0360$ & $0.0962$&\\
MHRC & & $-$ & $-$ & $-$ & & ${\color{red}{0.5217}}$ & $0.3350$ & $0.2580$&\\
MSRC & & $-$ & $-$ & $-$ & & $0.0354$ & $0.0485$ & $0.1091$&\\
MSH (FF3) & & $-$ & $-$ & $-$ & & $0.2158$ & $0.0491$ & $0.1161$&\\
MS & & $0.2741$ & ${\color{red}{0.8819}}$ & ${\color{red}{0.6032}}$ & & $0.1215$ & $0.1832$ & $0.2913$&\\
MH & & ${\color{red}{0.8319}}$ & $0.1896$ & ${\color{red}{0.7269}}$ & & $0.2771$ & $0.0755$ & $0.1426$&\\
M (CAPM) & & ${\color{red}{0.5457}}$ & $0.2193$ & ${\color{red}{0.6683}}$ & & $0.1062$ & $0.1913$ & $0.1276$&\\\hline\hline
  \end{tabular}}
 \end{center}
\end{minipage}\\ 

\vspace{-3mm}\\

\begin{minipage}[h]{0.45\textwidth}
 \begin{center}
  \scalebox{0.5}{\begin{tabular}{cccccccccc} \hline\hline
In the 2010's & & \multicolumn{3}{c}{FF3} & & \multicolumn{3}{c}{FF5} &\\\cline{3-5}\cline{7-9}
 & & MB25 & MO25 & MI25 & & MB25 & MO25 & MI25 &\\\hline
MSHRC (FF5) & &$-$ & $-$ & $-$ & & $-$ & $-$ & $-$ &\\
MSHR & & $-$ & $-$ & $-$ & & ${\color{red}{0.5870}}$ & ${\color{red}{0.5759}}$ & $0.4149$ &\\
MSHC & & $-$ & $-$ & $-$ & & $0.4935$ & $0.1545$ & $0.3842$&\\
MHRC & & $-$ & $-$ & $-$ & & $0.3298$ & $0.4640$ & $0.3282$&\\
MSRC & & $-$ & $-$ & $-$ & & $0.1835$ & $0.1600$ & $0.1412$&\\
MSH (FF3) & & $-$ & $-$ & $-$ & & ${\color{red}{0.6158}}$ & $0.2450$ & $0.4363$&\\
MS & & $0.1771$ & ${\color{red}{0.7156}}$ & ${\color{red}{0.5799}}$ & & ${\color{red}{0.5314}}$ & $0.2837$ & $0.3125$&\\
MH & & ${\color{red}{0.5351}}$ & $0.1139$ & $0.4009$ & & ${\color{red}{0.5729}}$ & $0.2324$ & $0.3439$&\\
M (CAPM) & & $0.2488$ & $0.1604$ &  $0.3948$ & &$0.4732$ & $0.2749$ & $0.3248$&\\\hline\hline
  \end{tabular}}
 \end{center}
\end{minipage}
&
\begin{minipage}[h]{0.45\textwidth}
 \begin{center}
  \scalebox{0.5}{\begin{tabular}{cccccccccc} \hline\hline
In the 2020's & & \multicolumn{3}{c}{FF3} & & \multicolumn{3}{c}{FF5} &\\\cline{3-5}\cline{7-9}
 & & MB25 & MO25 & MI25 & & MB25 & MO25 & MI25 &\\\hline
MSHRC (FF5) & & $-$ & $-$ & $-$ & & $-$ & $-$ & $-$&\\
MSHR & & $-$ & $-$ & $-$ & & ${\color{red}{0.9989}}$ & $0.4863$ & $0.0186$&\\
MSHC & & $-$ & $-$ & $-$ & & ${\color{red}{0.5137}}$ & $0.2552$ & $0.3713$&\\
MHRC & & $-$ & $-$ & $-$ & & $0.1512$ & $0.3308$ & $0.4425$&\\
MSRC & & $-$ & $-$ & $-$ & & $0.0201$ & $0.0534$ & $0.0393$&\\
MSH (FF3) & & $-$ & $-$ & $-$ & &${\color{red}{0.8379}}$ & $0.3100$ & $0.0646$&\\
MS & & $0.4992$ & ${\color{red}{0.5675}}$ & $0.4452$ & & $0.3406$ & $0.2508$ & $0.0350$&\\
MH & & $0.4103$ & $0.1223$ & $0.2615$ & & ${\color{red}{0.5367}}$ & $0.2322$ & $0.1610$&\\
M (CAPM) & & $0.4433$ & $0.1561$ & $0.3129$ & & $0.1643$ & $0.1347$ & $0.0526$&\\\hline\hline
  \end{tabular}}
 \end{center}
\end{minipage}\\ 

\vspace{-3mm}\\

\begin{minipage}[h]{0.45\textwidth}
 \begin{center}
  \scalebox{0.5}{\begin{tabular}{cccccccccc} \hline\hline
Whole period & & \multicolumn{3}{c}{FF3} & & \multicolumn{3}{c}{FF5} &\\\cline{3-5}\cline{7-9}
 & & MB25 & MO25 & MI25 & & MB25 & MO25 & MI25 &\\\hline
MSHRC (FF5) & & $-$ & $-$ & $-$ & & $-$ & $-$ & $-$ &\\
MSHR & & $-$ & $-$ & $-$ & & ${\color{red}{0.6136}}$ & $0.4781$ & $0.3501$ &\\
MSHC & & $-$ & $-$ & $-$ & & $0.4644$ & $0.0922$ & $0.3513$ &\\
MHRC & & $-$ & $-$ & $-$ & & $0.4633$ & $0.4517$ & $0.3722$ &\\
MSRC & & $-$ & $-$ & $-$ & & $0.2410$ & $0.4400$ & $0.4532$&\\
MSH (FF3) & & $-$ & $-$ & $-$ & & ${\color{red}{0.5341}}$ & $0.1450$ & $0.2855$&\\
MS & & $0.0110$ & $0.2716$ & $0.1643$ & & $0.3752$ & $0.3106$ & $0.3363$&\\
MH & & $0.1424$ & $0.0295$ & $0.3264$ & & ${\color{red}{0.5340}}$ & $0.1603$ & $0.2995$&\\
M (CAPM) & & $0.0153$ & $0.0186$ & $0.0986$ & & $0.3676$ & $0.3244$ & $0.2897$ &\\\hline\hline
  \end{tabular}}
 \end{center}
\end{minipage}
&
\end{tabular}
\vspace*{5pt}
{
\begin{minipage}{420pt}
\scriptsize
\underline{Notes}:
\begin{itemize}
 \item[(1)] The highlighted values indicate that the model outperforms the benchmark model over the half period.
 \item[(2)] R version 4.4.0 was used to compute the statistics.
\end{itemize}
\end{minipage}}
\end{table}